\documentclass[twocolumn,twocolappendix]{aastex701}

\usepackage{academicons}
\usepackage{amsmath}
\usepackage{bbm}
\usepackage{hyperref}
\usepackage{natbib}
\usepackage{xcolor}
\usepackage{xspace}
\usepackage{subfigure}
\usepackage{longtable}
\usepackage{booktabs}
\usepackage{threeparttable}
\usepackage{multirow}

\newcommand{\model}[1]{#1}
\newcommand{\e}{{\langle}e{\rangle}}

\newcommand{\Mstar}{\ensuremath{M_{\star}}\xspace}

\newcommand{\Msini}{\ensuremath{M_c \sin i}\xspace}

\newcommand{\Msun}{\ensuremath{M_{\odot}}\xspace}
\newcommand{\Mjup}{\ensuremath{M_{\mathrm{Jup}}}\xspace}

\newcommand{\K}{\ensuremath{\mathrm{K}}\xspace}

\newcommand{\Mearth}{\ensuremath{M_{\oplus}}\xspace}

\shortauthors{Van Zandt \& Bowler}

\graphicspath{{./}{}}

\newcommand{\McAllstarsPaperBoundsNstars}{\ensuremath{719}}

\newcommand{\McAllstarsPaperBoundsMedianMstar}{\ensuremath{0.90}}

\newcommand{\McAllstarsPaperBoundsNeffBinAZero}{\ensuremath{99.1}}
\newcommand{\McAllstarsPaperBoundsAvgComplBinAZero}{\ensuremath{0.77}}

\newcommand{\McAllstarsPaperBoundsPiecewiseIntOccBinaZero}{\ensuremath{0.19^{+0.02}_{-0.02}}}

\newcommand{\McAllstarsPaperBoundsSigmoidParamBOneBinaZero}{\ensuremath{0.075^{+0.012}_{-0.010}}}
\newcommand{\McAllstarsPaperBoundsSigmoidParamBTwoBinaZero}{\ensuremath{0.003^{+0.003}_{-0.002}}}
\newcommand{\McAllstarsPaperBoundsSigmoidParamCenterBinaZero}{\ensuremath{0.9^{+0.1}_{-0.1}}}
\newcommand{\McAllstarsPaperBoundsSigmoidParamWidthBinaZero}{\ensuremath{0.14^{+0.07}_{-0.06}}}
\newcommand{\McAllstarsPaperBoundsSigmoidIntOccBinaZero}{\ensuremath{0.20^{+0.02}_{-0.02}}}

\newcommand{\McAllstarsPaperBoundsLogGParamABinaZero}{\ensuremath{0.09^{+0.01}_{-0.01}}}
\newcommand{\McAllstarsPaperBoundsLogGParamMuBinaZero}{\ensuremath{0.23^{+0.08}_{-0.10}}}
\newcommand{\McAllstarsPaperBoundsLogGParamSigmaBinaZero}{\ensuremath{0.48^{+0.08}_{-0.05}}}
\newcommand{\McAllstarsPaperBoundsLogGIntOccBinaZero}{\ensuremath{0.19^{+0.02}_{-0.02}}}

\newcommand{\McAllstarsPaperBoundsLogLinearParamCOneBinaZero}{\ensuremath{0.08^{+0.02}_{-0.02}}}
\newcommand{\McAllstarsPaperBoundsLogLinearParamCTwoBinaZero}{\ensuremath{0.02^{+0.01}_{-0.01}}}

\newcommand{\McAllstarsPaperBoundsSigmoidBicBinaZero}{\ensuremath{832.6}}
\newcommand{\McAllstarsPaperBoundsLogGBicBinaZero}{\ensuremath{824.3}}

\newcommand{\McAllstarsPaperBoundsSigmoidLogGDbicBinaZero}{\ensuremath{8.3}}

\newcommand{\McAllstarsPaperBoundsNoGPEscarpmentParamDOneBinaZero}{\ensuremath{0.08^{+0.01}_{-0.01}}}
\newcommand{\McAllstarsPaperBoundsNoGPEscarpmentParamDTwoBinaZero}{\ensuremath{0.004^{+0.003}_{-0.002}}}
\newcommand{\McAllstarsPaperBoundsNoGPEscarpmentParamBPOneBinaZero}{\ensuremath{0.4^{+0.3}_{-0.2}}}
\newcommand{\McAllstarsPaperBoundsNoGPEscarpmentParamBPTwoBinaZero}{\ensuremath{1.3^{+0.1}_{-0.1}}}

\newcommand{\McAllstarsPaperBoundsNoGPEscarpmentIntOccBinaZero}{\ensuremath{0.20^{+0.02}_{-0.02}}}

\newcommand{\McAllstarsPaperBoundsNoGPEscarpmentBicBinaZero}{\ensuremath{830.9}}

\newcommand{\McHighMstarPaperBoundsNstars}{\ensuremath{229}}
\newcommand{\McHighMstarPaperBoundsNeff}{\ensuremath{60.1}}
\newcommand{\McHighMstarPaperBoundsAvgCompl}{\ensuremath{0.73}}
\newcommand{\McHighMstarPaperBoundsMedianMstar}{\ensuremath{1.11}}
\newcommand{\McHighMstarPaperBoundsMedianFeH}{\ensuremath{0.14}}
\newcommand{\McHighMstarPaperBoundsNeffBinAZero}{\ensuremath{60.1}}
\newcommand{\McHighMstarPaperBoundsAvgComplBinAZero}{\ensuremath{0.73}}

\newcommand{\McHighMstarPaperBoundsPiecewiseIntOccBinaZero}{\ensuremath{0.39^{+0.06}_{-0.05}}}

\newcommand{\McHighMstarPaperBoundsSigmoidParamBOneBinaZero}{\ensuremath{0.15^{+0.02}_{-0.02}}}
\newcommand{\McHighMstarPaperBoundsSigmoidParamBTwoBinaZero}{\ensuremath{0.007^{+0.008}_{-0.005}}}
\newcommand{\McHighMstarPaperBoundsSigmoidParamCenterBinaZero}{\ensuremath{0.94^{+0.08}_{-0.10}}}
\newcommand{\McHighMstarPaperBoundsSigmoidParamWidthBinaZero}{\ensuremath{0.10^{+0.06}_{-0.05}}}
\newcommand{\McHighMstarPaperBoundsSigmoidIntOccBinaZero}{\ensuremath{0.41^{+0.06}_{-0.05}}}

\newcommand{\McHighMstarPaperBoundsLogGParamABinaZero}{\ensuremath{0.20^{+0.04}_{-0.03}}}
\newcommand{\McHighMstarPaperBoundsLogGParamMuBinaZero}{\ensuremath{0.37^{+0.07}_{-0.08}}}
\newcommand{\McHighMstarPaperBoundsLogGParamSigmaBinaZero}{\ensuremath{0.40^{+0.07}_{-0.05}}}
\newcommand{\McHighMstarPaperBoundsLogGIntOccBinaZero}{\ensuremath{0.38^{+0.05}_{-0.05}}}

\newcommand{\McHighMstarPaperBoundsSigmoidBicBinaZero}{\ensuremath{421.1}}
\newcommand{\McHighMstarPaperBoundsLogGBicBinaZero}{\ensuremath{413.1}}

\newcommand{\McHighMstarPaperBoundsSigmoidLogGDbicBinaZero}{\ensuremath{8.1}}

\newcommand{\McLowMstarPaperBoundsNstars}{\ensuremath{490}}
\newcommand{\McLowMstarPaperBoundsNeff}{\ensuremath{39.0}}
\newcommand{\McLowMstarPaperBoundsAvgCompl}{\ensuremath{0.79}}
\newcommand{\McLowMstarPaperBoundsMedianMstar}{\ensuremath{0.79}}
\newcommand{\McLowMstarPaperBoundsMedianFeH}{\ensuremath{-0.03}}
\newcommand{\McLowMstarPaperBoundsNeffBinAZero}{\ensuremath{39.0}}
\newcommand{\McLowMstarPaperBoundsAvgComplBinAZero}{\ensuremath{0.79}}

\newcommand{\McLowMstarPaperBoundsPiecewiseIntOccBinaZero}{\ensuremath{0.11^{+0.02}_{-0.02}}}

\newcommand{\McLowMstarPaperBoundsSigmoidParamBOneBinaZero}{\ensuremath{0.08^{+0.04}_{-0.02}}}
\newcommand{\McLowMstarPaperBoundsSigmoidParamBTwoBinaZero}{\ensuremath{0.004^{+0.004}_{-0.003}}}
\newcommand{\McLowMstarPaperBoundsSigmoidParamCenterBinaZero}{\ensuremath{0.3^{+0.2}_{-0.4}}}
\newcommand{\McLowMstarPaperBoundsSigmoidParamWidthBinaZero}{\ensuremath{0.3^{+0.2}_{-0.2}}}
\newcommand{\McLowMstarPaperBoundsSigmoidIntOccBinaZero}{\ensuremath{0.12^{+0.02}_{-0.02}}}

\newcommand{\McLowMstarPaperBoundsLogGParamABinaZero}{\ensuremath{0.06^{+0.01}_{-0.01}}}
\newcommand{\McLowMstarPaperBoundsLogGParamMuBinaZero}{\ensuremath{-0.1^{+0.2}_{-0.2}}}
\newcommand{\McLowMstarPaperBoundsLogGParamSigmaBinaZero}{\ensuremath{0.6^{+0.1}_{-0.1}}}
\newcommand{\McLowMstarPaperBoundsLogGIntOccBinaZero}{\ensuremath{0.11^{+0.02}_{-0.02}}}

\newcommand{\McLowMstarPaperBoundsSigmoidBicBinaZero}{\ensuremath{376.9}}
\newcommand{\McLowMstarPaperBoundsLogGBicBinaZero}{\ensuremath{372.8}}

\newcommand{\McLowMstarPaperBoundsSigmoidLogGDbicBinaZero}{\ensuremath{4.1}}

\newcommand{\McHighFeHPaperBoundsNstars}{\ensuremath{395}}
\newcommand{\McHighFeHPaperBoundsNeff}{\ensuremath{83.2}}
\newcommand{\McHighFeHPaperBoundsAvgCompl}{\ensuremath{0.77}}
\newcommand{\McHighFeHPaperBoundsMedianMstar}{\ensuremath{0.98}}
\newcommand{\McHighFeHPaperBoundsMedianFeH}{\ensuremath{0.16}}
\newcommand{\McHighFeHPaperBoundsNeffBinAZero}{\ensuremath{83.2}}
\newcommand{\McHighFeHPaperBoundsAvgComplBinAZero}{\ensuremath{0.77}}

\newcommand{\McHighFeHPaperBoundsPiecewiseIntOccBinaZero}{\ensuremath{0.30^{+0.04}_{-0.03}}}

\newcommand{\McHighFeHPaperBoundsSigmoidParamBOneBinaZero}{\ensuremath{0.11^{+0.02}_{-0.02}}}
\newcommand{\McHighFeHPaperBoundsSigmoidParamBTwoBinaZero}{\ensuremath{0.004^{+0.005}_{-0.003}}}
\newcommand{\McHighFeHPaperBoundsSigmoidParamCenterBinaZero}{\ensuremath{0.92^{+0.10}_{-0.12}}}
\newcommand{\McHighFeHPaperBoundsSigmoidParamWidthBinaZero}{\ensuremath{0.13^{+0.06}_{-0.05}}}
\newcommand{\McHighFeHPaperBoundsSigmoidIntOccBinaZero}{\ensuremath{0.30^{+0.04}_{-0.03}}}

\newcommand{\McHighFeHPaperBoundsLogGParamABinaZero}{\ensuremath{0.15^{+0.02}_{-0.02}}}
\newcommand{\McHighFeHPaperBoundsLogGParamMuBinaZero}{\ensuremath{0.29^{+0.07}_{-0.09}}}
\newcommand{\McHighFeHPaperBoundsLogGParamSigmaBinaZero}{\ensuremath{0.43^{+0.07}_{-0.05}}}
\newcommand{\McHighFeHPaperBoundsLogGIntOccBinaZero}{\ensuremath{0.30^{+0.03}_{-0.03}}}

\newcommand{\McHighFeHPaperBoundsSigmoidBicBinaZero}{\ensuremath{625.8}}
\newcommand{\McHighFeHPaperBoundsLogGBicBinaZero}{\ensuremath{616.9}}

\newcommand{\McHighFeHPaperBoundsSigmoidLogGDbicBinaZero}{\ensuremath{8.8}}

\newcommand{\McLowFeHPaperBoundsNstars}{\ensuremath{324}}
\newcommand{\McLowFeHPaperBoundsNeff}{\ensuremath{15.7}}
\newcommand{\McLowFeHPaperBoundsAvgCompl}{\ensuremath{0.78}}
\newcommand{\McLowFeHPaperBoundsMedianMstar}{\ensuremath{0.80}}
\newcommand{\McLowFeHPaperBoundsMedianFeH}{\ensuremath{-0.15}}
\newcommand{\McLowFeHPaperBoundsNeffBinAZero}{\ensuremath{15.7}}
\newcommand{\McLowFeHPaperBoundsAvgComplBinAZero}{\ensuremath{0.78}}

\newcommand{\McLowFeHPaperBoundsPiecewiseIntOccBinaZero}{\ensuremath{0.07^{+0.02}_{-0.02}}}

\newcommand{\McLowFeHPaperBoundsSigmoidParamBOneBinaZero}{\ensuremath{0.05^{+0.04}_{-0.02}}}
\newcommand{\McLowFeHPaperBoundsSigmoidParamBTwoBinaZero}{\ensuremath{0.005^{+0.005}_{-0.003}}}
\newcommand{\McLowFeHPaperBoundsSigmoidParamCenterBinaZero}{\ensuremath{0.1^{+0.5}_{-0.3}}}
\newcommand{\McLowFeHPaperBoundsSigmoidParamWidthBinaZero}{\ensuremath{0.4^{+0.4}_{-0.2}}}
\newcommand{\McLowFeHPaperBoundsSigmoidIntOccBinaZero}{\ensuremath{0.07^{+0.02}_{-0.02}}}

\newcommand{\McLowFeHPaperBoundsLogGParamABinaZero}{\ensuremath{0.030^{+0.012}_{-0.009}}}
\newcommand{\McLowFeHPaperBoundsLogGParamMuBinaZero}{\ensuremath{-0.1^{+0.3}_{-0.2}}}
\newcommand{\McLowFeHPaperBoundsLogGParamSigmaBinaZero}{\ensuremath{0.7^{+0.3}_{-0.2}}}
\newcommand{\McLowFeHPaperBoundsLogGIntOccBinaZero}{\ensuremath{0.07^{+0.02}_{-0.02}}}

\newcommand{\McLowFeHPaperBoundsSigmoidBicBinaZero}{\ensuremath{173.4}}
\newcommand{\McLowFeHPaperBoundsLogGBicBinaZero}{\ensuremath{171.2}}

\newcommand{\McLowFeHPaperBoundsSigmoidLogGDbicBinaZero}{\ensuremath{2.2}}

\newcommand{\McHighActPaperBoundsNstars}{\ensuremath{188}}
\newcommand{\McHighActPaperBoundsNeff}{\ensuremath{18.3}}

\newcommand{\McHighActPaperBoundsNeffBinAZero}{\ensuremath{18.3}}
\newcommand{\McHighActPaperBoundsAvgComplBinAZero}{\ensuremath{0.75}}

\newcommand{\McHighActPaperBoundsPiecewiseIntOccBinaZero}{\ensuremath{0.14^{+0.03}_{-0.03}}}

\newcommand{\McHighActPaperBoundsSigmoidParamBOneBinaZero}{\ensuremath{0.06^{+0.04}_{-0.02}}}
\newcommand{\McHighActPaperBoundsSigmoidParamBTwoBinaZero}{\ensuremath{0.010^{+0.016}_{-0.008}}}
\newcommand{\McHighActPaperBoundsSigmoidParamCenterBinaZero}{\ensuremath{0.6^{+0.6}_{-0.7}}}
\newcommand{\McHighActPaperBoundsSigmoidParamWidthBinaZero}{\ensuremath{0.7^{+0.8}_{-0.5}}}
\newcommand{\McHighActPaperBoundsSigmoidIntOccBinaZero}{\ensuremath{0.15^{+0.04}_{-0.03}}}

\newcommand{\McHighActPaperBoundsLogGParamABinaZero}{\ensuremath{0.06^{+0.02}_{-0.02}}}
\newcommand{\McHighActPaperBoundsLogGParamMuBinaZero}{\ensuremath{0.3^{+0.2}_{-0.4}}}
\newcommand{\McHighActPaperBoundsLogGParamSigmaBinaZero}{\ensuremath{0.6^{+0.4}_{-0.2}}}
\newcommand{\McHighActPaperBoundsLogGIntOccBinaZero}{\ensuremath{0.14^{+0.03}_{-0.03}}}

\newcommand{\McHighActPaperBoundsSigmoidBicBinaZero}{\ensuremath{167.1}}
\newcommand{\McHighActPaperBoundsLogGBicBinaZero}{\ensuremath{168.1}}

\newcommand{\McLowActPaperBoundsNstars}{\ensuremath{190}}
\newcommand{\McLowActPaperBoundsNeff}{\ensuremath{53.0}}

\newcommand{\McLowActPaperBoundsNeffBinAZero}{\ensuremath{53.0}}
\newcommand{\McLowActPaperBoundsAvgComplBinAZero}{\ensuremath{0.81}}

\newcommand{\McLowActPaperBoundsPiecewiseIntOccBinaZero}{\ensuremath{0.36^{+0.05}_{-0.04}}}

\newcommand{\McLowActPaperBoundsSigmoidParamBOneBinaZero}{\ensuremath{0.14^{+0.03}_{-0.02}}}
\newcommand{\McLowActPaperBoundsSigmoidParamBTwoBinaZero}{\ensuremath{0.007^{+0.009}_{-0.005}}}
\newcommand{\McLowActPaperBoundsSigmoidParamCenterBinaZero}{\ensuremath{0.92^{+0.10}_{-0.14}}}
\newcommand{\McLowActPaperBoundsSigmoidParamWidthBinaZero}{\ensuremath{0.12^{+0.09}_{-0.06}}}
\newcommand{\McLowActPaperBoundsSigmoidIntOccBinaZero}{\ensuremath{0.38^{+0.05}_{-0.05}}}

\newcommand{\McLowActPaperBoundsLogGParamABinaZero}{\ensuremath{0.18^{+0.03}_{-0.03}}}
\newcommand{\McLowActPaperBoundsLogGParamMuBinaZero}{\ensuremath{0.33^{+0.07}_{-0.10}}}
\newcommand{\McLowActPaperBoundsLogGParamSigmaBinaZero}{\ensuremath{0.42^{+0.09}_{-0.06}}}
\newcommand{\McLowActPaperBoundsLogGIntOccBinaZero}{\ensuremath{0.36^{+0.05}_{-0.05}}}

\newcommand{\McLowActPaperBoundsSigmoidBicBinaZero}{\ensuremath{383.0}}
\newcommand{\McLowActPaperBoundsLogGBicBinaZero}{\ensuremath{375.8}}

\newcommand{\McCuiCutsCuiComparisonDiscussionNstars}{\ensuremath{576}}

\newcommand{\McCuiCutsCuiComparisonDiscussionPiecewiseIntOccBinmZero}{\ensuremath{3.5^{+1.1}_{-0.9}\%}}

\newcommand{\McCuiCutsCuiComparisonDiscussionPiecewiseIntOccBinmTwo}{\ensuremath{0.2^{+0.3}_{-0.2}\%}}

\newcommand{\QAllstarsPaperBoundsNstars}{\ensuremath{719}}

\newcommand{\QAllstarsPaperBoundsNeffBinAZero}{\ensuremath{101.2}}
\newcommand{\QAllstarsPaperBoundsAvgComplBinAZero}{\ensuremath{0.75}}

\newcommand{\QAllstarsPaperBoundsPiecewiseIntOccBinaZero}{\ensuremath{0.21^{+0.02}_{-0.02}}}

\newcommand{\QAllstarsPaperBoundsSigmoidParamBOneBinaZero}{\ensuremath{0.08^{+0.01}_{-0.01}}}
\newcommand{\QAllstarsPaperBoundsSigmoidParamBTwoBinaZero}{\ensuremath{0.002^{+0.002}_{-0.001}}}
\newcommand{\QAllstarsPaperBoundsSigmoidParamCenterBinaZero}{\ensuremath{-2.12^{+0.09}_{-0.12}}}
\newcommand{\QAllstarsPaperBoundsSigmoidParamWidthBinaZero}{\ensuremath{0.12^{+0.06}_{-0.06}}}
\newcommand{\QAllstarsPaperBoundsSigmoidIntOccBinaZero}{\ensuremath{0.21^{+0.02}_{-0.02}}}

\newcommand{\QAllstarsPaperBoundsLogGParamABinaZero}{\ensuremath{0.10^{+0.01}_{-0.01}}}
\newcommand{\QAllstarsPaperBoundsLogGParamMuBinaZero}{\ensuremath{-2.80^{+0.07}_{-0.10}}}
\newcommand{\QAllstarsPaperBoundsLogGParamSigmaBinaZero}{\ensuremath{0.45^{+0.07}_{-0.05}}}
\newcommand{\QAllstarsPaperBoundsLogGIntOccBinaZero}{\ensuremath{0.21^{+0.02}_{-0.02}}}

\newcommand{\QAllstarsPaperBoundsSigmoidBicBinaZero}{\ensuremath{827.2}}
\newcommand{\QAllstarsPaperBoundsLogGBicBinaZero}{\ensuremath{820.8}}

\newcommand{\QAllstarsPaperBoundsNoGPEscarpmentParamDOneBinaZero}{\ensuremath{0.08^{+0.01}_{-0.01}}}
\newcommand{\QAllstarsPaperBoundsNoGPEscarpmentParamDTwoBinaZero}{\ensuremath{0.002^{+0.003}_{-0.001}}}
\newcommand{\QAllstarsPaperBoundsNoGPEscarpmentParamBPOneBinaZero}{\ensuremath{-2.6^{+0.3}_{-0.2}}}
\newcommand{\QAllstarsPaperBoundsNoGPEscarpmentParamBPTwoBinaZero}{\ensuremath{-1.8^{+0.1}_{-0.1}}}

\newcommand{\QAllstarsPaperBoundsNoGPEscarpmentIntOccBinaZero}{\ensuremath{0.21^{+0.02}_{-0.02}}}

\newcommand{\QAllstarsPaperBoundsNoGPEscarpmentBicBinaZero}{\ensuremath{826.6}}

\newcommand{\QHighMstarPaperBoundsNstars}{\ensuremath{229}}

\newcommand{\QHighMstarPaperBoundsMedianMstar}{\ensuremath{1.11}}

\newcommand{\QHighMstarPaperBoundsNeffBinAZero}{\ensuremath{59.3}}
\newcommand{\QHighMstarPaperBoundsAvgComplBinAZero}{\ensuremath{0.75}}

\newcommand{\QHighMstarPaperBoundsPiecewiseIntOccBinaZero}{\ensuremath{0.38^{+0.05}_{-0.05}}}

\newcommand{\QHighMstarPaperBoundsSigmoidParamBOneBinaZero}{\ensuremath{0.15^{+0.02}_{-0.02}}}
\newcommand{\QHighMstarPaperBoundsSigmoidParamBTwoBinaZero}{\ensuremath{0.006^{+0.007}_{-0.004}}}
\newcommand{\QHighMstarPaperBoundsSigmoidParamCenterBinaZero}{\ensuremath{-2.10^{+0.07}_{-0.10}}}
\newcommand{\QHighMstarPaperBoundsSigmoidParamWidthBinaZero}{\ensuremath{0.09^{+0.07}_{-0.05}}}
\newcommand{\QHighMstarPaperBoundsSigmoidIntOccBinaZero}{\ensuremath{0.39^{+0.05}_{-0.04}}}

\newcommand{\QHighMstarPaperBoundsLogGParamABinaZero}{\ensuremath{0.19^{+0.03}_{-0.03}}}
\newcommand{\QHighMstarPaperBoundsLogGParamMuBinaZero}{\ensuremath{-2.71^{+0.07}_{-0.10}}}
\newcommand{\QHighMstarPaperBoundsLogGParamSigmaBinaZero}{\ensuremath{0.42^{+0.08}_{-0.05}}}
\newcommand{\QHighMstarPaperBoundsLogGIntOccBinaZero}{\ensuremath{0.37^{+0.05}_{-0.04}}}

\newcommand{\QHighMstarPaperBoundsSigmoidBicBinaZero}{\ensuremath{414.8}}
\newcommand{\QHighMstarPaperBoundsLogGBicBinaZero}{\ensuremath{411.0}}

\newcommand{\QLowMstarPaperBoundsNstars}{\ensuremath{490}}

\newcommand{\QLowMstarPaperBoundsMedianMstar}{\ensuremath{0.79}}

\newcommand{\QLowMstarPaperBoundsNeffBinAZero}{\ensuremath{41.9}}
\newcommand{\QLowMstarPaperBoundsAvgComplBinAZero}{\ensuremath{0.75}}

\newcommand{\QLowMstarPaperBoundsPiecewiseIntOccBinaZero}{\ensuremath{0.13^{+0.02}_{-0.02}}}

\newcommand{\QLowMstarPaperBoundsSigmoidParamBOneBinaZero}{\ensuremath{0.08^{+0.05}_{-0.02}}}
\newcommand{\QLowMstarPaperBoundsSigmoidParamBTwoBinaZero}{\ensuremath{0.003^{+0.004}_{-0.002}}}
\newcommand{\QLowMstarPaperBoundsSigmoidParamCenterBinaZero}{\ensuremath{-2.6^{+0.3}_{-0.5}}}
\newcommand{\QLowMstarPaperBoundsSigmoidParamWidthBinaZero}{\ensuremath{0.3^{+0.2}_{-0.1}}}
\newcommand{\QLowMstarPaperBoundsSigmoidIntOccBinaZero}{\ensuremath{0.14^{+0.02}_{-0.02}}}

\newcommand{\QLowMstarPaperBoundsLogGParamABinaZero}{\ensuremath{0.06^{+0.01}_{-0.01}}}
\newcommand{\QLowMstarPaperBoundsLogGParamMuBinaZero}{\ensuremath{-3.0^{+0.2}_{-0.2}}}
\newcommand{\QLowMstarPaperBoundsLogGParamSigmaBinaZero}{\ensuremath{0.6^{+0.1}_{-0.1}}}
\newcommand{\QLowMstarPaperBoundsLogGIntOccBinaZero}{\ensuremath{0.13^{+0.02}_{-0.02}}}

\newcommand{\QLowMstarPaperBoundsSigmoidBicBinaZero}{\ensuremath{391.7}}
\newcommand{\QLowMstarPaperBoundsLogGBicBinaZero}{\ensuremath{386.2}}

\newcommand{\QHighFeHPaperBoundsNstars}{\ensuremath{395}}

\newcommand{\QHighFeHPaperBoundsNeffBinAZero}{\ensuremath{86.1}}
\newcommand{\QHighFeHPaperBoundsAvgComplBinAZero}{\ensuremath{0.75}}

\newcommand{\QHighFeHPaperBoundsPiecewiseIntOccBinaZero}{\ensuremath{0.32^{+0.04}_{-0.03}}}

\newcommand{\QHighFeHPaperBoundsSigmoidParamBOneBinaZero}{\ensuremath{0.12^{+0.02}_{-0.02}}}
\newcommand{\QHighFeHPaperBoundsSigmoidParamBTwoBinaZero}{\ensuremath{0.003^{+0.005}_{-0.002}}}
\newcommand{\QHighFeHPaperBoundsSigmoidParamCenterBinaZero}{\ensuremath{-2.11^{+0.09}_{-0.11}}}
\newcommand{\QHighFeHPaperBoundsSigmoidParamWidthBinaZero}{\ensuremath{0.12^{+0.06}_{-0.05}}}
\newcommand{\QHighFeHPaperBoundsSigmoidIntOccBinaZero}{\ensuremath{0.33^{+0.04}_{-0.04}}}

\newcommand{\QHighFeHPaperBoundsLogGParamABinaZero}{\ensuremath{0.15^{+0.02}_{-0.02}}}
\newcommand{\QHighFeHPaperBoundsLogGParamMuBinaZero}{\ensuremath{-2.77^{+0.07}_{-0.10}}}
\newcommand{\QHighFeHPaperBoundsLogGParamSigmaBinaZero}{\ensuremath{0.44^{+0.07}_{-0.05}}}
\newcommand{\QHighFeHPaperBoundsLogGIntOccBinaZero}{\ensuremath{0.32^{+0.03}_{-0.03}}}

\newcommand{\QHighFeHPaperBoundsSigmoidBicBinaZero}{\ensuremath{634.1}}
\newcommand{\QHighFeHPaperBoundsLogGBicBinaZero}{\ensuremath{628.0}}

\newcommand{\QLowFeHPaperBoundsNstars}{\ensuremath{324}}

\newcommand{\QLowFeHPaperBoundsNeffBinAZero}{\ensuremath{15.1}}
\newcommand{\QLowFeHPaperBoundsAvgComplBinAZero}{\ensuremath{0.75}}

\newcommand{\QLowFeHPaperBoundsPiecewiseIntOccBinaZero}{\ensuremath{0.07^{+0.02}_{-0.02}}}

\newcommand{\QLowFeHPaperBoundsSigmoidParamBOneBinaZero}{\ensuremath{0.05^{+0.04}_{-0.02}}}
\newcommand{\QLowFeHPaperBoundsSigmoidParamBTwoBinaZero}{\ensuremath{0.003^{+0.004}_{-0.002}}}
\newcommand{\QLowFeHPaperBoundsSigmoidParamCenterBinaZero}{\ensuremath{-2.9^{+0.4}_{-0.4}}}
\newcommand{\QLowFeHPaperBoundsSigmoidParamWidthBinaZero}{\ensuremath{0.3^{+0.3}_{-0.2}}}
\newcommand{\QLowFeHPaperBoundsSigmoidIntOccBinaZero}{\ensuremath{0.08^{+0.02}_{-0.02}}}

\newcommand{\QLowFeHPaperBoundsLogGParamABinaZero}{\ensuremath{0.04^{+0.01}_{-0.01}}}
\newcommand{\QLowFeHPaperBoundsLogGParamMuBinaZero}{\ensuremath{-3.1^{+0.2}_{-0.2}}}
\newcommand{\QLowFeHPaperBoundsLogGParamSigmaBinaZero}{\ensuremath{0.6^{+0.2}_{-0.1}}}
\newcommand{\QLowFeHPaperBoundsLogGIntOccBinaZero}{\ensuremath{0.07^{+0.02}_{-0.02}}}

\newcommand{\QLowFeHPaperBoundsSigmoidBicBinaZero}{\ensuremath{162.4}}
\newcommand{\QLowFeHPaperBoundsLogGBicBinaZero}{\ensuremath{158.6}}

\newcommand{\QHighActPaperBoundsNstars}{\ensuremath{188}}

\newcommand{\QHighActPaperBoundsNeffBinAZero}{\ensuremath{18.4}}
\newcommand{\QHighActPaperBoundsAvgComplBinAZero}{\ensuremath{0.75}}

\newcommand{\QHighActPaperBoundsPiecewiseIntOccBinaZero}{\ensuremath{0.15^{+0.04}_{-0.03}}}

\newcommand{\QHighActPaperBoundsSigmoidParamBOneBinaZero}{\ensuremath{0.06^{+0.04}_{-0.02}}}
\newcommand{\QHighActPaperBoundsSigmoidParamBTwoBinaZero}{\ensuremath{0.011^{+0.017}_{-0.008}}}
\newcommand{\QHighActPaperBoundsSigmoidParamCenterBinaZero}{\ensuremath{-2.4^{+0.6}_{-0.7}}}
\newcommand{\QHighActPaperBoundsSigmoidParamWidthBinaZero}{\ensuremath{0.7^{+0.8}_{-0.5}}}
\newcommand{\QHighActPaperBoundsSigmoidIntOccBinaZero}{\ensuremath{0.15^{+0.04}_{-0.03}}}

\newcommand{\QHighActPaperBoundsLogGParamABinaZero}{\ensuremath{0.06^{+0.02}_{-0.02}}}
\newcommand{\QHighActPaperBoundsLogGParamMuBinaZero}{\ensuremath{-2.7^{+0.2}_{-0.4}}}
\newcommand{\QHighActPaperBoundsLogGParamSigmaBinaZero}{\ensuremath{0.6^{+0.4}_{-0.2}}}
\newcommand{\QHighActPaperBoundsLogGIntOccBinaZero}{\ensuremath{0.14^{+0.03}_{-0.03}}}

\newcommand{\QHighActPaperBoundsSigmoidBicBinaZero}{\ensuremath{169.2}}
\newcommand{\QHighActPaperBoundsLogGBicBinaZero}{\ensuremath{169.5}}

\newcommand{\QLowActPaperBoundsNstars}{\ensuremath{190}}

\newcommand{\QLowActPaperBoundsNeffBinAZero}{\ensuremath{53.0}}
\newcommand{\QLowActPaperBoundsAvgComplBinAZero}{\ensuremath{0.82}}

\newcommand{\QLowActPaperBoundsPiecewiseIntOccBinaZero}{\ensuremath{0.36^{+0.06}_{-0.05}}}

\newcommand{\QLowActPaperBoundsSigmoidParamBOneBinaZero}{\ensuremath{0.14^{+0.03}_{-0.02}}}
\newcommand{\QLowActPaperBoundsSigmoidParamBTwoBinaZero}{\ensuremath{0.007^{+0.009}_{-0.005}}}
\newcommand{\QLowActPaperBoundsSigmoidParamCenterBinaZero}{\ensuremath{-2.1^{+0.1}_{-0.1}}}
\newcommand{\QLowActPaperBoundsSigmoidParamWidthBinaZero}{\ensuremath{0.12^{+0.09}_{-0.06}}}
\newcommand{\QLowActPaperBoundsSigmoidIntOccBinaZero}{\ensuremath{0.38^{+0.06}_{-0.05}}}

\newcommand{\QLowActPaperBoundsLogGParamABinaZero}{\ensuremath{0.18^{+0.03}_{-0.03}}}
\newcommand{\QLowActPaperBoundsLogGParamMuBinaZero}{\ensuremath{-2.71^{+0.07}_{-0.10}}}
\newcommand{\QLowActPaperBoundsLogGParamSigmaBinaZero}{\ensuremath{0.42^{+0.08}_{-0.06}}}
\newcommand{\QLowActPaperBoundsLogGIntOccBinaZero}{\ensuremath{0.36^{+0.05}_{-0.05}}}

\newcommand{\QLowActPaperBoundsSigmoidBicBinaZero}{\ensuremath{382.2}}
\newcommand{\QLowActPaperBoundsLogGBicBinaZero}{\ensuremath{374.5}}

\newcommand{\McMstarPaperBoundsPiecewiseIntOccSignificanceBinaZero}{\ensuremath{5.3\,\sigma}}

\newcommand{\McMstarPaperBoundsSigmoidParamCenterDiffBinaZero}{\ensuremath{0.68}}
\newcommand{\McMstarPaperBoundsSigmoidParamCenterFactorBinaZero}{\ensuremath{4.8}}
\newcommand{\McMstarPaperBoundsSigmoidParamCenterSignificanceBinaZero}{\ensuremath{2.4\,\sigma}}

\newcommand{\McMstarPaperBoundsLogGParamMuDiffBinaZero}{\ensuremath{0.51}}
\newcommand{\McMstarPaperBoundsLogGParamMuFactorBinaZero}{\ensuremath{3.2}}
\newcommand{\McMstarPaperBoundsLogGParamMuSignificanceBinaZero}{\ensuremath{2.5\,\sigma}}

\newcommand{\McMstarPaperBoundsMedianMstarRatio}{\ensuremath{1.4}}
\newcommand{\McMstarPaperBoundsMedianFeHRatio}{\ensuremath{1.5}}

\newcommand{\McFeHPaperBoundsPiecewiseIntOccSignificanceBinaZero}{\ensuremath{6.0\,\sigma}}

\newcommand{\McFeHPaperBoundsSigmoidParamCenterFactorBinaZero}{\ensuremath{7}}
\newcommand{\McFeHPaperBoundsSigmoidParamCenterSignificanceBinaZero}{\ensuremath{1.5\,\sigma}}

\newcommand{\McFeHPaperBoundsLogGParamMuFactorBinaZero}{\ensuremath{2.7}}
\newcommand{\McFeHPaperBoundsLogGParamMuSignificanceBinaZero}{\ensuremath{1.5\,\sigma}}

\newcommand{\McFeHPaperBoundsMedianMstarRatio}{\ensuremath{1.2}}
\newcommand{\McFeHPaperBoundsMedianFeHRatio}{\ensuremath{2.0}}

\newcommand{\QMstarPaperBoundsSigmoidParamCenterDiffBinaZero}{\ensuremath{0.52}}

\newcommand{\QMstarPaperBoundsSigmoidParamCenterSignificanceBinaZero}{\ensuremath{1.6\,\sigma}}

\newcommand{\QMstarPaperBoundsLogGParamMuDiffBinaZero}{\ensuremath{0.33}}

\newcommand{\QMstarPaperBoundsLogGParamMuSignificanceBinaZero}{\ensuremath{1.6\,\sigma}}

\newcommand{\McStellarThreeParamsHighMstarHighFeHYoungNstars}{\ensuremath{47}}
\newcommand{\McStellarThreeParamsHighMstarHighFeHYoungNeff}{\ensuremath{4.0}}
\newcommand{\McStellarThreeParamsHighMstarHighFeHYoungAvgCompl}{\ensuremath{0.64}}
\newcommand{\McStellarThreeParamsHighMstarHighFeHYoungPiecewiseIntOcc}{\ensuremath{0.17^{+0.09}_{-0.06}}}
\newcommand{\McStellarThreeParamsHighMstarHighFeHYoungMedianMstar}{\ensuremath{1.08}}
\newcommand{\McStellarThreeParamsHighMstarHighFeHYoungMedianFeH}{\ensuremath{0.14}}
\newcommand{\McStellarThreeParamsHighMstarHighFeHYoungMedianAge}{\ensuremath{3.0}}

\newcommand{\McStellarThreeParamsHighMstarHighFeHOldNstars}{\ensuremath{86}}
\newcommand{\McStellarThreeParamsHighMstarHighFeHOldNeff}{\ensuremath{36.4}}
\newcommand{\McStellarThreeParamsHighMstarHighFeHOldAvgCompl}{\ensuremath{0.81}}
\newcommand{\McStellarThreeParamsHighMstarHighFeHOldPiecewiseIntOcc}{\ensuremath{0.55^{+0.10}_{-0.09}}}
\newcommand{\McStellarThreeParamsHighMstarHighFeHOldMedianMstar}{\ensuremath{1.10}}
\newcommand{\McStellarThreeParamsHighMstarHighFeHOldMedianFeH}{\ensuremath{0.23}}
\newcommand{\McStellarThreeParamsHighMstarHighFeHOldMedianAge}{\ensuremath{6.7}}

\newcommand{\McStellarThreeParamsHighMstarLowFeHYoungNstars}{\ensuremath{12}}
\newcommand{\McStellarThreeParamsHighMstarLowFeHYoungNeff}{\ensuremath{3.0}}
\newcommand{\McStellarThreeParamsHighMstarLowFeHYoungAvgCompl}{\ensuremath{0.73}}
\newcommand{\McStellarThreeParamsHighMstarLowFeHYoungPiecewiseIntOcc}{\ensuremath{0.4^{+0.2}_{-0.2}}}
\newcommand{\McStellarThreeParamsHighMstarLowFeHYoungMedianMstar}{\ensuremath{1.06}}
\newcommand{\McStellarThreeParamsHighMstarLowFeHYoungMedianFeH}{\ensuremath{-0.02}}
\newcommand{\McStellarThreeParamsHighMstarLowFeHYoungMedianAge}{\ensuremath{3.5}}

\newcommand{\McStellarThreeParamsHighMstarLowFeHOldNstars}{\ensuremath{25}}
\newcommand{\McStellarThreeParamsHighMstarLowFeHOldNeff}{\ensuremath{2.0}}
\newcommand{\McStellarThreeParamsHighMstarLowFeHOldAvgCompl}{\ensuremath{0.75}}
\newcommand{\McStellarThreeParamsHighMstarLowFeHOldPiecewiseIntOcc}{\ensuremath{0.14^{+0.10}_{-0.07}}}
\newcommand{\McStellarThreeParamsHighMstarLowFeHOldMedianMstar}{\ensuremath{1.05}}
\newcommand{\McStellarThreeParamsHighMstarLowFeHOldMedianFeH}{\ensuremath{-0.09}}
\newcommand{\McStellarThreeParamsHighMstarLowFeHOldMedianAge}{\ensuremath{7.0}}

\newcommand{\McStellarThreeParamsLowMstarHighFeHYoungNstars}{\ensuremath{72}}
\newcommand{\McStellarThreeParamsLowMstarHighFeHYoungNeff}{\ensuremath{9.3}}
\newcommand{\McStellarThreeParamsLowMstarHighFeHYoungAvgCompl}{\ensuremath{0.79}}
\newcommand{\McStellarThreeParamsLowMstarHighFeHYoungPiecewiseIntOcc}{\ensuremath{0.18^{+0.06}_{-0.05}}}
\newcommand{\McStellarThreeParamsLowMstarHighFeHYoungMedianMstar}{\ensuremath{0.92}}
\newcommand{\McStellarThreeParamsLowMstarHighFeHYoungMedianFeH}{\ensuremath{0.10}}
\newcommand{\McStellarThreeParamsLowMstarHighFeHYoungMedianAge}{\ensuremath{3.6}}

\newcommand{\McStellarThreeParamsLowMstarHighFeHOldNstars}{\ensuremath{39}}
\newcommand{\McStellarThreeParamsLowMstarHighFeHOldNeff}{\ensuremath{11.6}}
\newcommand{\McStellarThreeParamsLowMstarHighFeHOldAvgCompl}{\ensuremath{0.82}}
\newcommand{\McStellarThreeParamsLowMstarHighFeHOldPiecewiseIntOcc}{\ensuremath{0.4^{+0.1}_{-0.1}}}
\newcommand{\McStellarThreeParamsLowMstarHighFeHOldMedianMstar}{\ensuremath{0.95}}
\newcommand{\McStellarThreeParamsLowMstarHighFeHOldMedianFeH}{\ensuremath{0.17}}
\newcommand{\McStellarThreeParamsLowMstarHighFeHOldMedianAge}{\ensuremath{6.3}}

\newcommand{\McStellarThreeParamsLowMstarLowFeHYoungNstars}{\ensuremath{57}}
\newcommand{\McStellarThreeParamsLowMstarLowFeHYoungNeff}{\ensuremath{2.0}}
\newcommand{\McStellarThreeParamsLowMstarLowFeHYoungAvgCompl}{\ensuremath{0.80}}
\newcommand{\McStellarThreeParamsLowMstarLowFeHYoungPiecewiseIntOcc}{\ensuremath{0.06^{+0.04}_{-0.03}}}
\newcommand{\McStellarThreeParamsLowMstarLowFeHYoungMedianMstar}{\ensuremath{0.88}}
\newcommand{\McStellarThreeParamsLowMstarLowFeHYoungMedianFeH}{\ensuremath{-0.17}}
\newcommand{\McStellarThreeParamsLowMstarLowFeHYoungMedianAge}{\ensuremath{3.7}}

\newcommand{\McStellarThreeParamsLowMstarLowFeHOldNstars}{\ensuremath{40}}
\newcommand{\McStellarThreeParamsLowMstarLowFeHOldNeff}{\ensuremath{3.0}}
\newcommand{\McStellarThreeParamsLowMstarLowFeHOldAvgCompl}{\ensuremath{0.86}}
\newcommand{\McStellarThreeParamsLowMstarLowFeHOldPiecewiseIntOcc}{\ensuremath{0.11^{+0.06}_{-0.05}}}
\newcommand{\McStellarThreeParamsLowMstarLowFeHOldMedianMstar}{\ensuremath{0.92}}
\newcommand{\McStellarThreeParamsLowMstarLowFeHOldMedianFeH}{\ensuremath{-0.15}}
\newcommand{\McStellarThreeParamsLowMstarLowFeHOldMedianAge}{\ensuremath{6.1}}

\newcommand{\McStellarThreeParamsMassHighFeHOldSignificance}{\ensuremath{1.0\,\sigma}}
\newcommand{\McStellarThreeParamsMassHighFeHOldDynamicRange}{\ensuremath{1.2}}
\newcommand{\McStellarThreeParamsMassHighFeHYoungSignificance}{\ensuremath{0.1\,\sigma}}
\newcommand{\McStellarThreeParamsMassHighFeHYoungDynamicRange}{\ensuremath{1.2}}
\newcommand{\McStellarThreeParamsMassLowFeHOldSignificance}{\ensuremath{0.3\,\sigma}}
\newcommand{\McStellarThreeParamsMassLowFeHOldDynamicRange}{\ensuremath{1.1}}
\newcommand{\McStellarThreeParamsMassLowFeHYoungSignificance}{\ensuremath{2.4\,\sigma}}
\newcommand{\McStellarThreeParamsMassLowFeHYoungDynamicRange}{\ensuremath{1.2}}
\newcommand{\McStellarThreeParamsFeHHighMstarOldSignificance}{\ensuremath{2.6\,\sigma}}
\newcommand{\McStellarThreeParamsFeHHighMstarOldDynamicRange}{\ensuremath{2.1}}
\newcommand{\McStellarThreeParamsFeHHighMstarYoungSignificance}{\ensuremath{1.2\,\sigma}}
\newcommand{\McStellarThreeParamsFeHHighMstarYoungDynamicRange}{\ensuremath{1.4}}
\newcommand{\McStellarThreeParamsFeHLowMstarOldSignificance}{\ensuremath{2.4\,\sigma}}
\newcommand{\McStellarThreeParamsFeHLowMstarOldDynamicRange}{\ensuremath{2.1}}
\newcommand{\McStellarThreeParamsFeHLowMstarYoungSignificance}{\ensuremath{1.8\,\sigma}}
\newcommand{\McStellarThreeParamsFeHLowMstarYoungDynamicRange}{\ensuremath{1.9}}
\newcommand{\McStellarThreeParamsAgeHighMstarHighFeHSignificance}{\ensuremath{2.7\,\sigma}}
\newcommand{\McStellarThreeParamsAgeHighMstarHighFeHDynamicRange}{\ensuremath{2.2}}
\newcommand{\McStellarThreeParamsAgeHighMstarLowFeHSignificance}{\ensuremath{1.3\,\sigma}}
\newcommand{\McStellarThreeParamsAgeHighMstarLowFeHDynamicRange}{\ensuremath{2.0}}
\newcommand{\McStellarThreeParamsAgeLowMstarHighFeHSignificance}{\ensuremath{1.9\,\sigma}}
\newcommand{\McStellarThreeParamsAgeLowMstarHighFeHDynamicRange}{\ensuremath{1.8}}
\newcommand{\McStellarThreeParamsAgeLowMstarLowFeHSignificance}{\ensuremath{0.8\,\sigma}}
\newcommand{\McStellarThreeParamsAgeLowMstarLowFeHDynamicRange}{\ensuremath{1.6}}

\newcommand{\McStellarTwoParamsHighMstarHighFeHNstars}{\ensuremath{176}}
\newcommand{\McStellarTwoParamsHighMstarHighFeHNeff}{\ensuremath{53.9}}
\newcommand{\McStellarTwoParamsHighMstarHighFeHAvgCompl}{\ensuremath{0.74}}
\newcommand{\McStellarTwoParamsHighMstarHighFeHPiecewiseIntOcc}{\ensuremath{0.45^{+0.07}_{-0.06}}}
\newcommand{\McStellarTwoParamsHighMstarHighFeHMedianMstar}{\ensuremath{1.12}}
\newcommand{\McStellarTwoParamsHighMstarHighFeHMedianFeH}{\ensuremath{0.19}}

\newcommand{\McStellarTwoParamsHighMstarLowFeHNstars}{\ensuremath{53}}
\newcommand{\McStellarTwoParamsHighMstarLowFeHNeff}{\ensuremath{6.0}}
\newcommand{\McStellarTwoParamsHighMstarLowFeHAvgCompl}{\ensuremath{0.71}}
\newcommand{\McStellarTwoParamsHighMstarLowFeHPiecewiseIntOcc}{\ensuremath{0.18^{+0.08}_{-0.06}}}
\newcommand{\McStellarTwoParamsHighMstarLowFeHMedianMstar}{\ensuremath{1.10}}
\newcommand{\McStellarTwoParamsHighMstarLowFeHMedianFeH}{\ensuremath{-0.09}}

\newcommand{\McStellarTwoParamsLowMstarHighFeHNstars}{\ensuremath{219}}
\newcommand{\McStellarTwoParamsLowMstarHighFeHNeff}{\ensuremath{29.2}}
\newcommand{\McStellarTwoParamsLowMstarHighFeHAvgCompl}{\ensuremath{0.79}}
\newcommand{\McStellarTwoParamsLowMstarHighFeHPiecewiseIntOcc}{\ensuremath{0.19^{+0.04}_{-0.03}}}
\newcommand{\McStellarTwoParamsLowMstarHighFeHMedianMstar}{\ensuremath{0.82}}
\newcommand{\McStellarTwoParamsLowMstarHighFeHMedianFeH}{\ensuremath{0.11}}

\newcommand{\McStellarTwoParamsLowMstarLowFeHNstars}{\ensuremath{271}}
\newcommand{\McStellarTwoParamsLowMstarLowFeHNeff}{\ensuremath{9.8}}
\newcommand{\McStellarTwoParamsLowMstarLowFeHAvgCompl}{\ensuremath{0.80}}
\newcommand{\McStellarTwoParamsLowMstarLowFeHPiecewiseIntOcc}{\ensuremath{0.06^{+0.02}_{-0.02}}}
\newcommand{\McStellarTwoParamsLowMstarLowFeHMedianMstar}{\ensuremath{0.77}}
\newcommand{\McStellarTwoParamsLowMstarLowFeHMedianFeH}{\ensuremath{-0.17}}

\newcommand{\McStellarTwoParamsMassHighFeHSignificance}{\ensuremath{3.9\,\sigma}}

\newcommand{\McStellarTwoParamsMassHighFeHMedianMstarRatio}{\ensuremath{1.4}}
\newcommand{\McStellarTwoParamsMassHighFeHMedianFeHRatio}{\ensuremath{1.2}}
\newcommand{\McStellarTwoParamsMassLowFeHSignificance}{\ensuremath{2.2\,\sigma}}

\newcommand{\McStellarTwoParamsMassLowFeHMedianMstarRatio}{\ensuremath{1.4}}
\newcommand{\McStellarTwoParamsMassLowFeHMedianFeHRatio}{\ensuremath{1.2}}
\newcommand{\McStellarTwoParamsFeHHighMstarSignificance}{\ensuremath{2.5\,\sigma}}

\newcommand{\McStellarTwoParamsFeHHighMstarMedianMstarRatio}{\ensuremath{1.0}}
\newcommand{\McStellarTwoParamsFeHHighMstarMedianFeHRatio}{\ensuremath{1.9}}
\newcommand{\McStellarTwoParamsFeHLowMstarSignificance}{\ensuremath{3.6\,\sigma}}

\newcommand{\McStellarTwoParamsFeHLowMstarMedianMstarRatio}{\ensuremath{1.1}}
\newcommand{\McStellarTwoParamsFeHLowMstarMedianFeHRatio}{\ensuremath{1.9}}

\begin{document}

\title{The Giant Planet Mass Function Across Stellar Mass and Metallicity:\\ Massive Stars Make Massive Planets}

\correspondingauthor{Judah Van Zandt}
\email{judahvz@ucsb.edu}

\author[0000-0002-4290-6826]{Judah Van Zandt}
\affiliation{Department of Physics \& Astronomy, University of California Santa Barbara, Santa Barbara, CA 93106, USA}
\email{judahvz@ucsb.edu}

\author[0000-0003-2649-2288]{Brendan P. Bowler}
\affiliation{Department of Physics \& Astronomy, University of California Santa Barbara, Santa Barbara, CA 93106, USA}
\email{bpbowler@ucsb.edu}

\begin{abstract}
Although giant planet occurrence correlates with stellar mass and metallicity, it remains unclear whether these properties also shape the companion mass distribution. We investigate this question using 719 FGKM stars from the California Legacy Survey and companion true-mass constraints from radial velocities and $Hipparcos$--$Gaia$ astrometry. Our hierarchical occurrence analysis accounts for survey completeness and companion parameter uncertainties over giant planet and brown dwarf masses of $M_c=0.4$--$50~\Mjup$ and semi-major axes of $a=0.1$--$10$~AU. We find that, compared to the rate among our full sample of $\McAllstarsPaperBoundsPiecewiseIntOccBinaZero$ planets per star in this interval, giant planet occurrence is significantly enhanced around stars with high mass ($\Mstar>1~\Msun$; $\McHighMstarPaperBoundsPiecewiseIntOccBinaZero$ planets per star) and high metallicity ([Fe/H]$>0.0$; $\McHighFeHPaperBoundsPiecewiseIntOccBinaZero$ planets per star). We find evidence at the $2-3\sigma$ level that the planet mass function shape varies with stellar mass, with higher characteristic planet masses around stars with $\Mstar>1~\Msun$ than around lower-mass hosts, based on differences in both the CDFs of fitted models and constraints on model shape parameters. For the full sample, occurrence declines above a few Jupiter masses and is consistent with either a plateau below $\sim3~\Mjup$ or a peak near $1$--$2~\Mjup$. The brown dwarf desert begins at a companion-to-host mass ratio of $0.013$--$0.02$. These results suggest that stellar properties influence not only the frequency of giant companions but also their characteristic masses. Within the core accretion framework, this shift could reflect larger gas reservoirs around more massive stars or earlier core formation in metal-rich disks, allowing greater gas accretion before disk dispersal.
\end{abstract}

\section{Introduction}
\label{sec:intro}

The distribution of extrasolar companions in terms of their physical and orbital properties provides insight into the processes governing their formation and evolution. At close separation, transit and radial velocity (RV) studies revealed a pile-up of hot Jupiters with periods near 3 days (e.g., \citealt{Gaudi2005, Cumming2008}), while surveys sensitive to longer periods demonstrated the enhanced prevalence of giant companions beyond the ice line \citep{Wittenmyer2016, Fernandes2019, Fulton2021}, correlations between short- and long-period planets \citep{Bonomo2025, VanZandt2025}, and a scarcity of substellar objects between $\sim10-80~\Mjup$, known as the ``brown dwarf desert" \citep{Armitage2002, GretherLineweaver2006}. Studies of the widest-separation companions with direct imaging have shown that long-period giant planets are both more common \citep{Nielsen2019} and less eccentric \citep{Bowler2020} than brown dwarfs at similar separation.

Trends have likewise emerged linking planetary occurrence to the properties of the stars they orbit. One of the first connections made from the emerging exoplanet population was between stellar metallicity and the occurrence rate of giant exoplanets \citep{Gonzalez1997, Santos2004, FischerValenti2005}. This link is predicted by the core accretion theory of planet formation \citep{Pollack1996}, in which planets are assembled ``bottom-up" from the agglomeration of planetesimals, a process that takes place more efficiently in the solid-rich protoplanetary disks surrounding high-metallicity stars. Giant planet occurrence was also predicted to correlate with stellar mass based on the shorter dynamical timescales and higher overall mass budgets in the disks of higher-mass stars \citep{Laughlin2004, IdaLin2005, Alibert2011, Andrews2013, Burn2021}. Observational studies have linked stellar mass to both occurrence \citep{Endl2006, Johnson2010, Bowler2010} and maximum planet mass \citep{Howe2026}.

The discovery of these patterns was facilitated by the immense yields of space-borne observatories and long-term ground-based observing campaigns. The steady growth of the exoplanet population, together with improvements in homogeneously measured stellar parameters (e.g., \citealt{Petigura2015, Yee2017, Isaacson2024}), has made it possible to investigate population-level characteristics in ever-increasing detail. For example, the brown dwarf desert has been mapped in high resolution through the combination of RVs and astrometry \citep{VanZandt2026, Cui2026}, as well as independently through direct imaging \citep{Duchene2023} and gravitational microlensing \citep{Zhang2025}. Meanwhile, recent occurrence studies have controlled for the influence of stellar mass, metallicity, and age by calculating occurrence as a function of all three parameters \citep{Miyazaki2023, Sayeed2025}.

Giant planet occurrence is known to correlate with stellar mass and metallicity, but it remains unclear whether these stellar properties merely determine the total number of giant planets that form or the shape of their mass distribution as well. Identifying the factors that govern the mass function shape would have deep implications for our understanding of planet formation. For example, demonstrating that the distribution shape shifts toward higher mass as stellar mass increases would support the conclusion that giant planets reach final masses proportional to the host mass \citep{Mordasini2012, Pascucci2018}. A similar effect across stellar metallicity would show that the core accretion phase has strong downstream effects on the later stages of giant planet formation \citep{IdaLin2004}.

In this work, we calculate the occurrence of giant planets and brown dwarfs ($0.1-10$ AU, $0.4-50~\Mjup$) as a function of both companion mass $M_c$ and mass ratio $q=M_c/M_{\star}$ among a sample of 719 FGKM stars from the California Legacy Survey (CLS; \citealt{Rosenthal2021}). We additionally subdivide the CLS sample by stellar mass and metallicity to explore how the distributions vary in these dimensions. Our hierarchical Poisson occurrence model accounts for both survey sensitivity and the uncertainties in our companion posteriors. We describe the properties of the stars and companions in the CLS in Section \ref{sec:catalog}. In Section \ref{sec:analysis} we detail our hierarchical model and the occurrence functions we fit to the CLS catalog. We present our occurrence distribution fit results in Section \ref{sec:results} and discuss our findings in Section \ref{sec:discussion}. We summarize and conclude in Section \ref{sec:conclusion}.

\section{The California Legacy Survey}
\label{sec:catalog}

\subsection{Revised True Mass Companion Catalog}
\label{subsec:companion_catalog}
We derive occurrence rates in this work using the sample of the California Legacy Survey \citep{Rosenthal2021}, refit by \cite{VanZandt2026}, hereafter V26, using absolute astrometry from the $Hipparcos$-$Gaia$ Catalog of Accelerations \citep{Brandt2021}. The CLS collected RVs of 719 Sun-like stars over three decades, fitting 226 companions across a wide range of separations and minimum masses ($\Msini$). V26 identified the 128 systems in the CLS hosting at least one potentially substellar companion ($\Msini <80~\Mjup$), and refit the orbits of the 194 companions they hosted by combining CLS RVs with HGCA accelerations using the \texttt{Orvara} \citep{Orvara2021} orbit fitting code. The HGCA cross-calibrated the $Hipparcos$ and $Gaia$ catalogs to enable direct comparison of stellar proper motion over the 25-year span between them. Combining RVs and astrometry probes orbital motion in three dimensions, breaking the mass-inclination degeneracy from which RVs alone suffer and enabling measurements of true mass.

\cite{Rosenthal2021} also measured the sensitivity to orbital companions in each system by injecting synthetic signals into their RV data sets and attempting to recover them. They estimated survey sensitivity in a given region of ($a$, $\Msini$) space as the fraction of successfully recovered orbits in that region. We use the sensitivity maps published by \cite{Rosenthal2021} to compute two ingredients to our occurrence model. First, we convert the maps from $\Msini$ space to $M_c$ to correspond to the true masses in the V26 catalog. For each ($a$, $\Msini$) injection, we compute 10 realizations of true mass by dividing the minimum mass by $\sin i$, with $i$ randomly sampled from a distribution uniform in $\cos i$, and assign each realization the recovery outcome of the original injection. Although 10 inclinations sample $\sin i$ coarsely for any single injection, the sensitivity in each ($a$, $M_c$) cell is computed from the realizations of all injections that fall within it, which together sample the inclination distribution densely. Second, for each subset of stars among which we compute companion occurrence, we combine the corresponding maps to compute the average sensitivity across the given subsample.

\subsection{Stellar Catalog}
\label{subsec:stellar_catalog}

\begin{figure}[t]
    \centering
    \includegraphics[width=0.99\linewidth]{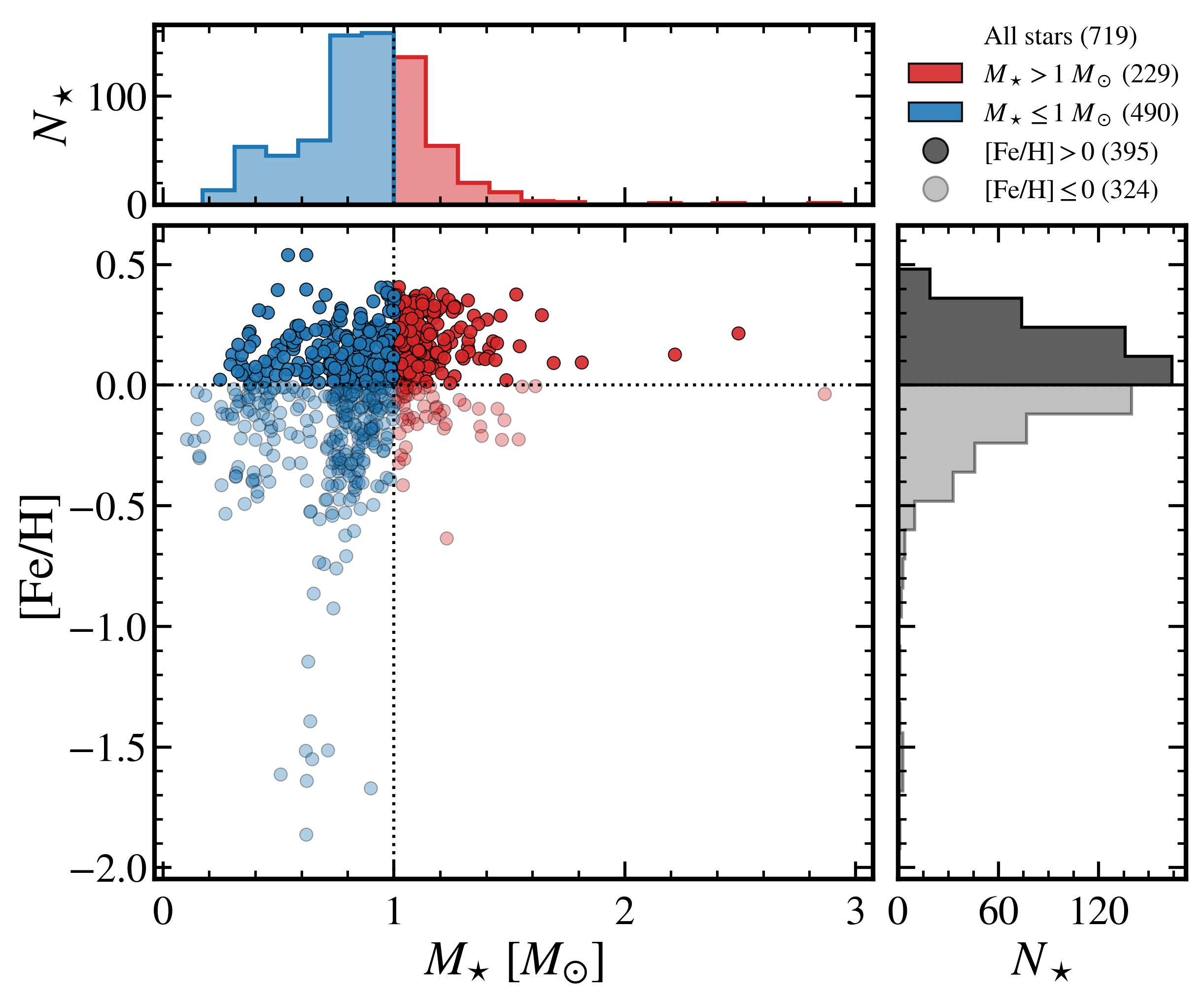}
    \caption{Stellar mass and metallicity of the CLS sample, with the marginal distributions of each. Points are colored by stellar mass sample (red: $\Mstar>1.0~\Msun$; blue: $\Mstar\leq1.0~\Msun$) with lower opacities for the metal-poor sample ([Fe/H]$\leq0.0$) than for the metal-rich sample ([Fe/H]$>0.0$). The top histogram shows the two mass samples in the same colors and the side histogram shows the two metallicity samples with the same difference in opacity. Dotted lines show the values we used to define our ``high" and ``low" subsamples, $1.0~\Msun$ and 0.0 dex, respectively. The number of stars in each sample is given in the legend.}
    \label{fig:stellar_histograms}
\end{figure}

\begin{figure}[t]
    \centering
    \begin{minipage}[b]{0.99\linewidth}
        \centering
        \includegraphics[width=\linewidth]{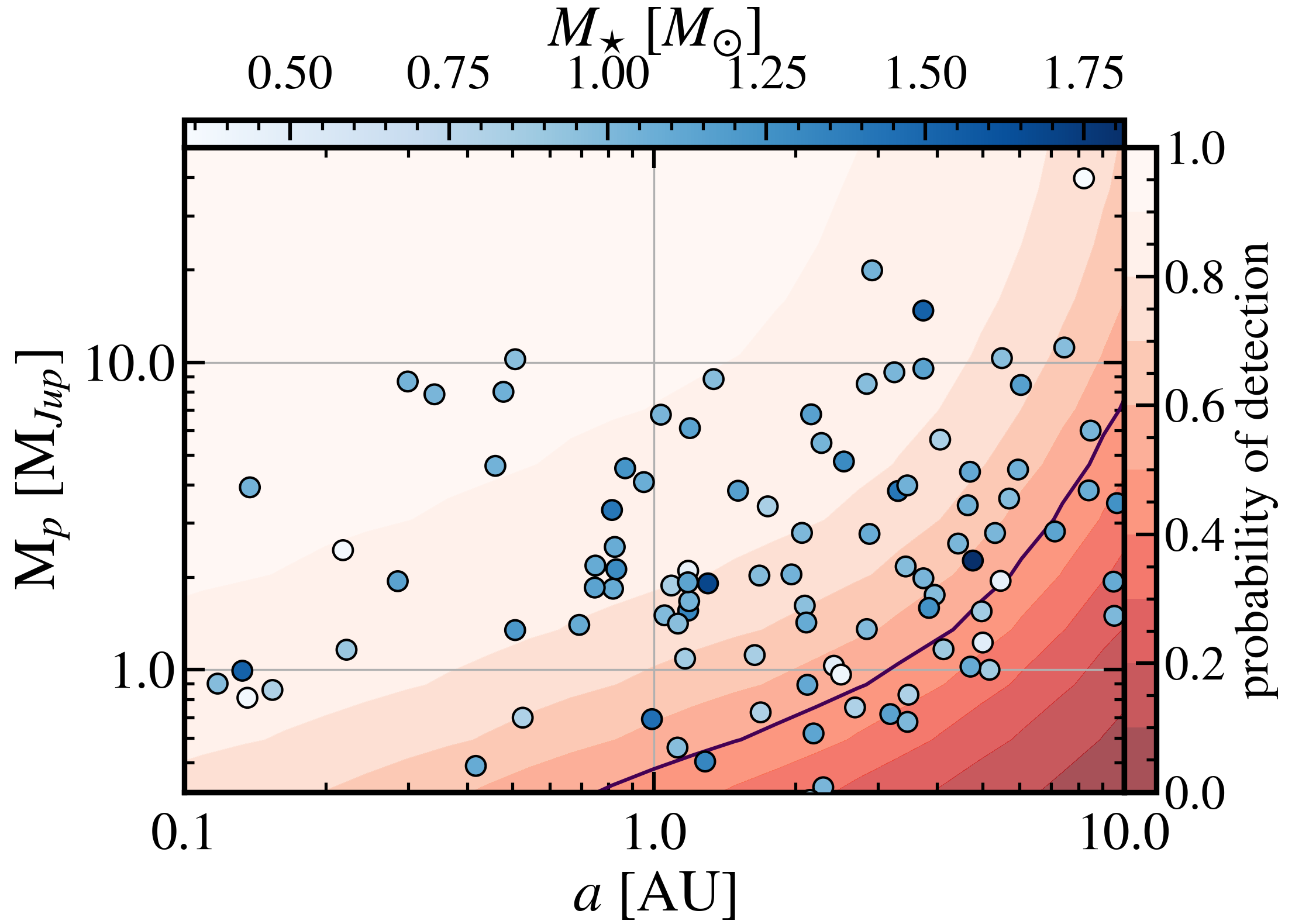}
    \end{minipage}
    \par\vspace{0.5em}
    \begin{minipage}[b]{0.99\linewidth}
        \centering
        \includegraphics[width=\linewidth]{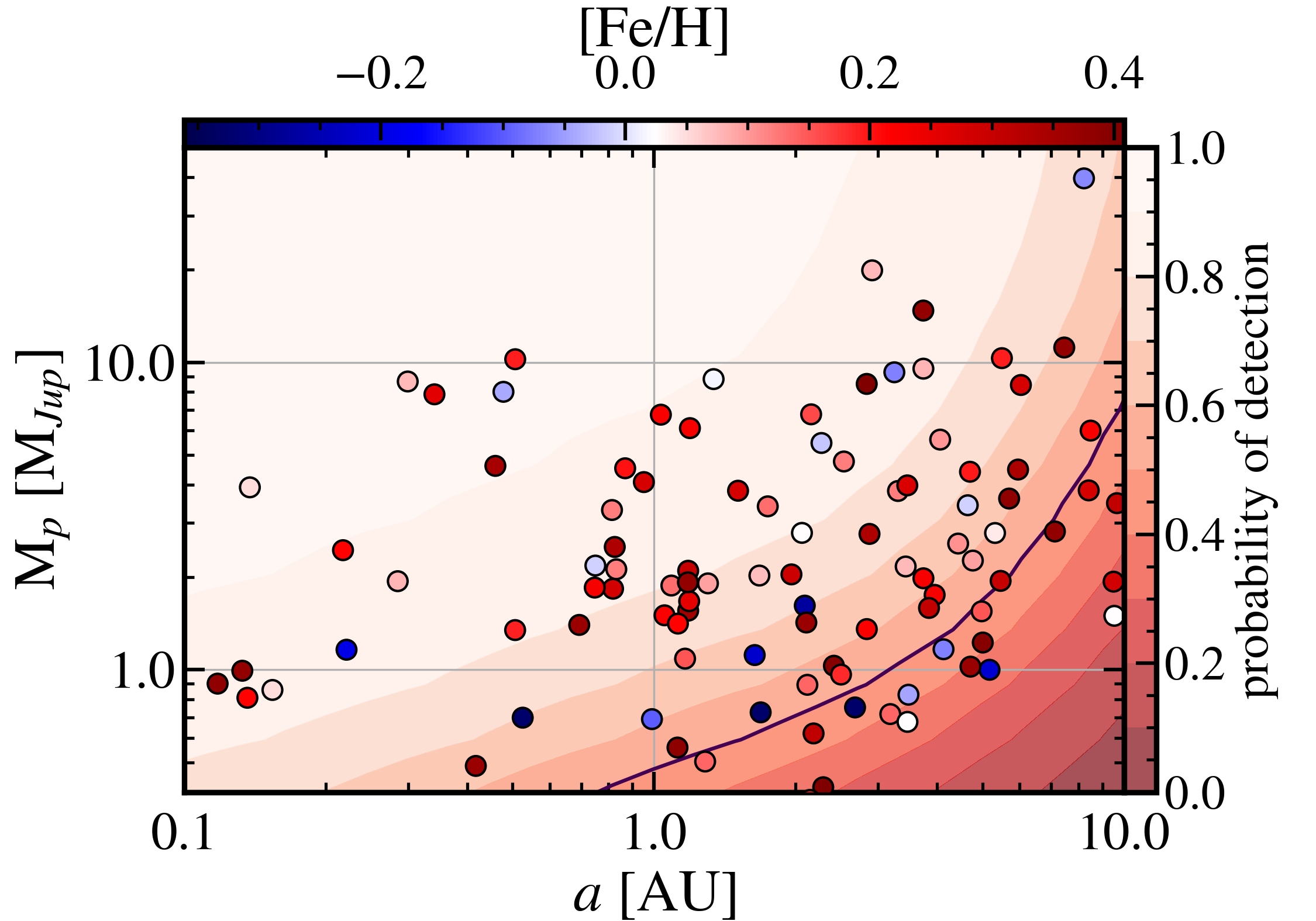}
    \end{minipage}
    \caption{The 98 companions in the catalog between $0.1-10$ AU and $0.4-50~\Mjup$. We represent each companion's position with its median mass and separation values for visual ease. In the top and bottom panels, marker colors correspond to host star mass and metallicity, respectively. The background map shows the sensitivity among the host stars of the plotted companions. The 50\% sensitivity contour is highlighted in black.}
    \label{fig:catalog_with_stellar_params}
\end{figure}

\cite{Rosenthal2021} derived stellar metallicity, effective temperature, surface gravity, and radius for all 719 CLS stars by matching stellar template spectra to interpolated grids of either synthetic \citep{Petigura2015} or empirical spectra \citep{Yee2017}. From these parameters they estimated stellar mass using \texttt{Isoclassify} \citep{Huber2017, Berger2020, Berger2023}. \cite{Isaacson2024} reported chromospheric activity for 710 of these stars, which we use to study stellar age in Appendix \ref{app:age}. We show the distributions of stellar mass and metallicity in Figure \ref{fig:stellar_histograms}, and the companion masses and separations, together with host star properties, in Figure \ref{fig:catalog_with_stellar_params}.

\section{Analysis}
\label{sec:analysis}

\subsection{Occurrence Framework}
\label{subsec:occurrence_framework}
We model the catalog as a draw from a censored Poisson point process \citep{Rogers2021, ForemanMackey2014}, defined by the companion occurrence rate density (ORD) $\lambda(\log a, \log M_c | \boldsymbol{\theta})$.\footnote{Throughout this study, we use $\log$ to indicate $\log_{10}.$} The intensity parameter $\lambda$ defines the number of companions per star per logarithmic interval in $a$ and $M_c$. The vector $\boldsymbol{\theta}$ contains parameters specific to each function we use to model occurrence, which we describe in Section \ref{subsec:model_fits}. We define $\boldsymbol{\omega}=(\log a, \log M_c)$ for brevity. We adopt a modified version of the hierarchical Poisson likelihood used by V26:

\begin{equation}
    \mathcal{L} \propto e^{-\Lambda} 
    \prod_{k=1}^{K} 
    \Bigg [
    \frac{1}{N_{k}}
    \sum_{n=1}^{N_{k}} \frac{\lambda(\boldsymbol{\omega}_{k}^{(n)}|\boldsymbol{\theta})Q_k(\boldsymbol{\omega}_{k}^{(n)})}{p(\boldsymbol{\omega}_{k}^{(n)}|\boldsymbol{\alpha})}
    \Bigg ]^{w_{k}},
    \label{eq:likelihood}
\end{equation}

\noindent where $n$ indexes the posterior samples that fall within the two-dimensional $a-M_c$ interval in which we calculate occurrence, hereafter the region of interest (ROI). The vector $\boldsymbol{\omega}_k ^{(n)}$ denotes the $n^{\text{th}}$ posterior sample of the $k^{\text{th}}$ companion. We draw $N=500$ samples from each companion posterior, and sum over the $N_{k}$ samples that fall within the ROI for each companion. We define the number of ``effective" companions in a given interval of parameter space as the total companion weight in that interval:

\begin{equation}
    N_{\text{eff}}=\sum_k w_k.
\end{equation}

\noindent The function $Q_k$ gives the sensitivity to companions in the system hosting the $k^{\text{th}}$ companion. By using the system-specific sensitivity we differ from V26, who used the survey-averaged sensitivity function in this term. For homogeneous surveys like CLS, we expect the difference in these approaches to be minor. The interim prior $p(\boldsymbol{\omega}_{k}^{(n)}|\boldsymbol{\alpha})$ encodes the prior weights applied in the original orbital fits, expressed with respect to our logarithmic parameter space. Because \texttt{Orvara} samples $a$ and $M_c$ log-uniformly, we set $p(\boldsymbol{\omega}_{k}^{(n)}|\boldsymbol{\alpha})$=1. Many of the posteriors in the V26 catalog have high variance compared to the ROI size, particularly at wide separations ($\gtrapprox10$ AU). We use a weighted likelihood to account for posteriors falling only partially within the ROI \citep{NewtonRaftery1994, Gilbert2025}. For each companion, we compute a weight, $w_{k}$, equal to the fraction of posterior samples from the $k^{\text{th}}$ companion that fall in the ROI.

In the leading term of Equation \ref{eq:likelihood}, the Poisson rate parameter gives the expected number of observed companions in the ROI among $N_\star$ host stars:

\begin{equation}
    \Lambda = N_{\star} \int \lambda(\boldsymbol{\omega}|\boldsymbol{\theta}) \, Q(\boldsymbol{\omega}) \, d\boldsymbol{\omega},
    \label{eq:Lambda_definition}
\end{equation}

\noindent where $Q$ is the companion sensitivity function averaged over all systems in the sample, and the integral is over the ROI.

\subsection{Model Fits}
\label{subsec:model_fits}

We choose three models to fit to the catalog in our primary analysis, seeking to balance complexity against goodness-of-fit in light of our incomplete knowledge of the astrophysical processes that shape the companion mass function. The first is a piecewise-constant \model{histogram} function with seven bins, chosen to be highly flexible and agnostic to the underlying physics of planet formation. We derive the ORD in each bin independently using Equation \ref{eq:likelihood}. We apply the assumption of a smoothly varying ORD function by drawing bin heights from a Gaussian Process (GP) prior \citep{ForemanMackey2014, Gilbert2026}, defined by a radial basis function kernel:

\begin{equation}
    K(M_{1}, M_{2}) = \sigma_{GP}^2 \cdot
    \exp\left [ -\frac{(\log M_{1}-\log M_{2})^2}{2l^2} \right ],
    \label{eq:rbf_kernel}
\end{equation}

\noindent where $M_1$ and $M_2$ are companion masses corresponding to bin centers and $l$ sets the smoothing scale. We apply the GP prior for all \model{histogram} fits except for the analysis of specific features of the mass function or for direct comparisons with other studies (see Section \ref{subsec:discussion_bd_desert_M_q}).

The second is a \model{sigmoid} function, which is well-suited to describing distributions with two near-constant components joined by a smooth transition (e.g., \citealt{Miyazaki2023, Gilbert2025}). Although the mass function shape is unknown, the brown dwarf desert produces constant occurrence at high masses, and the distribution may likewise plateau for sub-Jupiters. The \model{sigmoid} density is given by:

\begin{equation}
    \lambda(x|B_1, B_2, x_{t}, W) =
    B_1 + \frac{B_2-B_1}{1+e^{-\frac{\log(x/x_t)}{W}}},
    \label{eq:model_sigmoid}
\end{equation}

\noindent where $B_1$ and $B_2$ are the constant components at $x$-values far from the transition point $x_t$, and the width $W$ defines the steepness of the transition.

Finally, we fit a log-Gaussian function, \model{logG}, to allow for the possibility that the mass function reaches a peak at some $\log$ companion mass $\mu$, a feature predicted by various core accretion models (\citealt{IdaLin2005, Mordasini2009}; see Section \ref{subsec:dicussion_peak_in_mass_function}). The \model{logG} density is given by:

\begin{equation}
    \lambda(x|A, \mu, \sigma) =
    A\exp \left [-\frac{(\log x-\mu)^2}{2\sigma^2} \right ].
    \label{eq:model_logG}
\end{equation}

We introduce two additional models to test specific hypotheses about the behavior of the companion mass function. The models are defined below and the tests are described in detail in Section \ref{sec:discussion}.

The first is a log-linear function to test the hypothesis of monotonically increasing occurrence toward lower companion masses (see Section \ref{subsec:dicussion_peak_in_mass_function}). The \model{log-linear} model gives occurrence rate density as a linear function of logarithmic mass, parameterized by the endpoints of the interval over which the function is fit:

\begin{equation}
    \lambda(x|C_1, C_2) =
    C_1 + \frac{C_2-C_1}{\log(x_2/x_1)}\log(x/x_1),
    \label{eq:model_loglinear}
\end{equation}

\noindent where $x_1$ and $x_2$ are the endpoints of the fitting interval and $C_1$ and $C_2$ are the ORD values at those points.

We also fit a piecewise function consisting of two constant components joined by a linear segment. Analogous to the geological feature after which it is named, this \model{``escarpment"} model is similar to the \model{sigmoid} function, but with sharp transitions, imposing a more rigid distinction between different companion mass regimes:

\begin{equation}
    \lambda(x|D_1, D_2, x_{t,1}, x_{t,2}) =
    \begin{cases}
    D_1 & \text{if } x \leq x_{t,1},\\
    D_2 & \text{if } x > x_{t,2},\\
    m\cdot\log(\frac{x}{x_{t,1}})+D_1 & \text{otherwise},
    \end{cases}
    \label{eq:model_escarpment}
\end{equation}

\noindent where the slope $m$, equal to $\frac{D_2-D_1}{\log (x_{t,2}/x_{t,1})}$, is specified by the choice of model parameters.

We list the parameters of these models in Table \ref{tab:models}. We set priors on the model parameters as follows: amplitude parameters, including $A$, $B_1$, $B_2$, $C_1$, $C_2$, $D_1$, and $D_2$, are sampled uniformly between $10^{-6}-10$ planets per star per two-dimensional bin; logarithmic location parameters, including $\log(x_t)$, $\log(x_{t,1})$, $\log(x_{t,2})$, and $\mu$, are sampled uniformly over the mass range $0.4-50~\Mjup$; width parameters, including $W$ and $\sigma$, are sampled uniformly between $1\%$ of the mass range up to the full range. We set broad log-uniform hyper-priors on our GP parameters and constrain the integrated occurrence rate of every model to lie between 0 and 1.

For all models, we sample our hierarchical likelihood with \texttt{emcee} \citep{ForemanMackey2013}, using 50 walkers. We run 7000 total steps for the smooth models with 2000 discarded as burn-in, and 30000 steps for the \model{histogram} model with 5000 burn-in steps. We verify that each post-burn-in chain length is $\gtrsim50\times$ the autocorrelation time.

\begin{deluxetable*}{lcccc}
\tablecaption{Parametric Models}
\label{tab:models}
\tablehead{
\colhead{Model Name} &
\colhead{$\theta_1$} &
\colhead{$\theta_2$} &
\colhead{$\theta_3$} &
\colhead{$\theta_4$} 
}
\startdata
  \model{sigmoid} &  $B_1$  &  $B_2$  &  $\log (x_{t})$ &  $W$ \\
  \model{logG} &  $A$  &  $\mu$  &  $\sigma$ &  --- \\
  \model{log-linear} & $C_1$ & $C_2$ & --- & --- \\
  \model{escarpment} &  $D_1$  &  $D_2$  &  $\log (x_{t,1})$ &  $\log(x_{t,2})$ \\
\enddata

\tablenotetext{}{%
\textbf{Note.} We sample all parameters in this table uniformly, resulting in log-uniform priors on the transition points $x_t$ and uniform priors on all other parameters. The \model{histogram} model, omitted for brevity, includes seven bin heights, a normalization factor, and three GP parameters. Parametric model fits are summarized in Table \ref{tab:model_params_appendix}, and histogram bin heights are provided in Table \ref{tab:ord_appendix}.
}
\end{deluxetable*} 

\section{Results}
\label{sec:results}

We first fit the mass function of the full CLS sample and then explore its dependence on stellar mass and metallicity. We consider companions with masses and semi-major axes defined by the following boundaries: $\frac{a}{AU}=$\{0.1, 10\}, $\frac{M_c}{\Mjup}=$\{0.4, 0.8, 1.6, 3.2, 6.4, 13, 26, 50\}. We chose these bounds to include giant companions to which the CLS is sensitive. We set an upper bound of $50~\Mjup$ to exclude the stellar/substellar regime, where the CLS selection function may have introduced bias by removing suspected stellar companions. Figure \ref{fig:catalog_with_stellar_params} shows the companion catalog over this parameter interval, along with the survey completeness and host star parameters. For fits in mass ratio, we use the same $M_c$ bounds, divided by $1~\Msun$.

In the following sections, we quantify the significance of the difference between two parameters by comparing their posteriors as follows. We identify the ``higher'' posterior (size $A$) and the ``lower'' posterior (size $B$) based on their median values. We compare every possible pair of draws, a total of $A \times B$. We calculate the fraction $P$ of pairs for which the draw from the higher posterior is larger than the draw from the lower posterior, which estimates the probability that the higher parameter truly exceeds the lower one. We then convert the complementary probability $1-P$ to the equivalent one-sided Gaussian significance. Because a probability estimated from $n$ independent draws is reliable only down to $\sim1/n$, when $(1-P)<1/n$, with $n$ the minimum of the two posterior sizes, $\min(A,B)$, we instead divide the difference of the posterior medians by the $16^{\text{th}}$--$84^{\text{th}}$ percentile uncertainties of the two posteriors, added in quadrature.

\subsection{The Mass Distribution of the Full CLS Sample}
\label{subsec:results_full_sample}

\begin{figure*}[t]
    \centering
        \centering
        \includegraphics[width=\linewidth]{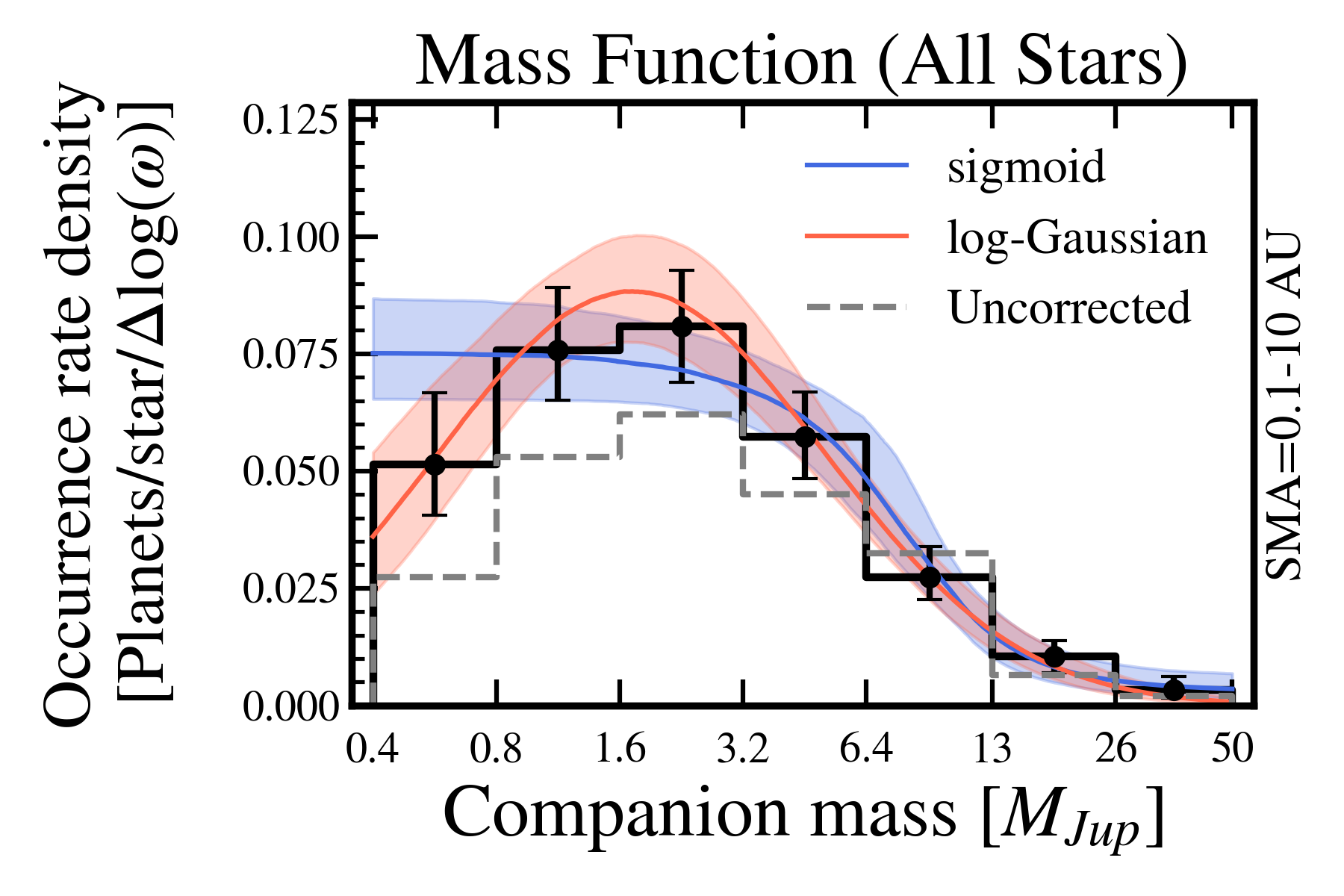}
    \caption{Occurrence rate density of companions between 0.4 and 50 $\Mjup$ and 0.1--10 AU for all $\McAllstarsPaperBoundsNstars$ stars in the sample. The colored lines show our fitted \model{sigmoid} (blue) and \model{logG} (orange) parametric models. The bold curves show the median model values from $10^4$ posterior draws and the bands show $16^{\text{th}}-84^{\text{th}}$ percentiles. The dashed gray histogram shows $N_{\text{eff}}/N_{\star}$ without correcting for completeness.}
    \label{fig:ORD_allstars}
\end{figure*}

We first calculate companion occurrence using the full CLS sample of 719 stars. We fit the \model{histogram}, \model{sigmoid}, and \model{logG} models presented in Section \ref{subsec:model_fits} using Equation \ref{eq:likelihood}, and calculate the Bayesian information criterion (BIC, \citealt{Schwarz1978}) for each smooth model (i.e., all models except the \model{histogram}). A smaller BIC indicates a better balance between model complexity and goodness of fit, and $\Delta$BIC, the difference between two model BIC values, quantifies their relative goodness of fit. Following \cite{Liddle2007}, we consider $\Delta$BIC$\gtrsim5$ to be ``strong" evidence for one model over another, and $\Delta$BIC$\gtrsim10$ to be ``decisive." We take the number of companions at least partially in the ROI as the number of observed data points. We find that the \model{logG} (BIC$_{\text{logG}}$$=\McAllstarsPaperBoundsLogGBicBinaZero$) model is significantly preferred over the \model{sigmoid} (BIC$_{\text{sigmoid}}$$=\McAllstarsPaperBoundsSigmoidBicBinaZero$) model ($\Delta$BIC$=\McAllstarsPaperBoundsSigmoidLogGDbicBinaZero$). Figure \ref{fig:ORD_allstars} shows our ORD distributions and model fits, and we report our derived parameters for the smooth models in Table \ref{tab:model_params}. Histogram bin heights are provided in Table \ref{tab:ord_appendix}. Our fitted \model{logG} mean corresponds to a peak in the mass function near $1-2~\Mjup$, and our \model{sigmoid} transition takes place at $6-10~\Mjup$. We also compute the integrated occurrence rate across the mass range using the \model{histogram} model, finding $P(M_c=0.4-50~\Mjup)=\McAllstarsPaperBoundsPiecewiseIntOccBinaZero$ companions per star. The rates for the two smooth models agree closely (Table \ref{tab:model_params_appendix}).

\begin{table}[t]
\centering
\caption{Derived Model Parameters for Full Sample}
\label{tab:model_params}
\begin{tabular}{lc}
\hline \hline
\textbf{Parameter} & \textbf{Value} \\
\hline
\multicolumn{2}{l}{logG} \\
\hline
$A$ & \McAllstarsPaperBoundsLogGParamABinaZero \\
$\mu$ & \McAllstarsPaperBoundsLogGParamMuBinaZero \\
$\sigma$ & \McAllstarsPaperBoundsLogGParamSigmaBinaZero \\
Occurrence & \McAllstarsPaperBoundsLogGIntOccBinaZero \\
\hline
\multicolumn{2}{l}{sigmoid} \\
\hline
$B_1$ & \McAllstarsPaperBoundsSigmoidParamBOneBinaZero \\
$B_2$ & \McAllstarsPaperBoundsSigmoidParamBTwoBinaZero \\
$\log(x_t)$ & \McAllstarsPaperBoundsSigmoidParamCenterBinaZero \\
$W$ & \McAllstarsPaperBoundsSigmoidParamWidthBinaZero \\
Occurrence & \McAllstarsPaperBoundsSigmoidIntOccBinaZero \\
\hline
\hline
\end{tabular}
\end{table}
 
\subsection{Companion Occurrence as a Function of Stellar Mass}
\label{subsec:results_Mstar}

To test the correlation between stellar host mass and companion occurrence, we divide the CLS sample at $M_{\star}=1.0~\Msun$ and calculate occurrence among both populations. The high-mass subsample hosts $N_{\text{eff}}=\McHighMstarPaperBoundsNeff$ effective companions (the summed posterior weight of companions within the ROI; Section \ref{subsec:occurrence_framework}) between $0.1-10$ AU and $0.4-50~\Mjup$ among $\McHighMstarPaperBoundsNstars$ stars, while the low-mass sample hosts $\McLowMstarPaperBoundsNeff$ effective companions in the same range among $\McLowMstarPaperBoundsNstars$ stars.\footnote{The uncertainties in our derived occurrence rates scale with the number of companions considered rather than the number of host stars.} 

\begin{figure}[t]
    \centering

    \begin{minipage}[t]{0.49\textwidth}
        \centering
        
        \includegraphics[width=\linewidth]{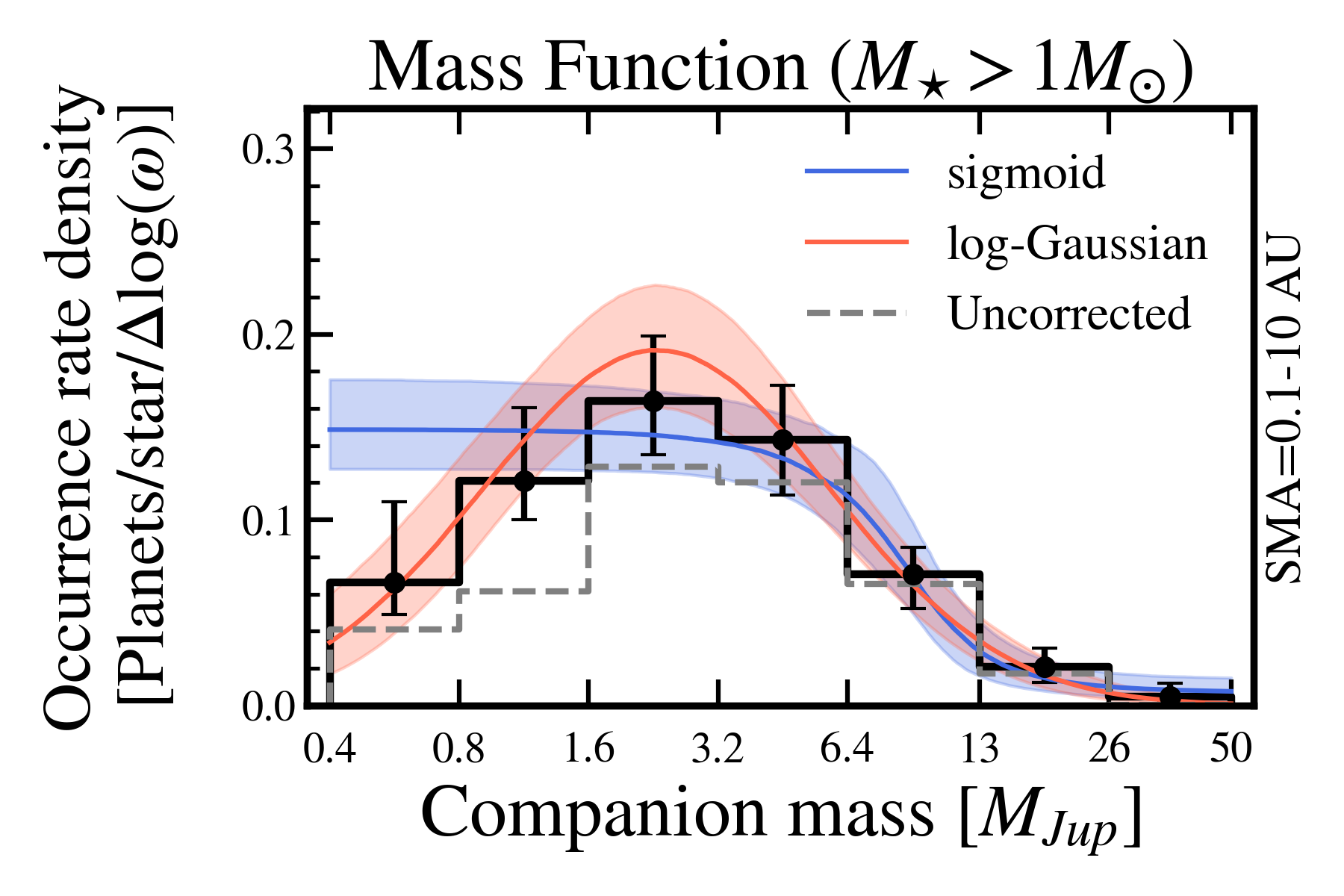}
        
        \vspace{0.5em}
        
        \includegraphics[width=\linewidth]{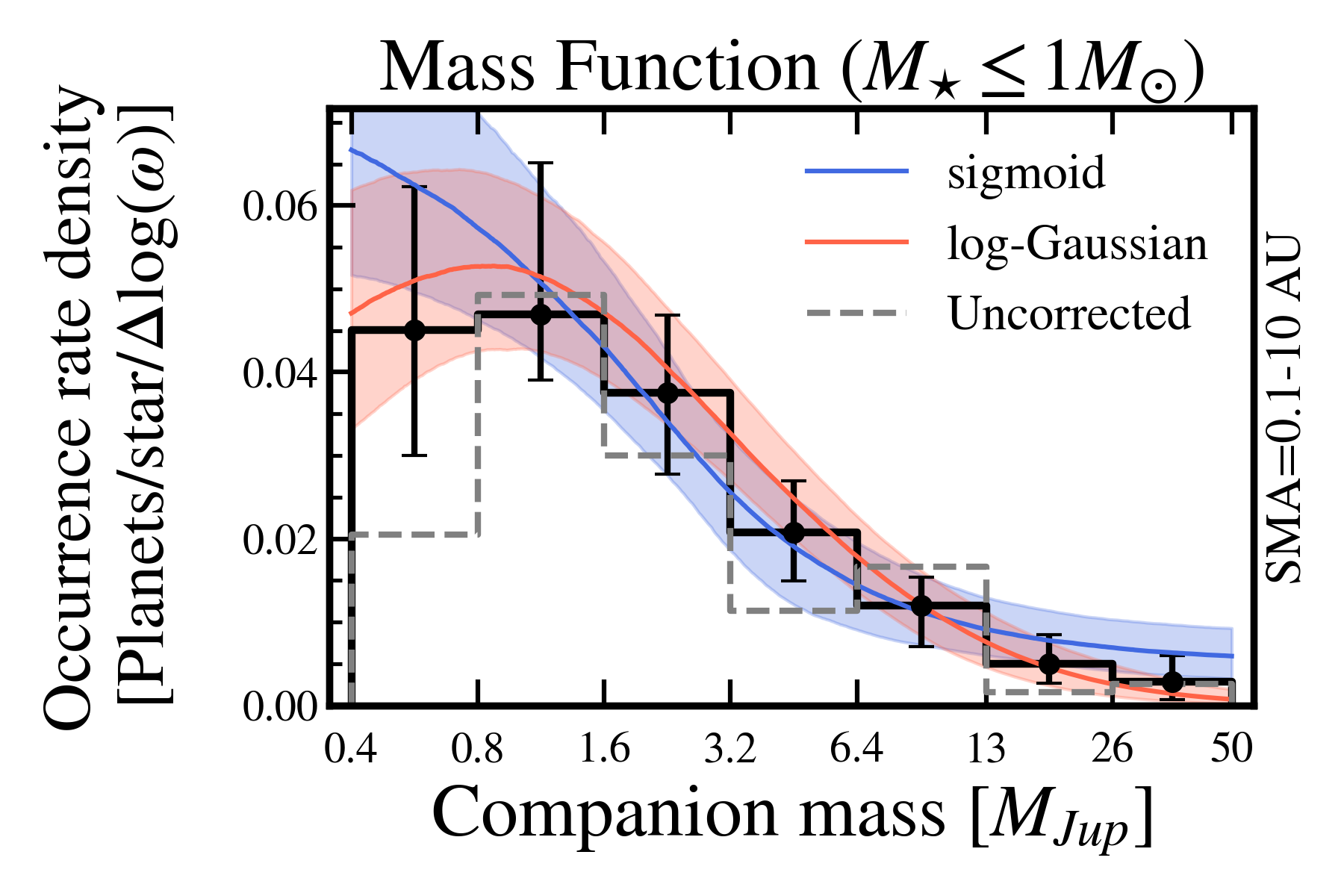}
        
    \end{minipage}

    \caption{Companion mass functions for the $\McHighMstarPaperBoundsNstars$ high-mass and $\McLowMstarPaperBoundsNstars$ low-mass stars in the sample, divided at $M_{\star}=1~M_{\odot}$. Solid black curves show the \model{histogram} model, dashed gray lines show the uncorrected rate densities, and colored curves and bands show parametric models fitted to the companion catalog. The transition point of the \model{sigmoid} model (blue) and the peak of the \model{logG} model (orange) both occur at higher companion mass among the high-$M_{\star}$ sample, suggesting that giant companions around higher-mass stars reach systematically higher masses.}
    
    \label{fig:ORD_Mstar}
\end{figure}

Our derived $M_{\star}$-dependent distributions are shown in Figure \ref{fig:ORD_Mstar}. We find differences between the high-$M_{\star}$ and low-$M_{\star}$ mass functions, both in shape and overall occurrence. For high-mass stars, we find a \model{sigmoid} transition point, that is, the companion mass at which the ORD is halfway between its maximum and minimum values, of $\log M_c=\McHighMstarPaperBoundsSigmoidParamCenterBinaZero$ ($6.9-10.5~\Mjup$), versus $\McLowMstarPaperBoundsSigmoidParamCenterBinaZero$ ($0.8-3.2~\Mjup$) for low-mass stars, a $\McMstarPaperBoundsSigmoidParamCenterSignificanceBinaZero$ difference. The peaks in the \model{logG} models differ as well: $\mu=\McHighMstarPaperBoundsLogGParamMuBinaZero$ ($1.9-2.8~\Mjup$) for high-mass stars versus $\McLowMstarPaperBoundsLogGParamMuBinaZero$ ($0.5-1.3~\Mjup$) among low-mass stars ($\McMstarPaperBoundsLogGParamMuSignificanceBinaZero$). The \model{logG} fit (BIC$_{\text{logG}}$$=\McHighMstarPaperBoundsLogGBicBinaZero$) is significantly preferred over the \model{sigmoid} (BIC$_{\text{sigmoid}}$$=\McHighMstarPaperBoundsSigmoidBicBinaZero$) for high-mass stars ($\Delta$BIC$=\McHighMstarPaperBoundsSigmoidLogGDbicBinaZero$), and slightly preferred for the low-mass sample: BIC$_{\text{logG}}$$=\McLowMstarPaperBoundsLogGBicBinaZero$, compared to BIC$_{\text{sigmoid}}$$=\McLowMstarPaperBoundsSigmoidBicBinaZero$ ($\Delta$BIC$=\McLowMstarPaperBoundsSigmoidLogGDbicBinaZero$). The integrated occurrence rates demonstrate the increased prevalence of giant planets in the high-$M_{\star}$ sample ($\McHighMstarPaperBoundsPiecewiseIntOccBinaZero$ companions per star) versus the low-$M_{\star}$ sample ($\McLowMstarPaperBoundsPiecewiseIntOccBinaZero$ companions per star). We provide integrated occurrence rates for our mass and metallicity samples, along with statistics for each stellar subsample, in Table \ref{tab:one_param_OR}. The same rates, along with fitted model parameters, are given in Table \ref{tab:model_params_appendix}. Because we did not control for metallicity while calculating the high- and low-$\Mstar$ mass functions, the extent to which metallicity influenced our findings is unclear. We control for both parameters simultaneously in Section \ref{sec:discussion}.

\subsection{Companion Occurrence as a Function of Stellar Metallicity}
\label{subsec:results_FeH}

\begin{figure}[t]
    \centering

    \begin{minipage}[t]{0.49\textwidth}
        \centering
        
        \includegraphics[width=\linewidth]{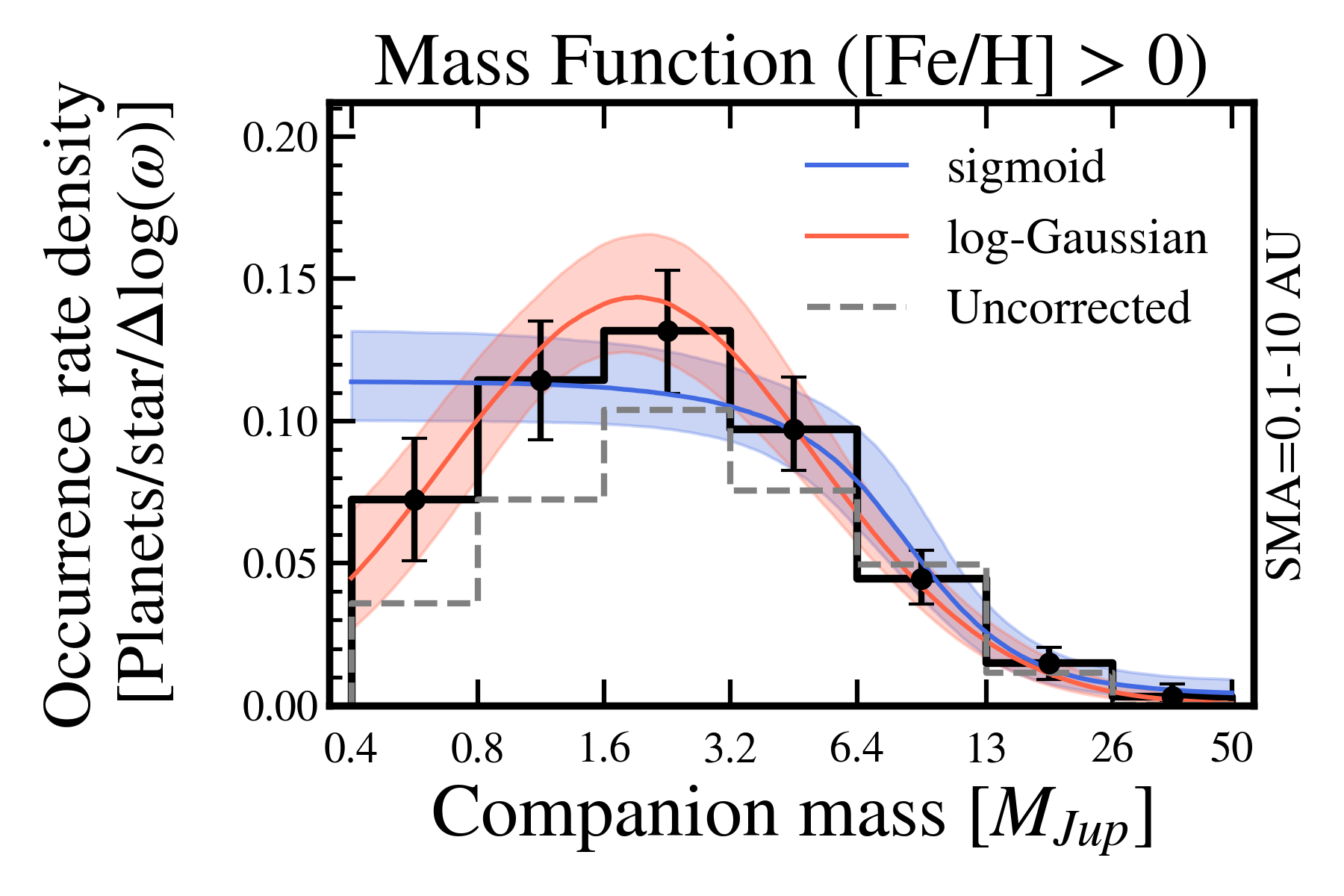}
        
        \vspace{0.5em}
        
        \includegraphics[width=\linewidth]{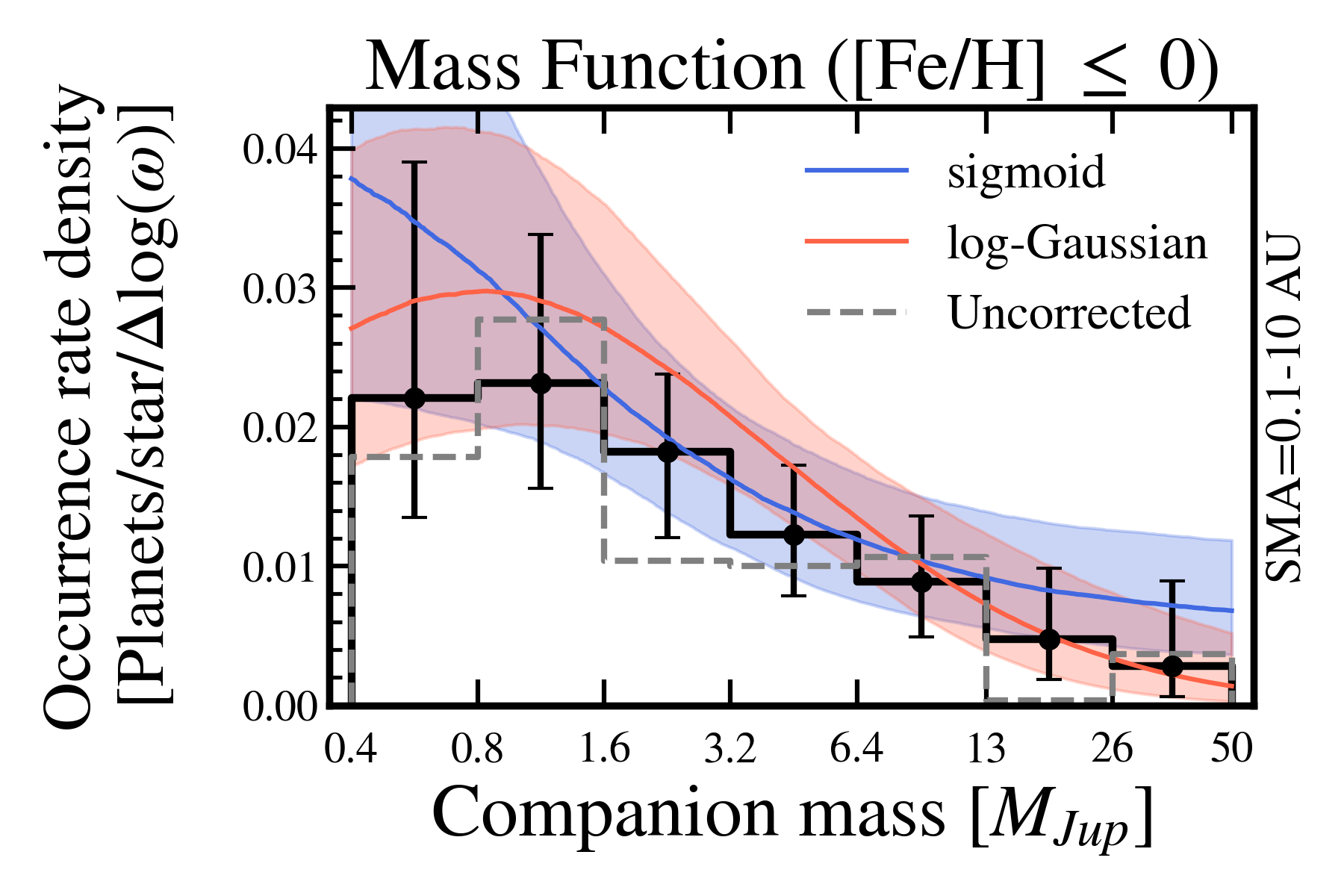}
    \end{minipage}

    \caption{
    The same as Figure \ref{fig:ORD_Mstar}, with the sample divided at [Fe/H]=0.0. The mass function for our $\McHighFeHPaperBoundsNstars$ metal-rich stars favors higher-mass companions, with an occurrence plateau out to $\sim3~\Mjup$ and a potential peak near $2~\Mjup$. Meanwhile, the distribution of the $\McLowFeHPaperBoundsNstars$ metal-poor stars is consistent with monotonically increasing occurrence toward lower companion masses.
    }
    \label{fig:ORD_FeH}
\end{figure}

We perform an analogous set of experiments to those in Section \ref{subsec:results_Mstar} for stellar metallicity. We divide the CLS at [Fe/H]=0.0 and calculate occurrence for both subsamples. The metal-rich subsample includes $\McHighFeHPaperBoundsNstars$ stars hosting $\McHighFeHPaperBoundsNeff$ effective companions between $0.1-10$ AU and $0.4-50~\Mjup$. The metal-poor sample includes $\McLowFeHPaperBoundsNstars$ stars hosting $\McLowFeHPaperBoundsNeff$ effective companions in the same domain. Our results are illustrated in Figure \ref{fig:ORD_FeH} and in Table \ref{tab:model_params_appendix}, and we summarize our key findings here. In brief, our results roughly match those of our stellar mass investigation (Section \ref{subsec:results_Mstar}), but with less significance in the distinction between the high- and low-metallicity samples than we found between the high- and low-mass samples. This similarity is partly due to the overlap between our high-$M_{\star}$ and metal-rich subsamples, which we discuss in Section \ref{sec:discussion}.

As with stellar mass, we find that the metal-rich and metal-poor companion mass functions differ in normalization, though less significantly in shape (see Figure \ref{fig:ORD_FeH}). Metal-rich stars in the sample host $\McHighFeHPaperBoundsPiecewiseIntOccBinaZero$ giants per star, compared with $\McLowFeHPaperBoundsPiecewiseIntOccBinaZero$ among metal-poor stars (Table \ref{tab:one_param_OR}), a $\McFeHPaperBoundsPiecewiseIntOccSignificanceBinaZero$ difference. The metal-rich sample exhibits a \model{sigmoid} transition at $\log M_c=\McHighFeHPaperBoundsSigmoidParamCenterBinaZero$, compared with $\McLowFeHPaperBoundsSigmoidParamCenterBinaZero$ for metal-poor stars ($\McFeHPaperBoundsSigmoidParamCenterSignificanceBinaZero$). Meanwhile, our \model{logG} fits are centered at $\mu=\McHighFeHPaperBoundsLogGParamMuBinaZero$ for metal-rich stars and $\McLowFeHPaperBoundsLogGParamMuBinaZero$ for metal-poor stars ($\McFeHPaperBoundsLogGParamMuSignificanceBinaZero$). The \model{logG} model significantly outperforms the \model{sigmoid} model for metal-rich stars (BIC$_{\text{logG}}$$=\McHighFeHPaperBoundsLogGBicBinaZero$, BIC$_{\text{sigmoid}}$$=\McHighFeHPaperBoundsSigmoidBicBinaZero$; $\Delta$BIC$=\McHighFeHPaperBoundsSigmoidLogGDbicBinaZero$), but the two are comparable for metal-poor stars (BIC$_{\text{logG}}$$=\McLowFeHPaperBoundsLogGBicBinaZero$, BIC$_{\text{sigmoid}}$$=\McLowFeHPaperBoundsSigmoidBicBinaZero$; $\Delta$BIC$=\McLowFeHPaperBoundsSigmoidLogGDbicBinaZero$).

\begin{deluxetable*}{lcccccccccc}
\tablecaption{Occurrence Rates by Stellar Mass or Metallicity}
\label{tab:one_param_OR}
\tablehead{
\colhead{} & \colhead{} & \colhead{Median} & \colhead{Median} & \colhead{} & \colhead{} & \colhead{Average} & \colhead{Occurrence} & \colhead{} & \colhead{High/Low} & \colhead{High/Low} \\[-1.6ex]
\colhead{Parameter} & \colhead{Sample} & \colhead{$M_{\star}$} & \colhead{$\text{[Fe/H]}$} & \colhead{$N_{\star}$} & \colhead{$N_{\mathrm{eff}}$} & \colhead{Completeness} & \colhead{Rate} & \colhead{Significance} & \colhead{$M_{\star}$} & \colhead{$\text{[Fe/H]}$}
}
\startdata
\multirow{2}{*}{Mass} & High & $\McHighMstarPaperBoundsMedianMstar$ & $\McHighMstarPaperBoundsMedianFeH$ & \McHighMstarPaperBoundsNstars & \McHighMstarPaperBoundsNeff & \McHighMstarPaperBoundsAvgCompl & \McHighMstarPaperBoundsPiecewiseIntOccBinaZero & \multirow{2}{*}{\McMstarPaperBoundsPiecewiseIntOccSignificanceBinaZero} & \multirow{2}{*}{\McMstarPaperBoundsMedianMstarRatio} & \multirow{2}{*}{\McMstarPaperBoundsMedianFeHRatio} \\
 & Low & $\McLowMstarPaperBoundsMedianMstar$ & $\McLowMstarPaperBoundsMedianFeH$ & \McLowMstarPaperBoundsNstars & \McLowMstarPaperBoundsNeff & \McLowMstarPaperBoundsAvgCompl & \McLowMstarPaperBoundsPiecewiseIntOccBinaZero &  &  &  \\
\multirow{2}{*}{$\text{[Fe/H]}$} & High & $\McHighFeHPaperBoundsMedianMstar$ & $\McHighFeHPaperBoundsMedianFeH$ & \McHighFeHPaperBoundsNstars & \McHighFeHPaperBoundsNeff & \McHighFeHPaperBoundsAvgCompl & \McHighFeHPaperBoundsPiecewiseIntOccBinaZero & \multirow{2}{*}{\McFeHPaperBoundsPiecewiseIntOccSignificanceBinaZero} & \multirow{2}{*}{\McFeHPaperBoundsMedianMstarRatio} & \multirow{2}{*}{\McFeHPaperBoundsMedianFeHRatio} \\
 & Low & $\McLowFeHPaperBoundsMedianMstar$ & $\McLowFeHPaperBoundsMedianFeH$ & \McLowFeHPaperBoundsNstars & \McLowFeHPaperBoundsNeff & \McLowFeHPaperBoundsAvgCompl & \McLowFeHPaperBoundsPiecewiseIntOccBinaZero &  &  &  \\
\enddata

\tablenotetext{}{%
\textbf{Note.} Stellar masses are given in $\Msun$ and metallicities are given in dex, with the dividing values for each at $1.0~\Msun$ and 0.0 dex, respectively. Occurrence rates have units of companions per star, and are calculated using the histogram model. The ``Significance" column provides the statistical significance of the difference between the occurrence rates in the High and Low subsamples. The last two columns give the ratio between the median values of the listed stellar parameter in the two samples. For metallicity, the ratios are between linear abundances, $10^{\Delta[\text{Fe/H}]}$.
}
\end{deluxetable*}
 
\section{Discussion}
\label{sec:discussion}

\subsection{Mass and Metallicity Both Enhance Occurrence}
\label{subsec:mass_metallicity_occurrence}

Stellar mass and metallicity are correlated in the CLS, and because we divide the sample on one parameter without holding the other fixed in Section \ref{sec:results}, each pair of samples differs in both (Table \ref{tab:one_param_OR}). The median masses of the high-mass ($\Mstar>1.0~\Msun$) and low-mass ($\Mstar\leq1.0~\Msun$) samples differ by a factor of $\McMstarPaperBoundsMedianMstarRatio$, but their median metal abundances differ by a comparable factor of $\McMstarPaperBoundsMedianFeHRatio$, so the stellar mass comparison probes metallicity at least as much as mass. Conversely, the metal-rich ([Fe/H]$>0.0$) and metal-poor ([Fe/H]$\leq0.0$) samples differ by a factor of $\McFeHPaperBoundsMedianFeHRatio$ in metal abundance and by a smaller factor of $\McFeHPaperBoundsMedianMstarRatio$ in median mass. All ratios of metallicity refer to linear metal abundance, $10^{\Delta[\mathrm{Fe/H}]}$.

To isolate the separate effects of mass and metallicity on occurrence, we subdivide the sample to control for both parameters simultaneously. Table \ref{tab:two_param_OR} shows integrated occurrence rates for each combination of high and low mass and metallicity. In all four cases, the parameter of interest spans a larger dynamic range than the confounding parameter (see Table \ref{tab:two_param_OR_reordered}). We find that increasing either parameter can enhance giant planet occurrence significantly, depending on the value of the controlled parameter.

\begin{deluxetable*}{lccccccc}
\tablecaption{Occurrence Rates by Stellar Mass and Metallicity}
\label{tab:two_param_OR}
\tablehead{
\colhead{} & \colhead{} & \colhead{Median} & \colhead{Median} & \colhead{} & \colhead{} & \colhead{Average} & \colhead{Occurrence} \\[-1.6ex]
\colhead{Mass} & \colhead{$\text{[Fe/H]}$} & \colhead{$M_{\star}$} & \colhead{$\text{[Fe/H]}$} & \colhead{$N_{\star}$} & \colhead{$N_{\mathrm{eff}}$} & \colhead{Completeness} & \colhead{Rate}
}
\startdata
High & High & $\McStellarTwoParamsHighMstarHighFeHMedianMstar$ & $\McStellarTwoParamsHighMstarHighFeHMedianFeH$ & \McStellarTwoParamsHighMstarHighFeHNstars & \McStellarTwoParamsHighMstarHighFeHNeff & \McStellarTwoParamsHighMstarHighFeHAvgCompl & \McStellarTwoParamsHighMstarHighFeHPiecewiseIntOcc \\
High & Low & $\McStellarTwoParamsHighMstarLowFeHMedianMstar$ & $\McStellarTwoParamsHighMstarLowFeHMedianFeH$ & \McStellarTwoParamsHighMstarLowFeHNstars & \McStellarTwoParamsHighMstarLowFeHNeff & \McStellarTwoParamsHighMstarLowFeHAvgCompl & \McStellarTwoParamsHighMstarLowFeHPiecewiseIntOcc \\
Low & High & $\McStellarTwoParamsLowMstarHighFeHMedianMstar$ & $\McStellarTwoParamsLowMstarHighFeHMedianFeH$ & \McStellarTwoParamsLowMstarHighFeHNstars & \McStellarTwoParamsLowMstarHighFeHNeff & \McStellarTwoParamsLowMstarHighFeHAvgCompl & \McStellarTwoParamsLowMstarHighFeHPiecewiseIntOcc \\
Low & Low & $\McStellarTwoParamsLowMstarLowFeHMedianMstar$ & $\McStellarTwoParamsLowMstarLowFeHMedianFeH$ & \McStellarTwoParamsLowMstarLowFeHNstars & \McStellarTwoParamsLowMstarLowFeHNeff & \McStellarTwoParamsLowMstarLowFeHAvgCompl & \McStellarTwoParamsLowMstarLowFeHPiecewiseIntOcc \\
\enddata

\tablenotetext{}{%
\textbf{Note.} Analogous to Table \ref{tab:one_param_OR}, controlling for both mass and metallicity. We use the same mass and metallicity divisions at $1.0~\Msun$ and 0.0 dex, respectively.
}
\end{deluxetable*}
\begin{deluxetable*}{cccccc}
\tablecaption{Occurrence Rates with One Stellar Parameter Held Fixed}
\label{tab:two_param_OR_reordered}
\tablehead{
\colhead{} & \multicolumn{2}{c}{} & \colhead{} & \colhead{High/Low} & \colhead{High/Low} \\[-1.6ex]
\colhead{Fixed Parameter} & \multicolumn{2}{c}{Occurrence Rate} & \colhead{Significance} & \colhead{$M_{\star}$} & \colhead{$\text{[Fe/H]}$}
}
\startdata
\textbf{$\text{[Fe/H]}$} & \textbf{Low Mass} & \textbf{High Mass} &  &  & \\
\hline
High & \McStellarTwoParamsLowMstarHighFeHPiecewiseIntOcc & \McStellarTwoParamsHighMstarHighFeHPiecewiseIntOcc & \McStellarTwoParamsMassHighFeHSignificance & \McStellarTwoParamsMassHighFeHMedianMstarRatio & \McStellarTwoParamsMassHighFeHMedianFeHRatio \\
Low & \McStellarTwoParamsLowMstarLowFeHPiecewiseIntOcc & \McStellarTwoParamsHighMstarLowFeHPiecewiseIntOcc & \McStellarTwoParamsMassLowFeHSignificance & \McStellarTwoParamsMassLowFeHMedianMstarRatio & \McStellarTwoParamsMassLowFeHMedianFeHRatio \\
\hline
\textbf{Mass} & \textbf{Low $\text{[Fe/H]}$} & \textbf{High $\text{[Fe/H]}$} &  &  & \\
\hline
High & \McStellarTwoParamsHighMstarLowFeHPiecewiseIntOcc & \McStellarTwoParamsHighMstarHighFeHPiecewiseIntOcc & \McStellarTwoParamsFeHHighMstarSignificance & \McStellarTwoParamsFeHHighMstarMedianMstarRatio & \McStellarTwoParamsFeHHighMstarMedianFeHRatio \\
Low & \McStellarTwoParamsLowMstarLowFeHPiecewiseIntOcc & \McStellarTwoParamsLowMstarHighFeHPiecewiseIntOcc & \McStellarTwoParamsFeHLowMstarSignificance & \McStellarTwoParamsFeHLowMstarMedianMstarRatio & \McStellarTwoParamsFeHLowMstarMedianFeHRatio \\
\hline
\enddata

\tablenotetext{}{%
\textbf{Note.} This table contains the same occurrence rates as Table \ref{tab:two_param_OR}, reordered to aid comparison across stellar parameters. Each row lists a fixed stellar parameter, the occurrence rate as a function of the parameter being tested, the statistical significance of the difference of those rates, and the ratio of median stellar mass and metallicity for the subsample in question.
}
\end{deluxetable*}

\subsection{Mass Influences Relative Giant Planet Occurrence}
\label{subsec:mass_metallicity_shape}

Our experiments in Section \ref{subsec:mass_metallicity_occurrence} demonstrate that giant companion occurrence scales with $M_{\star}$ and [Fe/H]. However, it is unclear from integrated occurrence rates alone how these parameters affect the \textit{relative} occurrence of giants, i.e., the shape of the companion mass function. Using population synthesis models, \cite{Mordasini2012} found that mass and metallicity contribute to giant planet occurrence in distinct ways: higher-metallicity stars produce giant planets in greater numbers, while higher-mass stars produce higher-mass giant planets. 

\begin{figure*}[t]
    \centering
        \centering
        \includegraphics[width=\linewidth]{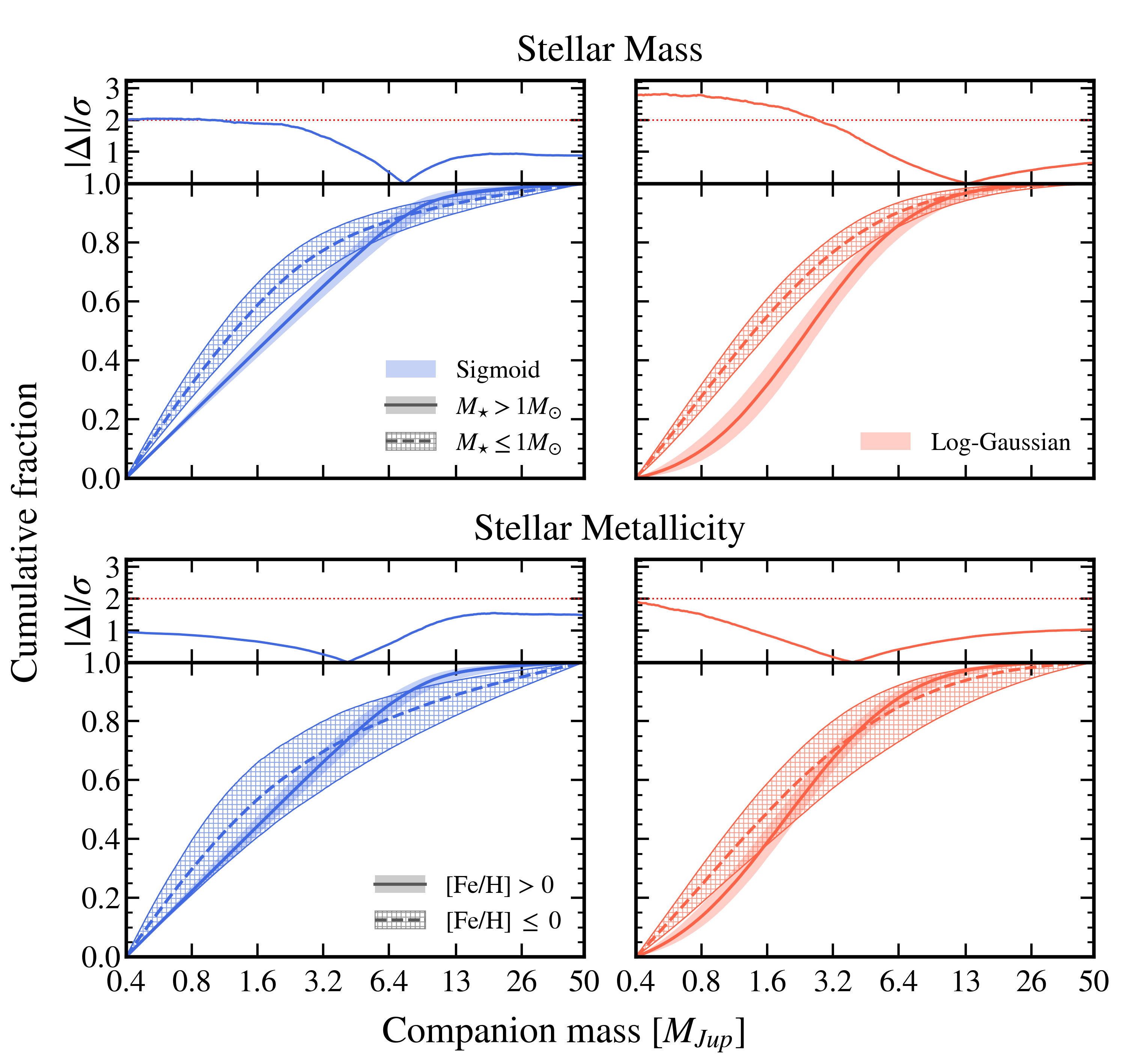}
    \caption{Cumulative distribution functions for the parametric model fits in Section \ref{sec:results}. The upper left panel shows the CDFs of the \model{sigmoid} fit to the high-mass (solid) and low-mass (hatched) stellar catalogs. The top sub-panel shows the absolute difference between the CDFs, normalized by the quadrature sum of the uncertainties. The red dotted line marks a $2\sigma$ difference. The upper right panel shows the same for the \model{logG} model. The bottom row shows analogous results for metallicity. Distribution shapes vary more with stellar mass ($\gtrsim2\sigma$) than with metallicity ($<2\sigma$).}
    \label{fig:CDF_comparison_mtrue}
\end{figure*}

To test whether stellar host mass is responsible for the shift in the companion mass function observed both for the high/low-$M_{\star}$ ($M_{\star}>1.0~\Msun$ and $\leq1.0~\Msun$; Section \ref{subsec:results_Mstar}) and high/low-[Fe/H] samples ([Fe/H]$>0.0$ and $\leq0.0$; Section \ref{subsec:results_FeH}), we compare the effect size as a function of each stellar parameter. Varying stellar mass from high to low shifts the \model{sigmoid} transition point by a factor of $\McMstarPaperBoundsSigmoidParamCenterFactorBinaZero$ in mass space ($\McMstarPaperBoundsSigmoidParamCenterSignificanceBinaZero$), and the \model{logG} center by a factor of $\McMstarPaperBoundsLogGParamMuFactorBinaZero$ ($\McMstarPaperBoundsLogGParamMuSignificanceBinaZero$). By comparison, varying between metal-rich and metal-poor stars produces shift factors of $\McFeHPaperBoundsSigmoidParamCenterFactorBinaZero$ ($\McFeHPaperBoundsSigmoidParamCenterSignificanceBinaZero$), and $\McFeHPaperBoundsLogGParamMuFactorBinaZero$ ($\McFeHPaperBoundsLogGParamMuSignificanceBinaZero$) for the sigmoid and logG models, respectively. The greater difference in shape between the high- and low-mass distributions is also apparent in the cumulative distribution functions (CDFs) of our model fits (see Figure \ref{fig:CDF_comparison_mtrue}).

\begin{figure*}[t]
    \centering
        \centering
        \includegraphics[width=\linewidth]{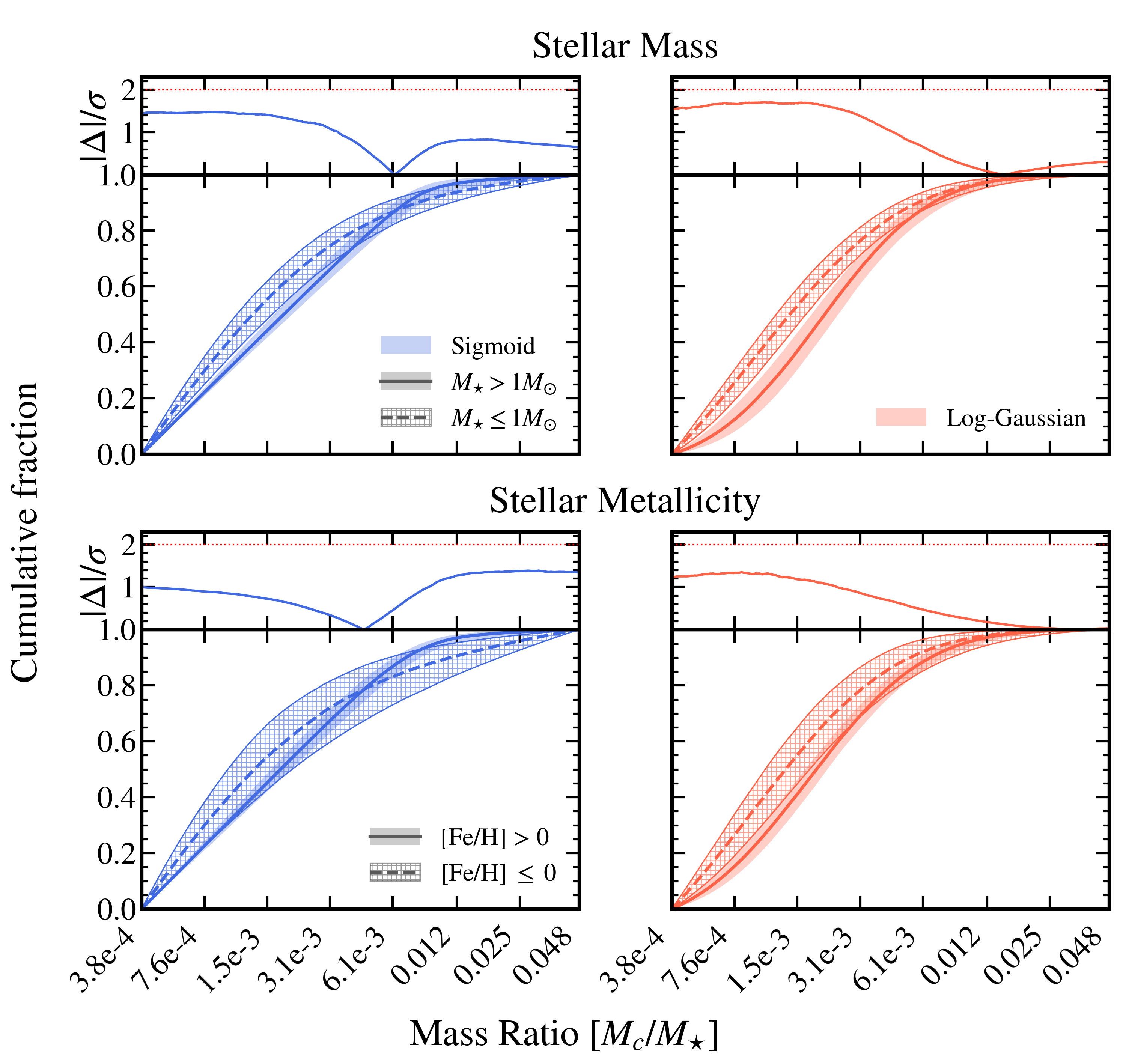}
    \caption{Same as Figure \ref{fig:CDF_comparison_mtrue} but for companion mass ratio. Normalizing for stellar mass shrinks the difference between the high- and low-mass distributions to below $2\sigma$, while the variation with metallicity remains $\gtrsim1\sigma$.}
    \label{fig:CDF_comparison_qtrue}
\end{figure*}

The mere fact that the shape of the companion mass distribution varies between the high/low-$M_{\star}$ samples is insufficient to conclude that stellar mass itself drives this effect because of the equally strong metallicity contrast between these samples (Table \ref{tab:one_param_OR}). However, the high/low-[Fe/H] samples exhibit both greater metallicity contrast and less significant variation in their distribution shapes, disfavoring metallicity as the sole source of the difference. To further investigate the role of stellar mass on the shape disparity, we calculate occurrence as a function of companion mass ratio. Figure \ref{fig:CDF_comparison_qtrue} shows that in mass ratio space, the difference between the high- and low-$\Mstar$ distributions is significantly reduced, and that they are consistent at $\QMstarPaperBoundsSigmoidParamCenterSignificanceBinaZero$. This finding favors the hypothesis that unlike the mass function, the mass ratio function maintains a consistent shape across stellar mass.

\subsection{A Peak in the Giant Planet Mass Function?}
\label{subsec:dicussion_peak_in_mass_function}

Our finding of a peak in the mass function for the full sample matches the predictions of some planet population synthesis models for Sun-like stars (e.g., \citealt{IdaLin2005, Mordasini2009, Emsenhuber2025}). In the core accretion framework, planets that achieve solid masses of $\sim30~\Mearth$ will undergo runaway gas accretion, meaning that few planets have final masses at this critical value. At high masses, planet formation is limited by the disk's mass budget and dissipation timescale \citep{TanigawaIkoma2007}. The minima at the critical core mass and the brown dwarf desert result in a peak at a few $\Mjup$. By contrast, simple formation models assuming equipartition of mass (e.g., \citealt{Schlichting2013}) result in monotonically increasing occurrence toward lower masses.

\begin{figure}[t]
    \centering
        \centering
        \includegraphics[width=\linewidth]{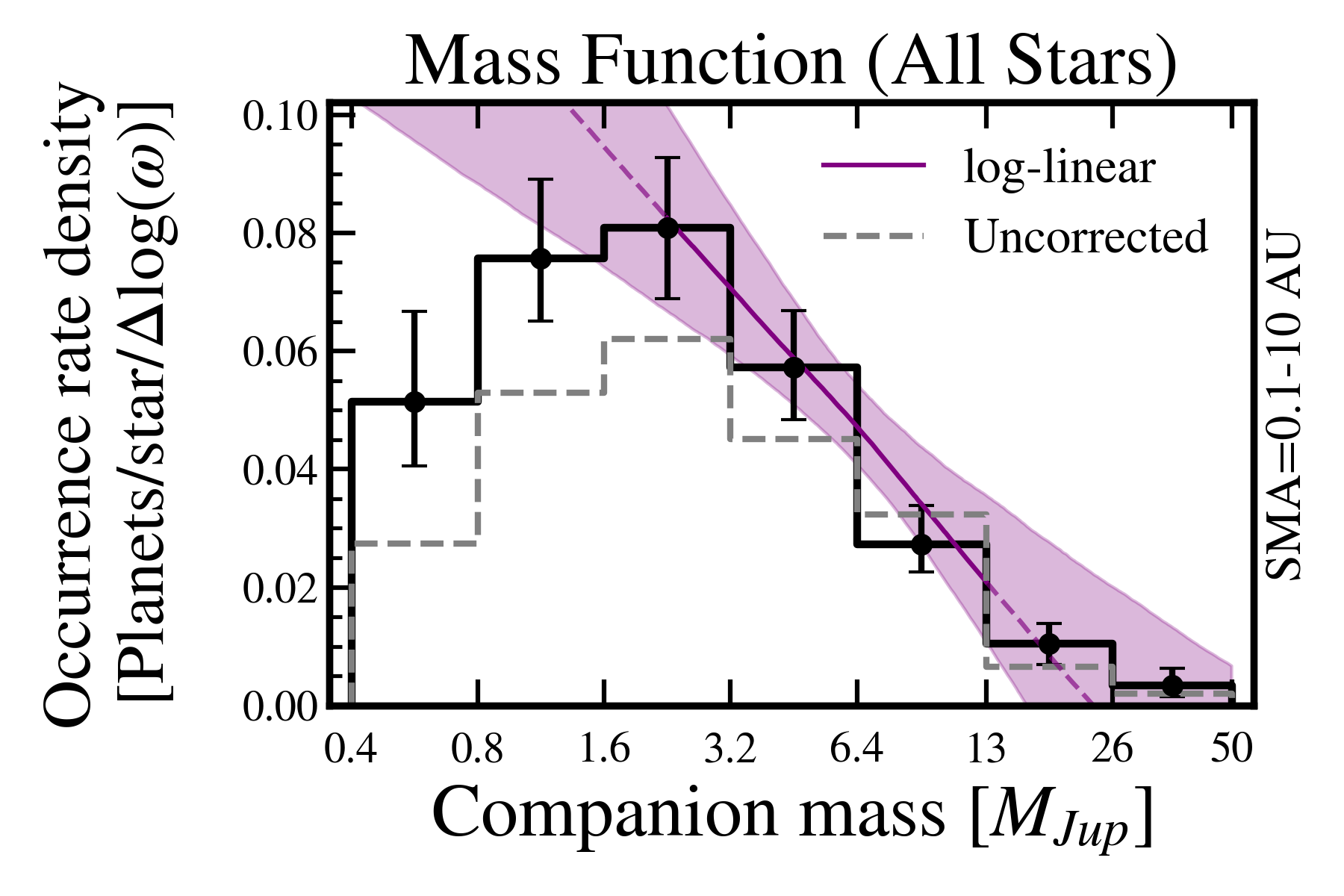}
    \caption{Same as Figure \ref{fig:ORD_allstars}, fit with the \model{log-linear} model (purple). We fit the \model{log-linear} model to a subset of companions with masses of $M_c=2.26-13~\Mjup$ (solid purple line), and extrapolate the model to higher and lower masses (dashed purple line) to test its predictive power. This model agrees with our histogram fit for intermediate masses ($\sim3-20~\Mjup$), but diverges below $\sim1.6~\Mjup$, differing by $2-3\sigma$ with the lowest-mass interval.}
    \label{fig:discussion_peak_in_mass_function}
\end{figure}

To test the consistency of the catalog with a peaked versus monotonic mass function, we fit a \model{log-linear} model to a subset of the catalog between $M_c=2.26-13~\Mjup$, approximating the interval over which the distribution appears log-linear (Figure \ref{fig:discussion_peak_in_mass_function}). The \model{log-linear}, \model{logG}, and \model{sigmoid} models agree between $\sim3-20~\Mjup$, but the \model{log-linear} fit diverges for low masses ($M_c\lessapprox1.6~\Mjup$): its modeled ORD at $0.6~\Mjup$ exceeds the corresponding \model{histogram} value by $2-3\sigma$. Thus, while we cannot definitively differentiate between a plateau, a peak, or even a shallower monotonic increase in companion occurrence at sub-Jovian masses, our results disfavor a log-linear increase from $13~\Mjup$ down to $0.4~\Mjup$ at $2-3\sigma$.

\subsection{The Brown Dwarf Desert in $M_c$ and $q$ Space}
\label{subsec:discussion_bd_desert_M_q}

\subsubsection{Comparison with Direct Imaging}
\begin{figure}[t]
    \centering
        \centering
        \includegraphics[width=\linewidth]{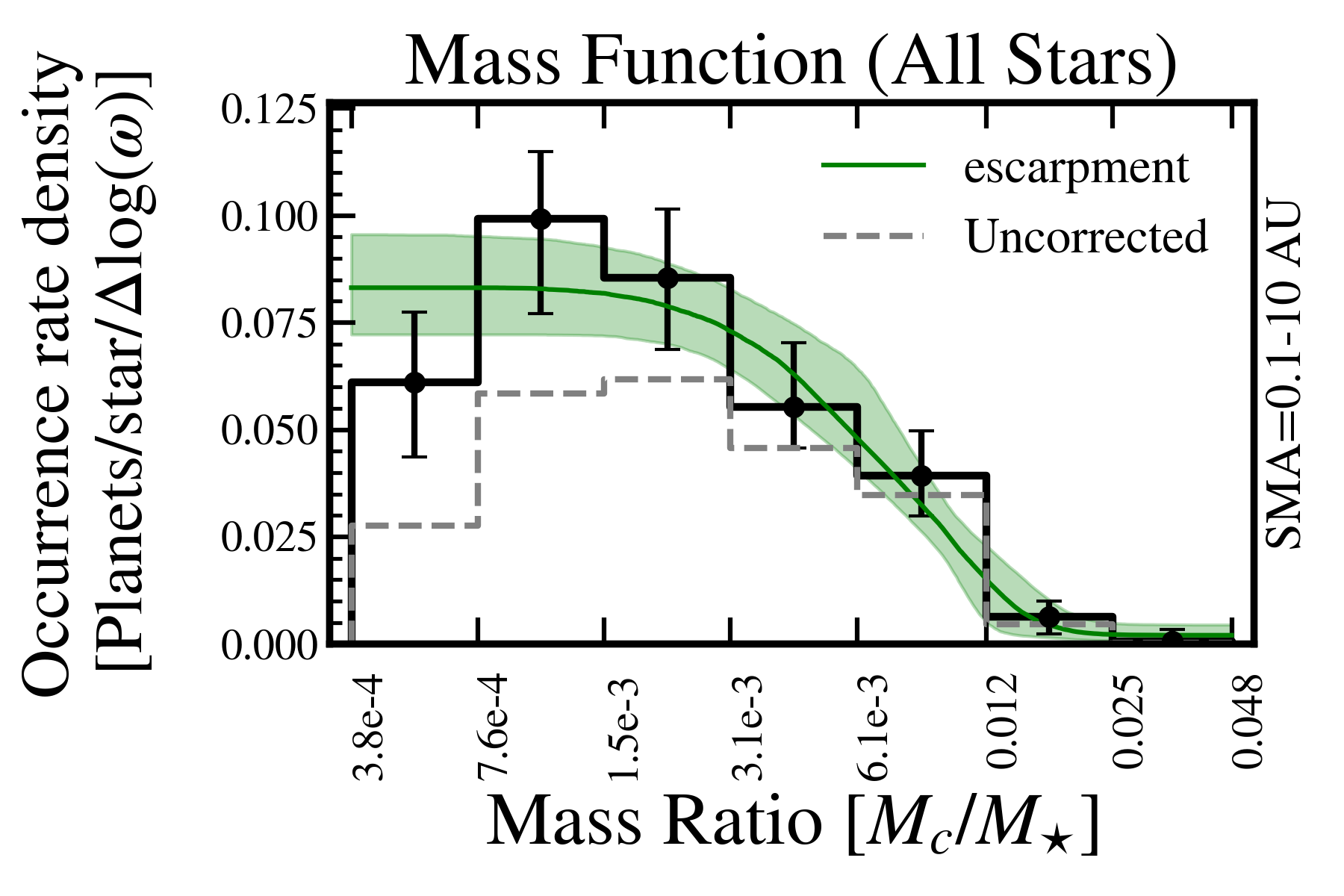}
    \caption{Occurrence rate density of companions in mass ratio space between $q\approx0.0004-0.05$ and 0.1--10 AU. We fit our \model{escarpment} model to this distribution because it allows for straightforward identification of features of the companion desert. For this fit we remove GP smoothing to probe the brown dwarf desert directly. This model suggests that the desert begins at $\log q=\QAllstarsPaperBoundsNoGPEscarpmentParamBPTwoBinaZero$ ($q=0.013-0.02$), consistent with previous findings of $q$=0.02. The abrupt drop in occurrence at the desert edge in our \model{histogram} model is somewhat more pronounced than the corresponding drop in the $M_c$ distribution.}
    \label{fig:ORD_allstars_q_discussion}
\end{figure}

\cite{Duchene2023} identified a companion desert in mass ratio space among a sample of 169 directly imaged A and B stars. They reported boundaries of $q=0.02-0.05$ for this desert, and predicted a ``steeply increasing number of giant planets towards lower masses around solar-type stars (at least down to $\approx1~\Mjup$)." We investigate the companion distribution in mass ratio space to assess whether the bounds of the desert are consistent across stellar type. Figure \ref{fig:ORD_allstars_q_discussion} shows our derived mass ratio distribution, which we calculate without GP smoothing to preserve the desert features. Although the sample sensitivity is insufficient to constrain the upper boundary of the brown dwarf desert ($q>0.05$), we find that the desert begins at $\log q=\QAllstarsPaperBoundsNoGPEscarpmentParamBPTwoBinaZero$ ($q=0.013-0.02$), consistent with \cite{Duchene2023}. Further, we find that the occurrence rate density function rises with decreasing mass below the desert, potentially plateauing below $\log q=\QAllstarsPaperBoundsNoGPEscarpmentParamBPOneBinaZero$ ($q=0.002-0.005$) or peaking at $\log M_c$$=\McAllstarsPaperBoundsLogGParamMuBinaZero$ ($M_c\approx1.7~\Mjup$; Section \ref{subsec:results_full_sample}), consistent with their prediction.

\subsubsection{Comparison with Microlensing}
We also compare our findings to those of \cite{Zhang2025}, who studied the brown dwarf desert as a function of companion mass ratio using two literature samples comprising a total of 134 companions detected via microlensing. They found a decreasing occurrence rate density between mass ratios of $-3.4<\log q<-1.75$ ($q=4.0\times 10^{-4}-1.8\times 10^{-2}$). They also found an abrupt truncation in the ORD at $\log q\approx-1.7$ ($q=0.02$), equivalent to a decrease in the ORD by a factor of eight compared to the previous bin. Finally, they calculated an integrated ORD of $5.8\pm0.8$ planets per 100 stars per $\Delta\log(a)$ in the range $q=4.0\times 10^{-4}-1.8\times 10^{-2}$. 

Our results are largely in agreement: our \model{escarpment} model fit supports decreasing occurrence between $\log q=\QAllstarsPaperBoundsNoGPEscarpmentParamBPOneBinaZero$ to $\QAllstarsPaperBoundsNoGPEscarpmentParamBPTwoBinaZero$ ($q=0.0025-0.016$), as shown in Figure \ref{fig:ORD_allstars_q_discussion}. We expect that our finding of decreasing occurrence beginning at a higher $q$ value --- $q=0.0025$ versus 0.0004 --- results from the greater flexibility of our \model{escarpment} model over the log-linear fit used by \cite{Zhang2025}.

The ORD values in our two brown dwarf desert bins ($q=0.012-0.048$) are $0.007^{+0.005}_{-0.003}$ and $0.002^{+0.003}_{-0.001}$ companions per star per decade in separation and mass ratio, compared to $0.041^{+0.011}_{-0.009}$ in the last pre-desert bin, yielding a decrease by a factor of $\sim9$ ($3.3\sigma$) at the desert boundary. By comparison, the boundary is marginally less pronounced in $M_c$ space, with a decrease by a factor of $\sim6$ ($2.7\sigma$) below versus within the desert.

We find an integrated occurrence rate of $\QAllstarsPaperBoundsPiecewiseIntOccBinaZero$ planets per star between $q=3.8\times 10^{-4}-4.8\times 10^{-2}$ and $a=0.1-10$ AU, corresponding to $0.21/\log\left(\frac{10}{0.1}\right)=$ 0.11 planets per star per decade of semi-major axis. Noting that our two highest-mass bins contribute only modestly to the total occurrence, we find that our integrated occurrence rate is approximately twice that of \cite{Zhang2025} over a similar range in companion mass ratio. This discrepancy may arise from the higher average stellar mass of the CLS sample. Although the host star mass distribution of \cite{Zhang2025}'s sample is not well-characterized, their derived occurrence rates are in good agreement with those calculated from \cite{Mignon2025}'s RV sample of 197 M dwarfs, whose median stellar mass ($0.48~\Msun$) was approximately half that of the CLS sample, $\McAllstarsPaperBoundsMedianMstar~\Msun$. If \cite{Zhang2025}'s sample has a similar median host mass, then the scaling of planet occurrence with stellar mass \citep{Johnson2010} implies that we would find an occurrence rate twice as high as that of \cite{Zhang2025} and \cite{Mignon2025} in our sample, consistent with our measurement.

Two features of microlensing samples complicate direct comparison with our results. First, microlensing measures companion mass ratios rather than host masses, so the host stars are characterized only statistically (typically $0.05-0.8~\Msun$), and some may be white dwarfs rather than main-sequence stars, as \cite{Zhang2025} notes for several companions within the desert. Second, a microlensing anomaly can often be fit by astrophysically distinct models, such as a planetary companion, a binary source, or a stellar-binary companion, which imply very different mass ratios. \cite{Zhang2025} excluded events with such ambiguities without correcting for those whose mass ratio was uncertain. Because these events preferentially lie near the boundary between planetary and stellar-binary mass ratios, their exclusion may cause the desert to appear emptier than it is. In contrast, each mass ratio in our catalog derives from an individual orbit fit to a star with a measured mass.

\subsubsection{Comparison with Astrometry}
Recently, Thompson et al. (submitted) mapped the mass ratio distribution among a sample of over 60,000 B--M stars characterized with Gaia DR2, DR3, and Hipparcos data, marking the first population-level demographics study using primarily astrometric constraints. They found a minimum in the brown dwarf desert at $1.76^{+0.26}_{-0.19}\%$ for host masses between $0.54-5.1~\Msun$. Although we do not detect the upper edge of the desert in our sample and therefore cannot measure the minimum, our finding that the desert begins at $\log q = \QAllstarsPaperBoundsNoGPEscarpmentParamBPTwoBinaZero$ ($q=1.3-2.0\%$) is compatible with their result.

\subsubsection{Comparison with Radial Velocities and Astrometry}
\cite{Cui2026} took a similar approach to ours by calculating occurrence as a function of companion true mass using a sample analyzed with both RVs and absolute astrometry. Their sample consists of 790 FGK stars hosting 55 companions, drawn from the samples of the Anglo-Australian Planet Search \citep{Jones2002, Tinney2003}, the HARPS Guaranteed Time Observations (GTO) program \citep{Mayor2003}, the Magellan Planet Search Program \citep{Arriagada2010}, and the California Planet Search \citep{Howard2010}. They found a lower mass boundary of $13.2^{+3.1}_{-2.3}~\Mjup$ for the brown dwarf desert, consistent with our finding of $\log M_c=\McAllstarsPaperBoundsNoGPEscarpmentParamBPTwoBinaZero$ ($M_c=20~\Mjup$) at the $1\sigma$ level. Further, they found an upper boundary of $70.5^{+25.3}_{-26.6}~\Mjup$, and a minimum in the desert at $30.9^{+5.5}_{-4.8}~\Mjup$. Although limitations in the sample sensitivity and composition prevent us from measuring an upper mass limit in the full sample, we measure a monotonic decrease in occurrence with increasing mass up to our largest mass bin of $26-50~\Mjup$, consistent with a minimum above $26~\Mjup$. 

We repeat their occurrence calculation between $2-20$ AU and $5-42~\Mjup$ and over the same stellar mass range they used, $\Mstar=0.6-1.4~\Msun$, which covers $\McCuiCutsCuiComparisonDiscussionNstars$ CLS stars. We do not apply GP smoothing to enable a more direct comparison. We find significantly lower occurrence rates: whereas \cite{Cui2026} found rates of $12.6^{+2.3}_{-1.9}\%$ between $5-14~\Mjup$ and $1.6^{+0.7}_{-0.5}\%$ between $24-42~\Mjup$, we find $\McCuiCutsCuiComparisonDiscussionPiecewiseIntOccBinmZero$ and $\McCuiCutsCuiComparisonDiscussionPiecewiseIntOccBinmTwo$, respectively, in these ranges. Further, they found a peak in the occurrence rate distribution near $10~\Mjup$, with decreasing occurrence toward lower masses, at odds with the trend of increasing occurrence down to $\sim1~\Mjup$ predicted by \cite{Duchene2023} and found in this work.

\subsubsection{Mass Ratio as a Fundamental Parameter}

\begin{figure}[t]
    \centering
    \begin{minipage}[t]{0.49\textwidth}
        \centering
        \includegraphics[width=\linewidth]{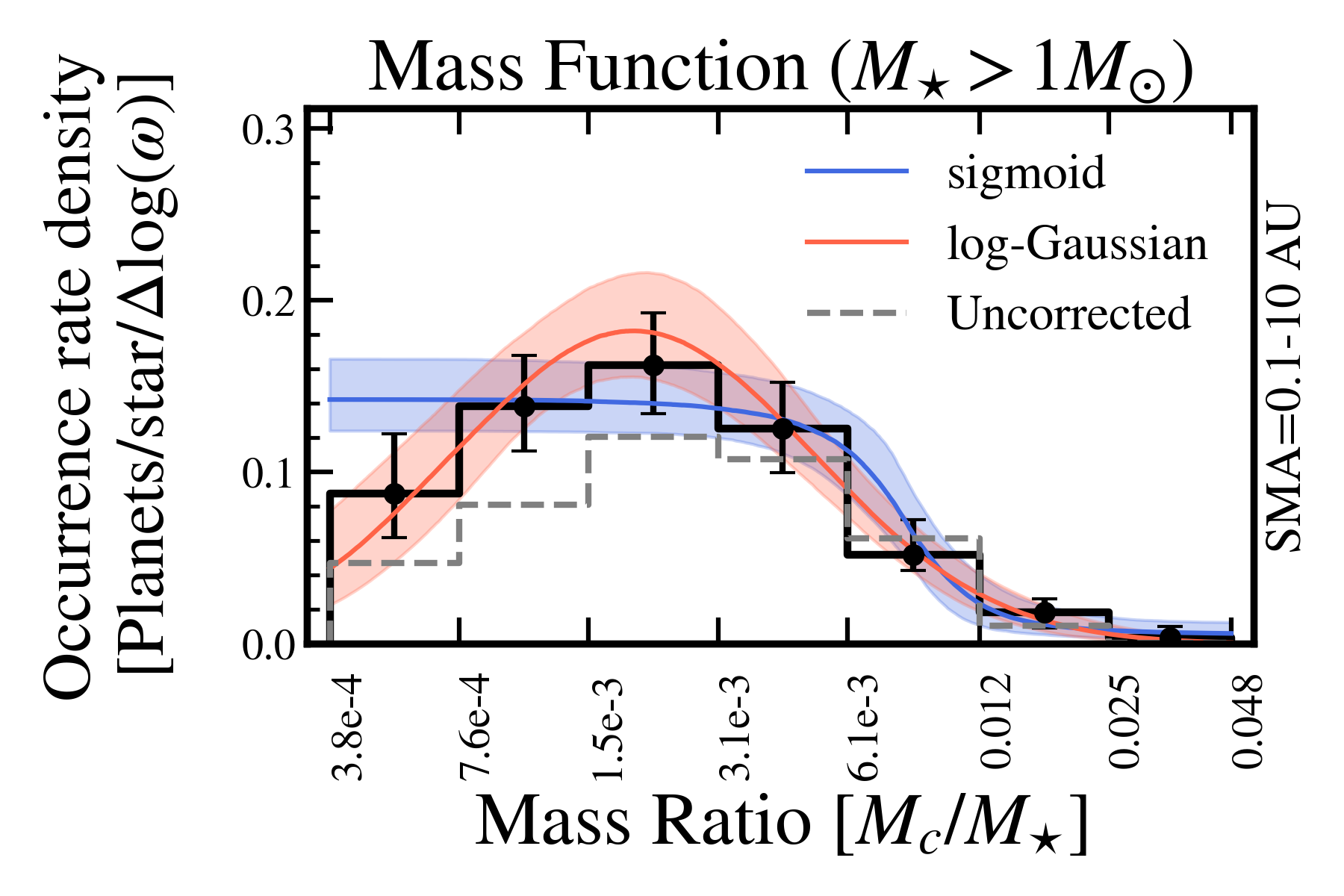}
        \vspace{0.5em}
        \includegraphics[width=\linewidth]{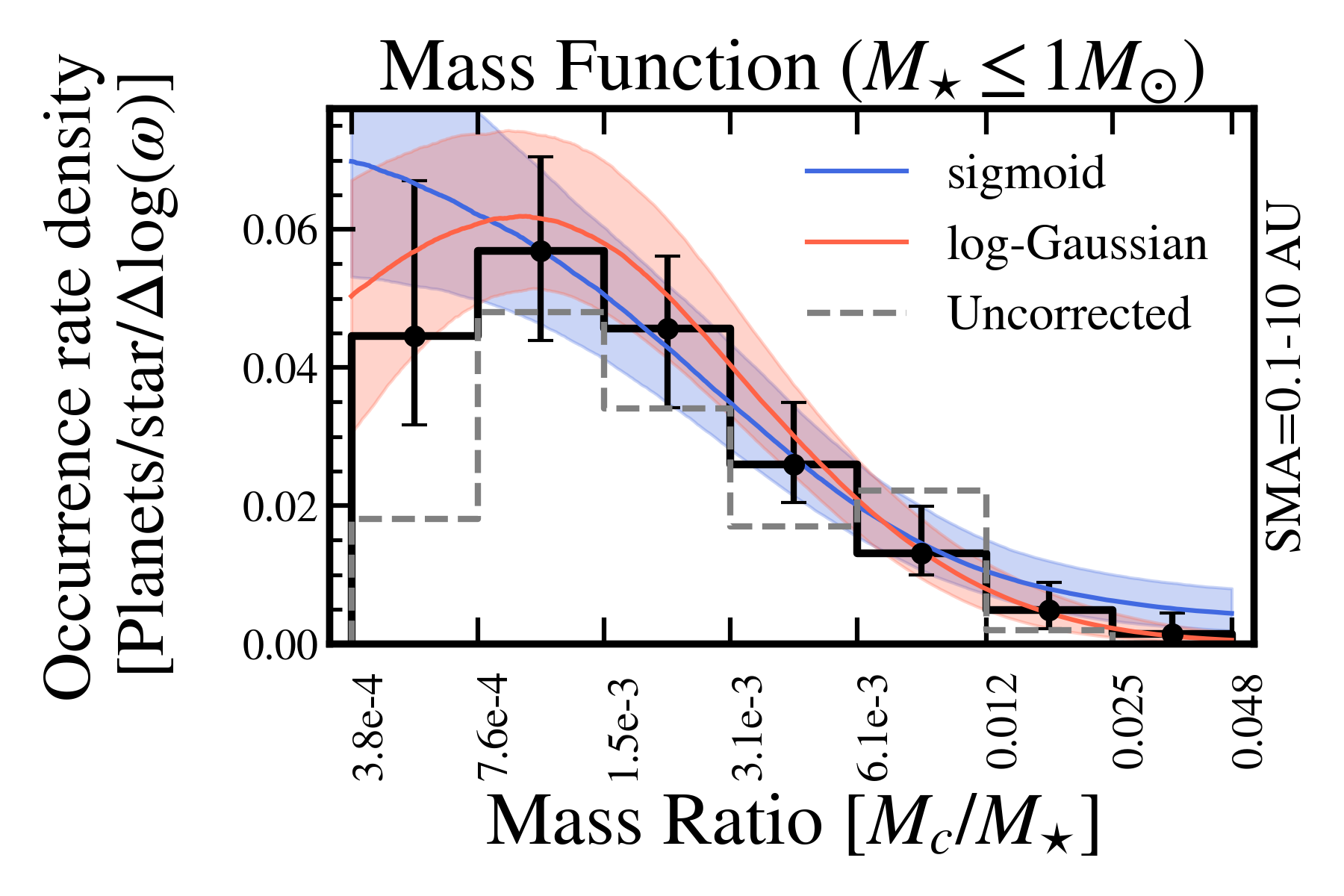}
    \end{minipage}
    \caption{Companion mass ratio functions for high-mass and low-mass stars, divided at $M_{\star}=1~M_{\odot}$. The difference in the locations of the \model{sigmoid} fits for the high-mass and low-mass samples ($\QMstarPaperBoundsSigmoidParamCenterDiffBinaZero$ dex) is somewhat less than in the true mass comparison ($\McMstarPaperBoundsSigmoidParamCenterDiffBinaZero$ dex). The same is true for the \model{logG} model ($\QMstarPaperBoundsLogGParamMuDiffBinaZero$ dex vs. $\McMstarPaperBoundsLogGParamMuDiffBinaZero$ dex).}
    \label{fig:ORD_Mstar_q_discussion}
\end{figure}

Recent studies of the brown dwarf desert across stellar mass have found that certain features persist in the companion mass ratio distribution, $\lambda(q)$, but not in the companion mass distribution, $\lambda(M_c)$ \citep{Duchene2023, Zhang2025, Meyer2025}. These findings have raised the question of whether $q$ is a more fundamental parameter to companion formation than $M_c$. Here we test this idea by exploring how the companion mass ratio function varies with stellar mass. Figure \ref{fig:ORD_Mstar_q_discussion} shows the companion distribution function in $q$ space for high-mass ($M_{\star}>1.0~\Msun$) and low-mass ($M_{\star}\leq1.0~\Msun$) stars. 

We fit a \model{sigmoid} transition point of $\log q=\QHighMstarPaperBoundsSigmoidParamCenterBinaZero$ for high-mass stars and $\log q = \QLowMstarPaperBoundsSigmoidParamCenterBinaZero$ for low-mass stars, a difference of $\QMstarPaperBoundsSigmoidParamCenterDiffBinaZero$ dex, compared to the difference of $\McMstarPaperBoundsSigmoidParamCenterDiffBinaZero$ dex we find for the true mass distributions in Section \ref{subsec:results_Mstar}. Our \model{logG} fit is centered at $\mu=\QHighMstarPaperBoundsLogGParamMuBinaZero$ for high-mass stars, compared to $\mu=\QLowMstarPaperBoundsLogGParamMuBinaZero$ for low-mass stars. This gives a difference of $\QMstarPaperBoundsLogGParamMuDiffBinaZero$ dex, compared to $\McMstarPaperBoundsLogGParamMuDiffBinaZero$ dex for the true mass distributions. The fact that the true mass distributions differ by more than the mass ratio distributions differ is expected based on the difference in stellar mass between the samples, $\log\frac{\QHighMstarPaperBoundsMedianMstar}{\QLowMstarPaperBoundsMedianMstar}\approx0.15$ dex.

Our results show that a universal true mass function is disfavored at the $\McMstarPaperBoundsSigmoidParamCenterSignificanceBinaZero$ (\model{sigmoid}) and $\McMstarPaperBoundsLogGParamMuSignificanceBinaZero$ (\model{logG}) levels, while the mass ratio functions are consistent within 
$\QMstarPaperBoundsSigmoidParamCenterSignificanceBinaZero$ (\model{sigmoid}) and $\QMstarPaperBoundsLogGParamMuSignificanceBinaZero$ (\model{logG}). The true mass transition shifts by $\McMstarPaperBoundsSigmoidParamCenterDiffBinaZero$ dex (\model{sigmoid}) and $\McMstarPaperBoundsLogGParamMuDiffBinaZero$ dex (\model{logG}) for a $\sim0.15$ dex difference in stellar mass, a much steeper scaling than the shift a universal mass ratio function would produce: because $\log M_c = \log q + \log M_\star$, a universal mass ratio function would shift the true-mass transition by only the $\sim0.15$ dex difference in stellar mass. The remainder appears as an offset between the mass ratio functions themselves ($\QMstarPaperBoundsSigmoidParamCenterDiffBinaZero$ and $\QMstarPaperBoundsLogGParamMuDiffBinaZero$ dex for the \model{sigmoid} and \model{logG} models), which is significant at only $\QMstarPaperBoundsSigmoidParamCenterSignificanceBinaZero$ (sigmoid) and $\QMstarPaperBoundsLogGParamMuSignificanceBinaZero$ (logG), so the data cannot distinguish it from zero. If confirmed, it would indicate that the characteristic companion mass grows more steeply than linearly with stellar mass, a scaling that neither a universal true mass function nor a universal mass ratio function accounts for. The relatively narrow stellar mass range of the sample limits our ability to differentiate the roles of true mass and mass ratio at high significance, and because both samples are fit over the same $q$ interval, which corresponds to different true mass ranges for high- and low-mass hosts, the interval edges may modestly affect the fitted locations; repeating this comparison across a larger range of stellar masses would clarify the role that mass ratio plays in planet formation.

\section{Conclusion}
\label{sec:conclusion}

In this work we use the California Legacy Survey, a sample of 719 FGKM stars with multi-decade RV monitoring hosting 194 companions with refit true masses, to measure the occurrence rates of giant companions between $0.1-10$ AU and $0.4-50~\Mjup$ as a function of companion and stellar parameters. Our Poisson occurrence framework simultaneously accounts for uncertainties in companion parameters and survey sensitivity in each stellar subset. We measure occurrence for the full sample as a function of both companion mass $M_c$ and mass ratio $q$, and make similar calculations for subsets of the sample divided according to each of stellar mass, metallicity, and activity, the last of which we use as a proxy for system age. We fit parametric models to our derived occurrence distributions to quantify differences in their shapes. To disentangle the influence of different stellar parameters, we calculate total occurrence rates as a function of one parameter while controlling for the other two. Our main results are as follows:

\begin{itemize}

    \item The companion mass function is itself a function of stellar mass. The mass function begins to fall monotonically at a higher characteristic companion mass for high-mass stars ($\Mstar>1.0~\Msun$) than for low-mass stars ($\Mstar\leq1.0~\Msun$). Among two samples that vary less in stellar mass and more in metallicity, this effect is less significant, suggesting that metallicity is not the driving factor.

    \item Our findings are consistent with stellar mass and metallicity both contributing to total giant planet occurrence, as found by \cite{Johnson2010}.

    \item Beyond $\sim3~\Mjup$, the companion mass function decreases monotonically up to the beginning of the brown dwarf desert at $\log M_c$$=\McAllstarsPaperBoundsNoGPEscarpmentParamBPTwoBinaZero$ ($16-25~\Mjup$). We find evidence of a peak or plateau in the mass function near $\sim1-2~\Mjup$ which, if validated, points to a turnover consistent with the findings of \cite{Duchene2023} and \cite{Mordasini2012}.

    \item The brown dwarf desert is more sharply defined in $q$ space than in $M_c$ space in our sample. The mass ratio functions of high- and low-mass stars agree within $\QMstarPaperBoundsSigmoidParamCenterSignificanceBinaZero$, while their true-mass functions differ at about $\McMstarPaperBoundsSigmoidParamCenterSignificanceBinaZero$. Our comparison spans a narrow mass range ($\sim0.15$ dex), and future studies over a larger range will provide more precise tests of how planet formation scales with stellar mass.

\end{itemize}

The California Legacy Survey is a rich resource for planet occurrence rate studies. Uniformly measured stellar parameters, companion orbital elements, and survey completeness reduce or eliminate many of the biases known to skew the results of population inference. Here we have used this sample to study companion occurrence in relation to one of the most fundamental companion properties, mass, and how this relation varies with stellar parameters.

The thousands of exoplanet discoveries around stars spanning a broad range of spectral types expected from the $Gaia$ mission's fourth data release \citep{Lammers2026} will drastically increase the scope and precision of giant planet occurrence studies both with the CLS and beyond. Sample sizes $1-2$ orders of magnitude larger than those analyzed in this study, especially when combined with RV observations, will enable clarification of stellar parameter correlations, reveal fine structure in occurrence rate distributions, and permit joint analyses of occurrence in the space of multiple orbital parameters at once.

\begin{acknowledgments}

This study makes use of the California Legacy Survey, a multi-decade effort representing thousands of hours of human planning, coordination, and observing time. The authors recognize and are grateful for the work that went into producing this rich data set, especially by Andrew Howard, Erik Petigura, Lauren Weiss, B.J. Fulton, and Howard Isaacson.

The authors wish to recognize and acknowledge the very significant cultural role and reverence that the summit of Maunakea has always had within the Native Hawaiian community. We are very fortunate to have the opportunity to make use of observations conducted from this mountain.
\end{acknowledgments}

\software{\texttt{astropy} \citep{astropy:2013, astropy:2018, astropy:2022}, \texttt{emcee} \citep{ForemanMackey2013}}

\appendix

\section{Stellar Age}
\label{app:age}

We present our treatment of stellar age in this appendix because of the large uncertainties in the age analysis and the resulting inconclusiveness of our findings. Only about half of the sample falls within the mass range supported by the activity-age relation of \cite{MamajekHillenbrand2008} (Appendix \ref{subsec:activity_ages}). The consequent reduction in sample size leaves the young sample with too few companions ($N_{\text{eff}}=\McHighActPaperBoundsNeff$) to constrain the shape of its mass function. Nevertheless, we provide our findings here for completeness and to serve as a starting point for more comprehensive studies in the future.

\subsection{Activity-Derived Ages}
\label{subsec:activity_ages}

Stellar mass and metallicity are essentially fixed over the main sequence lifetime, meaning that while they provide insights into the environment in which planets formed, they give no information about the system's post-formation evolution. By contrast, stellar activity decreases over the system lifetime, and can therefore serve as a proxy for system age (e.g., \citealt{Skumanich1972, Soderblom1991}). Various features of the exoplanet population have been shown or suggested to correlate with age, such as the tendency for hot Jupiters to orbit younger stars \citep{HamerSchlaufman2019}, the tendency for long-period giants to orbit older stars \citep{Swastik2023}, and the decreased average planet multiplicity and increased mutual inclinations of planets orbiting older stars \citep{Yang2023}. These trends suggest that secular processes such as tidal decay and planet-planet interactions can significantly alter system architectures on Gyr timescales. In this section, we examine whether such patterns are present in the CLS sample.

\cite{Isaacson2024} reported stellar activity values for 710 of the 719 CLS stars. They analyzed the excess emission in the cores of the Calcium H and K lines, $\log R^{\prime}_{HK}$, which serves as an indicator of chromospheric activity \citep{Noyes1984}, with larger $\log R^{\prime}_{HK}$ values corresponding to more active stars. Because the $\log R^{\prime}_{HK}$ from a single stellar spectrum captures the stellar activity only at the epoch of observation, \cite{Isaacson2024} reported activity values averaged over multiple epochs, with each star having at least five measurements.

\begin{figure}[t]
    \centering
    \includegraphics[width=0.99\linewidth]{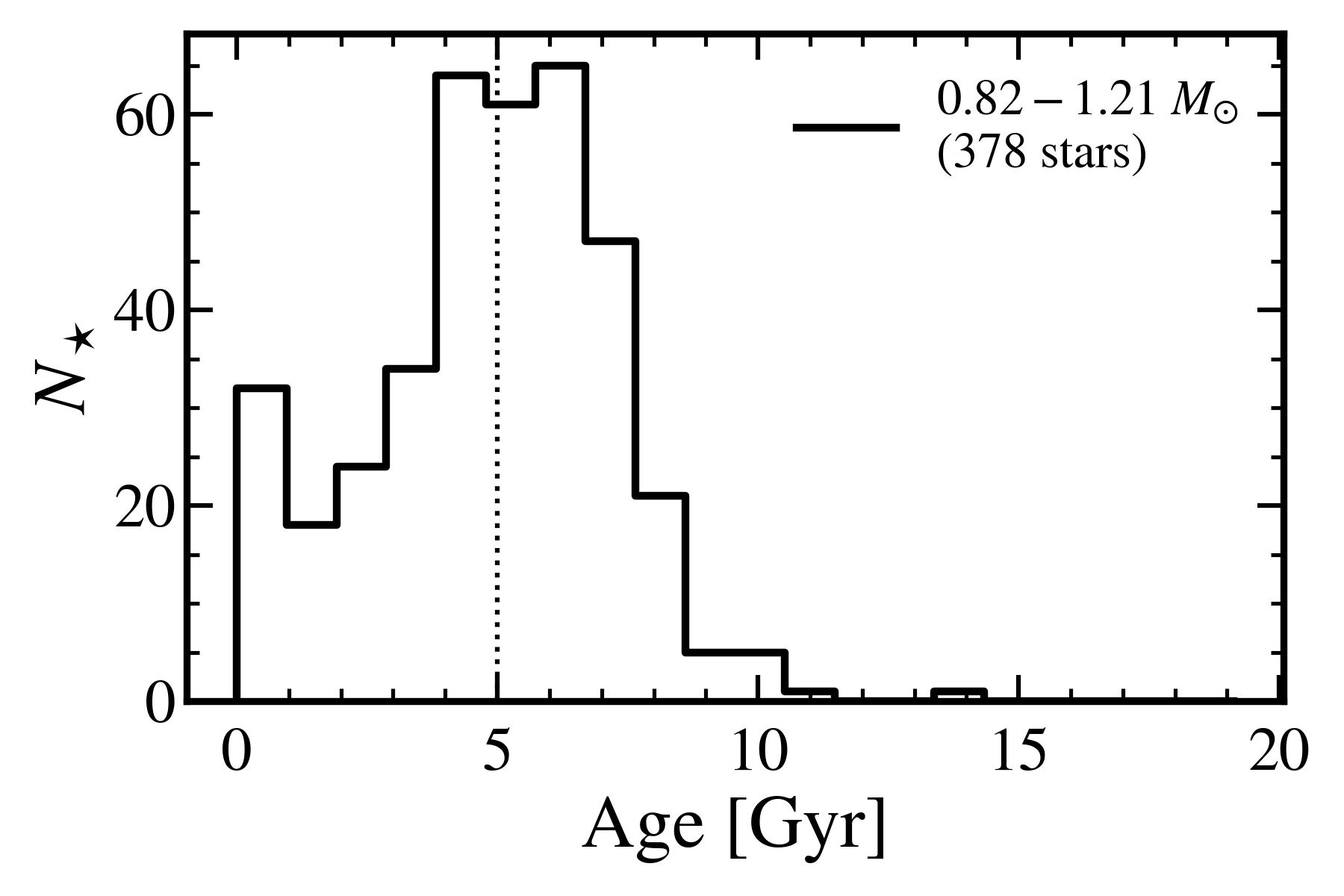}
    \caption{The distribution of activity-derived stellar ages for the $0.82-1.21~\Msun$ stars in the CLS, the range corresponding to the K2--F7 dwarfs used by \cite{MamajekHillenbrand2008} to convert activity to age. The vertical dotted line at 5 Gyr divides our ``old" and ``young" samples.}
    \label{fig:histogram_age}
\end{figure}

\begin{figure}[t]
    \centering
    \includegraphics[width=0.99\linewidth]{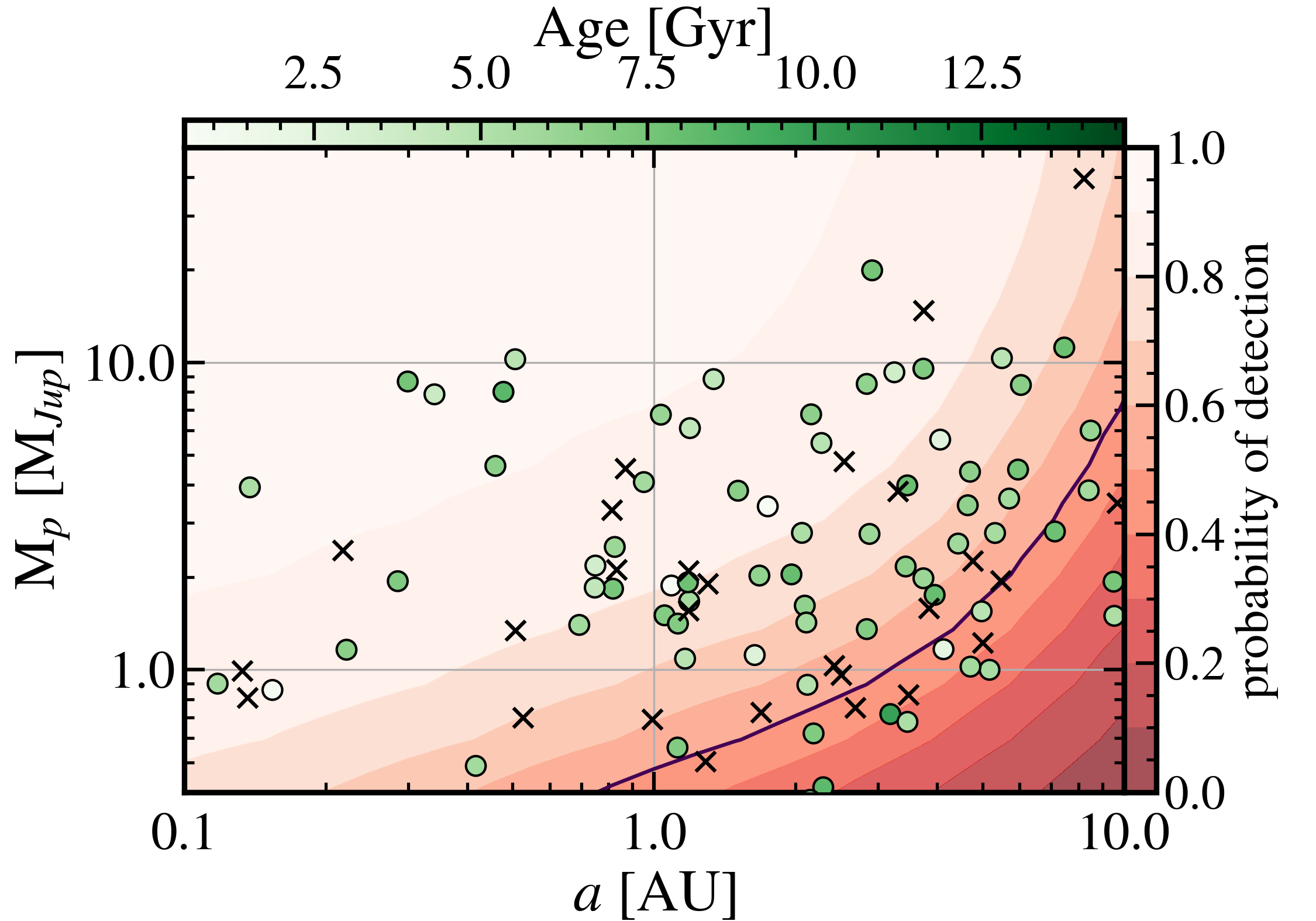}
    \caption{Same as Figure \ref{fig:catalog_with_stellar_params}, with marker colors corresponding to host star age. We mark with a black $\times$ companions with a host star mass outside the $0.82-1.21~\Msun$ range over which the activity-age relation of \cite{MamajekHillenbrand2008} is defined.}
    \label{fig:catalog_age}
\end{figure}

\cite{Isaacson2024} also calculated stellar ages using the empirical age-activity relation of \cite{MamajekHillenbrand2008}. Because this relation was derived using a sample of K2$-$F7 dwarfs, we restrict our analysis of stellar age to the same interval in spectral type. We use the relation between stellar mass and spectral type given by \cite{PecautMamajek2013} to estimate corresponding mass bounds of $0.82-1.21~\Msun$, a range that encompasses 378 CLS stars, about 53\% of the sample. This subsample has a median activity-derived age of 5.0 Gyr, with $16^{\text{th}}$ and $84^{\text{th}}$ percentiles of 2.5 Gyr and 7.1 Gyr, respectively. Figure \ref{fig:histogram_age} shows the distribution of modeled stellar ages in the CLS, and Figure \ref{fig:catalog_age} shows the companions in the catalog colored by the ages of their host stars.

\subsection{Companion Occurrence as a Function of Stellar Age}
\label{subsec:results_logrhk}

\begin{figure*}[t]
    \centering
    \begin{minipage}[t]{0.49\textwidth}
        \centering
        \includegraphics[width=\linewidth]{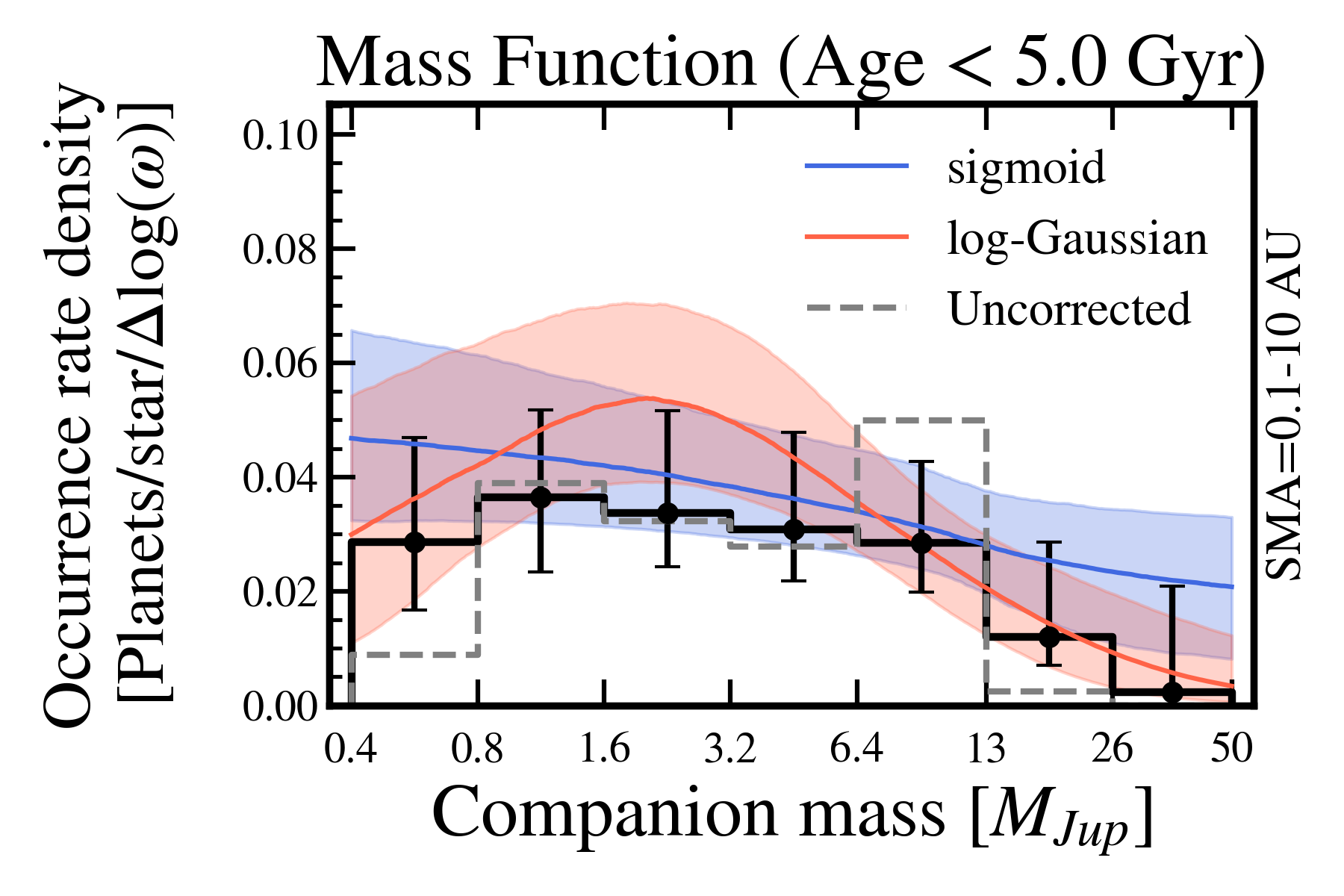}
    \end{minipage}
    \hfill
    \begin{minipage}[t]{0.49\textwidth}
        \centering
        \includegraphics[width=\linewidth]{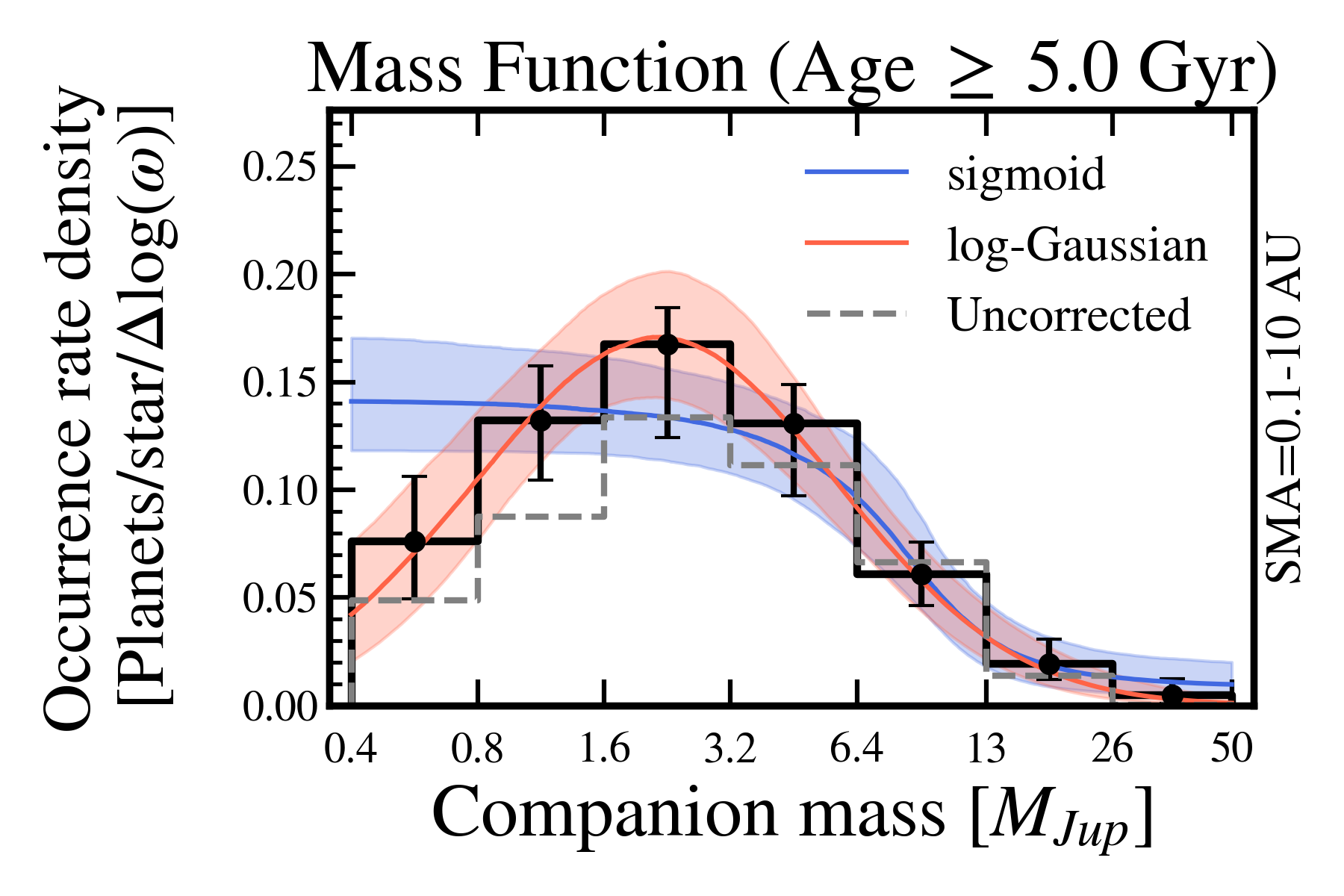}
    \end{minipage}

    \caption{
    The same as Figure \ref{fig:ORD_Mstar}, with the sample divided at an activity level of $\log R^{\prime}_{HK}=-4.969$, corresponding to an age of 5.0 Gyr through the activity-age relation of \cite{MamajekHillenbrand2008}. For our activity calculations, we consider only stars with $\Mstar=0.82-1.21~\Msun$, the mass range over which this relation was fit. Although we measure the shape of the mass function for the $\McLowActPaperBoundsNstars$ old stars in the sample, the small number of companions in the sample of $\McHighActPaperBoundsNstars$ young stars prevents a meaningful comparison.
    }    
    \label{fig:ORD_logrhk}
\end{figure*}

We use stellar activity, $\log R^{\prime}_{HK}$, as a proxy for stellar age through the age-activity correspondence given by \cite{MamajekHillenbrand2008}. For this experiment we use the subset of 378 CLS stars between $0.82-1.21~\Msun$, divided at the median activity of $\log R^{\prime}_{HK}=-4.969$, equivalent to 5.0 Gyr. The results of this experiment are shown in Figure \ref{fig:ORD_logrhk} and Table \ref{tab:model_params_appendix}. We examine age together with stellar mass and metallicity in Appendix \ref{subsec:discussion_three_parameters}.

The young sample, with $\McHighActPaperBoundsNeff$ effective companions among $\McHighActPaperBoundsNstars$ host stars, has significantly lower total occurrence, $\McHighActPaperBoundsPiecewiseIntOccBinaZero$ companions per star, than the old sample, which includes $\McLowActPaperBoundsNeff$ effective companions among $\McLowActPaperBoundsNstars$ hosts, and has an integrated occurrence rate of $\McLowActPaperBoundsPiecewiseIntOccBinaZero$ companions per star.

The low companion count in the young stellar sample prevents us from placing precise constraints on the mass function; our \model{sigmoid} fit is consistent with a flat occurrence rate, and all but one of the piecewise intervals include fewer than five effective companions. The distribution among the old stellar sample, by contrast, exhibits a shape consistent with the high-mass and metal-rich samples: the \model{sigmoid} model indicates high occurrence up to $\sim3~\Mjup$, followed by a decline at higher masses, while the \model{logG} function suggests a peak in occurrence near $2-3~\Mjup$.

\subsection{Occurrence as a Function of Mass, Metallicity, and Age}
\label{subsec:discussion_three_parameters}

\begin{deluxetable*}{lccccccccc}
\tablecaption{Occurrence Rates by Stellar Mass, Metallicity, and Age}
\label{tab:three_param_OR}
\tablehead{
\colhead{} & \colhead{} & \colhead{} & \colhead{Median} & \colhead{Median} & \colhead{Median} & \colhead{} & \colhead{} & \colhead{Average} & \colhead{Occurrence} \\[-1.6ex]
\colhead{Mass} & \colhead{$\text{[Fe/H]}$} & \colhead{Age} & \colhead{$M_{\star}$} & \colhead{$\text{[Fe/H]}$} & \colhead{Age (Gyr)} & \colhead{$N_{\star}$} & \colhead{$N_{\mathrm{eff}}$} & \colhead{Completeness} & \colhead{Rate}
}
\startdata
High & High & Young & $\McStellarThreeParamsHighMstarHighFeHYoungMedianMstar$ & $\McStellarThreeParamsHighMstarHighFeHYoungMedianFeH$ & $\McStellarThreeParamsHighMstarHighFeHYoungMedianAge$ & \McStellarThreeParamsHighMstarHighFeHYoungNstars & \McStellarThreeParamsHighMstarHighFeHYoungNeff & \McStellarThreeParamsHighMstarHighFeHYoungAvgCompl & \McStellarThreeParamsHighMstarHighFeHYoungPiecewiseIntOcc \\
High & High & Old & $\McStellarThreeParamsHighMstarHighFeHOldMedianMstar$ & $\McStellarThreeParamsHighMstarHighFeHOldMedianFeH$ & $\McStellarThreeParamsHighMstarHighFeHOldMedianAge$ & \McStellarThreeParamsHighMstarHighFeHOldNstars & \McStellarThreeParamsHighMstarHighFeHOldNeff & \McStellarThreeParamsHighMstarHighFeHOldAvgCompl & \McStellarThreeParamsHighMstarHighFeHOldPiecewiseIntOcc \\
High & Low & Young & $\McStellarThreeParamsHighMstarLowFeHYoungMedianMstar$ & $\McStellarThreeParamsHighMstarLowFeHYoungMedianFeH$ & $\McStellarThreeParamsHighMstarLowFeHYoungMedianAge$ & \McStellarThreeParamsHighMstarLowFeHYoungNstars & \McStellarThreeParamsHighMstarLowFeHYoungNeff & \McStellarThreeParamsHighMstarLowFeHYoungAvgCompl & \McStellarThreeParamsHighMstarLowFeHYoungPiecewiseIntOcc \\
High & Low & Old & $\McStellarThreeParamsHighMstarLowFeHOldMedianMstar$ & $\McStellarThreeParamsHighMstarLowFeHOldMedianFeH$ & $\McStellarThreeParamsHighMstarLowFeHOldMedianAge$ & \McStellarThreeParamsHighMstarLowFeHOldNstars & \McStellarThreeParamsHighMstarLowFeHOldNeff & \McStellarThreeParamsHighMstarLowFeHOldAvgCompl & \McStellarThreeParamsHighMstarLowFeHOldPiecewiseIntOcc \\
Low & High & Young & $\McStellarThreeParamsLowMstarHighFeHYoungMedianMstar$ & $\McStellarThreeParamsLowMstarHighFeHYoungMedianFeH$ & $\McStellarThreeParamsLowMstarHighFeHYoungMedianAge$ & \McStellarThreeParamsLowMstarHighFeHYoungNstars & \McStellarThreeParamsLowMstarHighFeHYoungNeff & \McStellarThreeParamsLowMstarHighFeHYoungAvgCompl & \McStellarThreeParamsLowMstarHighFeHYoungPiecewiseIntOcc \\
Low & High & Old & $\McStellarThreeParamsLowMstarHighFeHOldMedianMstar$ & $\McStellarThreeParamsLowMstarHighFeHOldMedianFeH$ & $\McStellarThreeParamsLowMstarHighFeHOldMedianAge$ & \McStellarThreeParamsLowMstarHighFeHOldNstars & \McStellarThreeParamsLowMstarHighFeHOldNeff & \McStellarThreeParamsLowMstarHighFeHOldAvgCompl & \McStellarThreeParamsLowMstarHighFeHOldPiecewiseIntOcc \\
Low & Low & Young & $\McStellarThreeParamsLowMstarLowFeHYoungMedianMstar$ & $\McStellarThreeParamsLowMstarLowFeHYoungMedianFeH$ & $\McStellarThreeParamsLowMstarLowFeHYoungMedianAge$ & \McStellarThreeParamsLowMstarLowFeHYoungNstars & \McStellarThreeParamsLowMstarLowFeHYoungNeff & \McStellarThreeParamsLowMstarLowFeHYoungAvgCompl & \McStellarThreeParamsLowMstarLowFeHYoungPiecewiseIntOcc \\
Low & Low & Old & $\McStellarThreeParamsLowMstarLowFeHOldMedianMstar$ & $\McStellarThreeParamsLowMstarLowFeHOldMedianFeH$ & $\McStellarThreeParamsLowMstarLowFeHOldMedianAge$ & \McStellarThreeParamsLowMstarLowFeHOldNstars & \McStellarThreeParamsLowMstarLowFeHOldNeff & \McStellarThreeParamsLowMstarLowFeHOldAvgCompl & \McStellarThreeParamsLowMstarLowFeHOldPiecewiseIntOcc \\
\enddata
\end{deluxetable*}
\begin{deluxetable*}{cccccc}
\tablecaption{Occurrence Rates with Two Stellar Parameters Held Fixed}
\label{tab:three_param_OR_reordered}
\tablehead{
\colhead{Fixed Parameter 1} &
\colhead{Fixed Parameter 2} &
\multicolumn{2}{c}{Occurrence Rate} &
\colhead{Significance} &
\colhead{High/Low}
}
\startdata
\textbf{$\text{[Fe/H]}$} & \textbf{Age} & \textbf{Low Mass} & \textbf{High Mass} &  & \\
\hline
High & Old & \McStellarThreeParamsLowMstarHighFeHOldPiecewiseIntOcc & \McStellarThreeParamsHighMstarHighFeHOldPiecewiseIntOcc & \McStellarThreeParamsMassHighFeHOldSignificance & \McStellarThreeParamsMassHighFeHOldDynamicRange \\
High & Young & \McStellarThreeParamsLowMstarHighFeHYoungPiecewiseIntOcc & \McStellarThreeParamsHighMstarHighFeHYoungPiecewiseIntOcc & \McStellarThreeParamsMassHighFeHYoungSignificance & \McStellarThreeParamsMassHighFeHYoungDynamicRange \\
Low & Old & \McStellarThreeParamsLowMstarLowFeHOldPiecewiseIntOcc & \McStellarThreeParamsHighMstarLowFeHOldPiecewiseIntOcc & \McStellarThreeParamsMassLowFeHOldSignificance & \McStellarThreeParamsMassLowFeHOldDynamicRange \\
Low & Young & \McStellarThreeParamsLowMstarLowFeHYoungPiecewiseIntOcc & \McStellarThreeParamsHighMstarLowFeHYoungPiecewiseIntOcc & \McStellarThreeParamsMassLowFeHYoungSignificance & \McStellarThreeParamsMassLowFeHYoungDynamicRange \\
\hline
\textbf{Mass} & \textbf{Age} & \textbf{Low $\text{[Fe/H]}$} & \textbf{High $\text{[Fe/H]}$} &  & \\
\hline
High & Old & \McStellarThreeParamsHighMstarLowFeHOldPiecewiseIntOcc & \McStellarThreeParamsHighMstarHighFeHOldPiecewiseIntOcc & \McStellarThreeParamsFeHHighMstarOldSignificance & \McStellarThreeParamsFeHHighMstarOldDynamicRange \\
High & Young & \McStellarThreeParamsHighMstarLowFeHYoungPiecewiseIntOcc & \McStellarThreeParamsHighMstarHighFeHYoungPiecewiseIntOcc & \McStellarThreeParamsFeHHighMstarYoungSignificance & \McStellarThreeParamsFeHHighMstarYoungDynamicRange \\
Low & Old & \McStellarThreeParamsLowMstarLowFeHOldPiecewiseIntOcc & \McStellarThreeParamsLowMstarHighFeHOldPiecewiseIntOcc & \McStellarThreeParamsFeHLowMstarOldSignificance & \McStellarThreeParamsFeHLowMstarOldDynamicRange \\
Low & Young & \McStellarThreeParamsLowMstarLowFeHYoungPiecewiseIntOcc & \McStellarThreeParamsLowMstarHighFeHYoungPiecewiseIntOcc & \McStellarThreeParamsFeHLowMstarYoungSignificance & \McStellarThreeParamsFeHLowMstarYoungDynamicRange \\
\hline
\textbf{Mass} & \textbf{$\text{[Fe/H]}$} & \textbf{Young} & \textbf{Old} &  & \\
\hline
High & High & \McStellarThreeParamsHighMstarHighFeHYoungPiecewiseIntOcc & \McStellarThreeParamsHighMstarHighFeHOldPiecewiseIntOcc & \McStellarThreeParamsAgeHighMstarHighFeHSignificance & \McStellarThreeParamsAgeHighMstarHighFeHDynamicRange \\
High & Low & \McStellarThreeParamsHighMstarLowFeHYoungPiecewiseIntOcc & \McStellarThreeParamsHighMstarLowFeHOldPiecewiseIntOcc & \McStellarThreeParamsAgeHighMstarLowFeHSignificance & \McStellarThreeParamsAgeHighMstarLowFeHDynamicRange \\
Low & High & \McStellarThreeParamsLowMstarHighFeHYoungPiecewiseIntOcc & \McStellarThreeParamsLowMstarHighFeHOldPiecewiseIntOcc & \McStellarThreeParamsAgeLowMstarHighFeHSignificance & \McStellarThreeParamsAgeLowMstarHighFeHDynamicRange \\
Low & Low & \McStellarThreeParamsLowMstarLowFeHYoungPiecewiseIntOcc & \McStellarThreeParamsLowMstarLowFeHOldPiecewiseIntOcc & \McStellarThreeParamsAgeLowMstarLowFeHSignificance & \McStellarThreeParamsAgeLowMstarLowFeHDynamicRange \\
\hline
\enddata

\tablenotetext{}{%
\textbf{Note.} Analogous to Table \ref{tab:two_param_OR_reordered} for stellar mass, metallicity, and age. The last column indicates the ratio between the median values of the high and low subsamples indicated in each row. Dividing on three stellar parameters reduces the typical number of effective companions per sample to the point that none of the listed experiments exceed $3\sigma$ significance.
}
\end{deluxetable*}
 
Subdividing the sample by all three stellar parameters minimizes the impact of confounding variables at the expense of increased Poisson noise. Table \ref{tab:three_param_OR_reordered} shows that at fixed metallicity and age, occurrence does not vary strongly (as much as $\McStellarThreeParamsMassLowFeHYoungSignificance$) between the low-$\Mstar$ and high-$\Mstar$ samples. At fixed mass and age, occurrence shows a similar dependence on metallicity, rising by $\McStellarThreeParamsFeHHighMstarOldSignificance$ among old high-mass stars, and $\McStellarThreeParamsFeHLowMstarOldSignificance$ among old low-mass stars. There is also marginal ($\McStellarThreeParamsFeHLowMstarYoungSignificance$) evidence for a correlation among young low-mass stars, though the low-metallicity occurrence rate in this comparison, having been derived from only two companion detections, is highly uncertain. At fixed mass and metallicity, the highest correlation between age and occurrence is among high-mass, metal-rich stars, where stars older than 5 Gyr show a $\McStellarThreeParamsAgeHighMstarHighFeHSignificance$ enhancement over their younger counterparts. For low-mass, metal-rich stars there is a weaker $\McStellarThreeParamsAgeLowMstarHighFeHSignificance$ enhancement, while the metal-poor stars exhibit no age-occurrence relation. Overall, we find no strong correlations between any single parameter and total occurrence when controlling for the other two parameters.

\subsection{The Effects of Age on Giant Planet Occurrence}
\label{subsubsec:discussion_age}

Our finding that stellar age may have either a neutral (around metal-poor stars) or slightly positive (around metal-rich stars) correlation with giant planet occurrence matches some previous studies of the giant planet-stellar age relationship. RV studies of intermediate-separation giant planets ($a\lesssim2.5$ AU) in young systems have found marginal evidence that the prevalence of giant planets increases over time \citep{Grandjean2023, Tran2025}. Meanwhile, yields from the K2 mission \citep{Howell2014} have shown that young stars host more hot sub-Neptunes than the field rate \citep{Christiansen2023}, while older stars host fewer \citep{Zink2023}. After controlling for stellar mass and metallicity, \cite{Sayeed2025} found no strong evidence that occurrence changes with system age among Kepler systems. These disagreements suggest that the age-occurrence correlation is a function of planetary mass and separation.

Using a subset of the CLS sample, \cite{Miyazaki2023} reported decreasing occurrence with system age for hot Jupiters (e.g., \citealt{Chen2023, Kontiainen2025}), and no strong relationship for cold Jupiters ($1-5$ AU, $0.3-10~\Mjup$). Nevertheless, their cold Jupiter occurrence results are consistent with the tentative enhancement we observe: our occurrence rates in Table \ref{tab:three_param_OR_reordered} show that giant occurrence among old stars (median age=$\McStellarThreeParamsHighMstarHighFeHOldMedianAge$ Gyr) is roughly triple that of the young sample (median age=$\McStellarThreeParamsHighMstarHighFeHYoungMedianAge$ Gyr) at $\McStellarThreeParamsAgeHighMstarHighFeHSignificance$ significance for the high-mass, high-metallicity bin. There is also a weaker $\McStellarThreeParamsAgeLowMstarHighFeHSignificance$ enhancement with age for low-mass, metal-rich stars. These potential increases are consistent with the prediction of the exponential occurrence model in \cite{Miyazaki2023} (see their Figure 9).

To make a more direct comparison, we adopt \cite{Miyazaki2023}'s definition of a cold Jupiter, albeit with companion true mass rather than $\Msini$, and recompute occurrence while controlling for stellar mass and metallicity. We find no strong correlation between age and occurrence in any of the stellar subsamples. It is possible that while the occurrence of close-in planets ($P\lesssim100$ days) decreases with system age over a wide range of masses, giant planets at wider separations ($P\gtrsim1$ year) exhibit no such trend, or even the reverse trend in metal-rich systems. This effect, if real, would require secular dynamical effects on Gyr timescales, such as inward migration of very long-period planets which are initially not detectable \citep{WuLithwick2011, PetrovichTremaine2016}. It is also possible that the low number of systems with modeled ages ($\sim50\%$ of the sample) and uncertainty in the age-activity relation itself produce this apparent effect.

\section{Fitted Model Parameters}
\label{app:model_params}
\begin{longrotatetable}
\begin{deluxetable*}{cccccccccccc}
\tablecaption{Fitted Model Parameters}
\label{tab:model_params_appendix}

\tablehead{
\shortstack{Mass\\Variable} &
\shortstack{Stellar\\Subsample} &
$N_\star$ &
\shortstack{$N_{\mathrm{eff}}$} &
$Q_{\text{avg}}$ &
Model &
\shortstack{Integrated\\Rate} &
$\theta_1$ & $\theta_2$ & $\theta_3$ & $\theta_4$ &
BIC
}
\startdata
$M_c$ & All Stars & \McAllstarsPaperBoundsNstars & \McAllstarsPaperBoundsNeffBinAZero & \McAllstarsPaperBoundsAvgComplBinAZero & sigmoid & \McAllstarsPaperBoundsSigmoidIntOccBinaZero & \shortstack{$B_1$\\\McAllstarsPaperBoundsSigmoidParamBOneBinaZero} & \shortstack{$B_2$\\\McAllstarsPaperBoundsSigmoidParamBTwoBinaZero} & \shortstack{$\log(x_t)$\\\McAllstarsPaperBoundsSigmoidParamCenterBinaZero} & \shortstack{$W$\\\McAllstarsPaperBoundsSigmoidParamWidthBinaZero} & \McAllstarsPaperBoundsSigmoidBicBinaZero \\
 &  &  &  &  & logG & \McAllstarsPaperBoundsLogGIntOccBinaZero & \shortstack{$A$\\\McAllstarsPaperBoundsLogGParamABinaZero} & \shortstack{$\mu$\\\McAllstarsPaperBoundsLogGParamMuBinaZero} & \shortstack{$\sigma$\\\McAllstarsPaperBoundsLogGParamSigmaBinaZero} & \nodata & \McAllstarsPaperBoundsLogGBicBinaZero \\
 &  &  &  &  & log-linear & \nodata & \shortstack{$C_1$\\\McAllstarsPaperBoundsLogLinearParamCOneBinaZero} & \shortstack{$C_2$\\\McAllstarsPaperBoundsLogLinearParamCTwoBinaZero} & \nodata & \nodata & \nodata \\
 &  &  &  &  & escarpment & \McAllstarsPaperBoundsNoGPEscarpmentIntOccBinaZero & \shortstack{$D_1$\\\McAllstarsPaperBoundsNoGPEscarpmentParamDOneBinaZero} & \shortstack{$D_2$\\\McAllstarsPaperBoundsNoGPEscarpmentParamDTwoBinaZero} & \shortstack{$\log(x_{t,1})$\\\McAllstarsPaperBoundsNoGPEscarpmentParamBPOneBinaZero} & \shortstack{$\log(x_{t,2})$\\\McAllstarsPaperBoundsNoGPEscarpmentParamBPTwoBinaZero} & \McAllstarsPaperBoundsNoGPEscarpmentBicBinaZero \\
\hline
$M_c$ & High $M_\star$ & \McHighMstarPaperBoundsNstars & \McHighMstarPaperBoundsNeffBinAZero & \McHighMstarPaperBoundsAvgComplBinAZero & sigmoid & \McHighMstarPaperBoundsSigmoidIntOccBinaZero & \shortstack{$B_1$\\\McHighMstarPaperBoundsSigmoidParamBOneBinaZero} & \shortstack{$B_2$\\\McHighMstarPaperBoundsSigmoidParamBTwoBinaZero} & \shortstack{$\log(x_t)$\\\McHighMstarPaperBoundsSigmoidParamCenterBinaZero} & \shortstack{$W$\\\McHighMstarPaperBoundsSigmoidParamWidthBinaZero} & \McHighMstarPaperBoundsSigmoidBicBinaZero \\
 &  &  &  &  & logG & \McHighMstarPaperBoundsLogGIntOccBinaZero & \shortstack{$A$\\\McHighMstarPaperBoundsLogGParamABinaZero} & \shortstack{$\mu$\\\McHighMstarPaperBoundsLogGParamMuBinaZero} & \shortstack{$\sigma$\\\McHighMstarPaperBoundsLogGParamSigmaBinaZero} & \nodata & \McHighMstarPaperBoundsLogGBicBinaZero \\
\hline
$M_c$ & Low $M_\star$ & \McLowMstarPaperBoundsNstars & \McLowMstarPaperBoundsNeffBinAZero & \McLowMstarPaperBoundsAvgComplBinAZero & sigmoid & \McLowMstarPaperBoundsSigmoidIntOccBinaZero & \shortstack{$B_1$\\\McLowMstarPaperBoundsSigmoidParamBOneBinaZero} & \shortstack{$B_2$\\\McLowMstarPaperBoundsSigmoidParamBTwoBinaZero} & \shortstack{$\log(x_t)$\\\McLowMstarPaperBoundsSigmoidParamCenterBinaZero} & \shortstack{$W$\\\McLowMstarPaperBoundsSigmoidParamWidthBinaZero} & \McLowMstarPaperBoundsSigmoidBicBinaZero \\
 &  &  &  &  & logG & \McLowMstarPaperBoundsLogGIntOccBinaZero & \shortstack{$A$\\\McLowMstarPaperBoundsLogGParamABinaZero} & \shortstack{$\mu$\\\McLowMstarPaperBoundsLogGParamMuBinaZero} & \shortstack{$\sigma$\\\McLowMstarPaperBoundsLogGParamSigmaBinaZero} & \nodata & \McLowMstarPaperBoundsLogGBicBinaZero \\
\hline
$M_c$ & High [Fe/H] & \McHighFeHPaperBoundsNstars & \McHighFeHPaperBoundsNeffBinAZero & \McHighFeHPaperBoundsAvgComplBinAZero & sigmoid & \McHighFeHPaperBoundsSigmoidIntOccBinaZero & \shortstack{$B_1$\\\McHighFeHPaperBoundsSigmoidParamBOneBinaZero} & \shortstack{$B_2$\\\McHighFeHPaperBoundsSigmoidParamBTwoBinaZero} & \shortstack{$\log(x_t)$\\\McHighFeHPaperBoundsSigmoidParamCenterBinaZero} & \shortstack{$W$\\\McHighFeHPaperBoundsSigmoidParamWidthBinaZero} & \McHighFeHPaperBoundsSigmoidBicBinaZero \\
 &  &  &  &  & logG & \McHighFeHPaperBoundsLogGIntOccBinaZero & \shortstack{$A$\\\McHighFeHPaperBoundsLogGParamABinaZero} & \shortstack{$\mu$\\\McHighFeHPaperBoundsLogGParamMuBinaZero} & \shortstack{$\sigma$\\\McHighFeHPaperBoundsLogGParamSigmaBinaZero} & \nodata & \McHighFeHPaperBoundsLogGBicBinaZero \\
\hline
$M_c$ & Low [Fe/H] & \McLowFeHPaperBoundsNstars & \McLowFeHPaperBoundsNeffBinAZero & \McLowFeHPaperBoundsAvgComplBinAZero & sigmoid & \McLowFeHPaperBoundsSigmoidIntOccBinaZero & \shortstack{$B_1$\\\McLowFeHPaperBoundsSigmoidParamBOneBinaZero} & \shortstack{$B_2$\\\McLowFeHPaperBoundsSigmoidParamBTwoBinaZero} & \shortstack{$\log(x_t)$\\\McLowFeHPaperBoundsSigmoidParamCenterBinaZero} & \shortstack{$W$\\\McLowFeHPaperBoundsSigmoidParamWidthBinaZero} & \McLowFeHPaperBoundsSigmoidBicBinaZero \\
 &  &  &  &  & logG & \McLowFeHPaperBoundsLogGIntOccBinaZero & \shortstack{$A$\\\McLowFeHPaperBoundsLogGParamABinaZero} & \shortstack{$\mu$\\\McLowFeHPaperBoundsLogGParamMuBinaZero} & \shortstack{$\sigma$\\\McLowFeHPaperBoundsLogGParamSigmaBinaZero} & \nodata & \McLowFeHPaperBoundsLogGBicBinaZero \\
\hline
$M_c$ & Young & \McHighActPaperBoundsNstars & \McHighActPaperBoundsNeffBinAZero & \McHighActPaperBoundsAvgComplBinAZero & sigmoid & \McHighActPaperBoundsSigmoidIntOccBinaZero & \shortstack{$B_1$\\\McHighActPaperBoundsSigmoidParamBOneBinaZero} & \shortstack{$B_2$\\\McHighActPaperBoundsSigmoidParamBTwoBinaZero} & \shortstack{$\log(x_t)$\\\McHighActPaperBoundsSigmoidParamCenterBinaZero} & \shortstack{$W$\\\McHighActPaperBoundsSigmoidParamWidthBinaZero} & \McHighActPaperBoundsSigmoidBicBinaZero \\
 &  &  &  &  & logG & \McHighActPaperBoundsLogGIntOccBinaZero & \shortstack{$A$\\\McHighActPaperBoundsLogGParamABinaZero} & \shortstack{$\mu$\\\McHighActPaperBoundsLogGParamMuBinaZero} & \shortstack{$\sigma$\\\McHighActPaperBoundsLogGParamSigmaBinaZero} & \nodata & \McHighActPaperBoundsLogGBicBinaZero \\
\hline
$M_c$ & Old & \McLowActPaperBoundsNstars & \McLowActPaperBoundsNeffBinAZero & \McLowActPaperBoundsAvgComplBinAZero & sigmoid & \McLowActPaperBoundsSigmoidIntOccBinaZero & \shortstack{$B_1$\\\McLowActPaperBoundsSigmoidParamBOneBinaZero} & \shortstack{$B_2$\\\McLowActPaperBoundsSigmoidParamBTwoBinaZero} & \shortstack{$\log(x_t)$\\\McLowActPaperBoundsSigmoidParamCenterBinaZero} & \shortstack{$W$\\\McLowActPaperBoundsSigmoidParamWidthBinaZero} & \McLowActPaperBoundsSigmoidBicBinaZero \\
 &  &  &  &  & logG & \McLowActPaperBoundsLogGIntOccBinaZero & \shortstack{$A$\\\McLowActPaperBoundsLogGParamABinaZero} & \shortstack{$\mu$\\\McLowActPaperBoundsLogGParamMuBinaZero} & \shortstack{$\sigma$\\\McLowActPaperBoundsLogGParamSigmaBinaZero} & \nodata & \McLowActPaperBoundsLogGBicBinaZero \\
\hline
\tablebreak
$q$ & All Stars & \QAllstarsPaperBoundsNstars & \QAllstarsPaperBoundsNeffBinAZero & \QAllstarsPaperBoundsAvgComplBinAZero & sigmoid & \QAllstarsPaperBoundsSigmoidIntOccBinaZero & \shortstack{$B_1$\\\QAllstarsPaperBoundsSigmoidParamBOneBinaZero} & \shortstack{$B_2$\\\QAllstarsPaperBoundsSigmoidParamBTwoBinaZero} & \shortstack{$\log(x_t)$\\\QAllstarsPaperBoundsSigmoidParamCenterBinaZero} & \shortstack{$W$\\\QAllstarsPaperBoundsSigmoidParamWidthBinaZero} & \QAllstarsPaperBoundsSigmoidBicBinaZero \\
 &  &  &  &  & logG & \QAllstarsPaperBoundsLogGIntOccBinaZero & \shortstack{$A$\\\QAllstarsPaperBoundsLogGParamABinaZero} & \shortstack{$\mu$\\\QAllstarsPaperBoundsLogGParamMuBinaZero} & \shortstack{$\sigma$\\\QAllstarsPaperBoundsLogGParamSigmaBinaZero} & \nodata & \QAllstarsPaperBoundsLogGBicBinaZero \\
 &  &  &  &  & escarpment & \QAllstarsPaperBoundsNoGPEscarpmentIntOccBinaZero & \shortstack{$D_1$\\\QAllstarsPaperBoundsNoGPEscarpmentParamDOneBinaZero} & \shortstack{$D_2$\\\QAllstarsPaperBoundsNoGPEscarpmentParamDTwoBinaZero} & \shortstack{$\log(x_{t,1})$\\\QAllstarsPaperBoundsNoGPEscarpmentParamBPOneBinaZero} & \shortstack{$\log(x_{t,2})$\\\QAllstarsPaperBoundsNoGPEscarpmentParamBPTwoBinaZero} & \QAllstarsPaperBoundsNoGPEscarpmentBicBinaZero \\
\hline
$q$ & High $M_\star$ & \QHighMstarPaperBoundsNstars & \QHighMstarPaperBoundsNeffBinAZero & \QHighMstarPaperBoundsAvgComplBinAZero & sigmoid & \QHighMstarPaperBoundsSigmoidIntOccBinaZero & \shortstack{$B_1$\\\QHighMstarPaperBoundsSigmoidParamBOneBinaZero} & \shortstack{$B_2$\\\QHighMstarPaperBoundsSigmoidParamBTwoBinaZero} & \shortstack{$\log(x_t)$\\\QHighMstarPaperBoundsSigmoidParamCenterBinaZero} & \shortstack{$W$\\\QHighMstarPaperBoundsSigmoidParamWidthBinaZero} & \QHighMstarPaperBoundsSigmoidBicBinaZero \\
 &  &  &  &  & logG & \QHighMstarPaperBoundsLogGIntOccBinaZero & \shortstack{$A$\\\QHighMstarPaperBoundsLogGParamABinaZero} & \shortstack{$\mu$\\\QHighMstarPaperBoundsLogGParamMuBinaZero} & \shortstack{$\sigma$\\\QHighMstarPaperBoundsLogGParamSigmaBinaZero} & \nodata & \QHighMstarPaperBoundsLogGBicBinaZero \\
\hline
$q$ & Low $M_\star$ & \QLowMstarPaperBoundsNstars & \QLowMstarPaperBoundsNeffBinAZero & \QLowMstarPaperBoundsAvgComplBinAZero & sigmoid & \QLowMstarPaperBoundsSigmoidIntOccBinaZero & \shortstack{$B_1$\\\QLowMstarPaperBoundsSigmoidParamBOneBinaZero} & \shortstack{$B_2$\\\QLowMstarPaperBoundsSigmoidParamBTwoBinaZero} & \shortstack{$\log(x_t)$\\\QLowMstarPaperBoundsSigmoidParamCenterBinaZero} & \shortstack{$W$\\\QLowMstarPaperBoundsSigmoidParamWidthBinaZero} & \QLowMstarPaperBoundsSigmoidBicBinaZero \\
 &  &  &  &  & logG & \QLowMstarPaperBoundsLogGIntOccBinaZero & \shortstack{$A$\\\QLowMstarPaperBoundsLogGParamABinaZero} & \shortstack{$\mu$\\\QLowMstarPaperBoundsLogGParamMuBinaZero} & \shortstack{$\sigma$\\\QLowMstarPaperBoundsLogGParamSigmaBinaZero} & \nodata & \QLowMstarPaperBoundsLogGBicBinaZero \\
\hline
$q$ & High [Fe/H] & \QHighFeHPaperBoundsNstars & \QHighFeHPaperBoundsNeffBinAZero & \QHighFeHPaperBoundsAvgComplBinAZero & sigmoid & \QHighFeHPaperBoundsSigmoidIntOccBinaZero & \shortstack{$B_1$\\\QHighFeHPaperBoundsSigmoidParamBOneBinaZero} & \shortstack{$B_2$\\\QHighFeHPaperBoundsSigmoidParamBTwoBinaZero} & \shortstack{$\log(x_t)$\\\QHighFeHPaperBoundsSigmoidParamCenterBinaZero} & \shortstack{$W$\\\QHighFeHPaperBoundsSigmoidParamWidthBinaZero} & \QHighFeHPaperBoundsSigmoidBicBinaZero \\
 &  &  &  &  & logG & \QHighFeHPaperBoundsLogGIntOccBinaZero & \shortstack{$A$\\\QHighFeHPaperBoundsLogGParamABinaZero} & \shortstack{$\mu$\\\QHighFeHPaperBoundsLogGParamMuBinaZero} & \shortstack{$\sigma$\\\QHighFeHPaperBoundsLogGParamSigmaBinaZero} & \nodata & \QHighFeHPaperBoundsLogGBicBinaZero \\
\hline
$q$ & Low [Fe/H] & \QLowFeHPaperBoundsNstars & \QLowFeHPaperBoundsNeffBinAZero & \QLowFeHPaperBoundsAvgComplBinAZero & sigmoid & \QLowFeHPaperBoundsSigmoidIntOccBinaZero & \shortstack{$B_1$\\\QLowFeHPaperBoundsSigmoidParamBOneBinaZero} & \shortstack{$B_2$\\\QLowFeHPaperBoundsSigmoidParamBTwoBinaZero} & \shortstack{$\log(x_t)$\\\QLowFeHPaperBoundsSigmoidParamCenterBinaZero} & \shortstack{$W$\\\QLowFeHPaperBoundsSigmoidParamWidthBinaZero} & \QLowFeHPaperBoundsSigmoidBicBinaZero \\
 &  &  &  &  & logG & \QLowFeHPaperBoundsLogGIntOccBinaZero & \shortstack{$A$\\\QLowFeHPaperBoundsLogGParamABinaZero} & \shortstack{$\mu$\\\QLowFeHPaperBoundsLogGParamMuBinaZero} & \shortstack{$\sigma$\\\QLowFeHPaperBoundsLogGParamSigmaBinaZero} & \nodata & \QLowFeHPaperBoundsLogGBicBinaZero \\
\hline
$q$ & Young & \QHighActPaperBoundsNstars & \QHighActPaperBoundsNeffBinAZero & \QHighActPaperBoundsAvgComplBinAZero & sigmoid & \QHighActPaperBoundsSigmoidIntOccBinaZero & \shortstack{$B_1$\\\QHighActPaperBoundsSigmoidParamBOneBinaZero} & \shortstack{$B_2$\\\QHighActPaperBoundsSigmoidParamBTwoBinaZero} & \shortstack{$\log(x_t)$\\\QHighActPaperBoundsSigmoidParamCenterBinaZero} & \shortstack{$W$\\\QHighActPaperBoundsSigmoidParamWidthBinaZero} & \QHighActPaperBoundsSigmoidBicBinaZero \\
 &  &  &  &  & logG & \QHighActPaperBoundsLogGIntOccBinaZero & \shortstack{$A$\\\QHighActPaperBoundsLogGParamABinaZero} & \shortstack{$\mu$\\\QHighActPaperBoundsLogGParamMuBinaZero} & \shortstack{$\sigma$\\\QHighActPaperBoundsLogGParamSigmaBinaZero} & \nodata & \QHighActPaperBoundsLogGBicBinaZero \\
\hline
$q$ & Old & \QLowActPaperBoundsNstars & \QLowActPaperBoundsNeffBinAZero & \QLowActPaperBoundsAvgComplBinAZero & sigmoid & \QLowActPaperBoundsSigmoidIntOccBinaZero & \shortstack{$B_1$\\\QLowActPaperBoundsSigmoidParamBOneBinaZero} & \shortstack{$B_2$\\\QLowActPaperBoundsSigmoidParamBTwoBinaZero} & \shortstack{$\log(x_t)$\\\QLowActPaperBoundsSigmoidParamCenterBinaZero} & \shortstack{$W$\\\QLowActPaperBoundsSigmoidParamWidthBinaZero} & \QLowActPaperBoundsSigmoidBicBinaZero \\
 &  &  &  &  & logG & \QLowActPaperBoundsLogGIntOccBinaZero & \shortstack{$A$\\\QLowActPaperBoundsLogGParamABinaZero} & \shortstack{$\mu$\\\QLowActPaperBoundsLogGParamMuBinaZero} & \shortstack{$\sigma$\\\QLowActPaperBoundsLogGParamSigmaBinaZero} & \nodata & \QLowActPaperBoundsLogGBicBinaZero \\
\hline
\enddata

\tablenotetext{}{%
\textbf{Note.} Stellar mass, metallicity, and age subsamples are divided at $1~\Msun$, 0.0 dex, and 5.0 Gyr, respectively. The old and young subsamples contain only the 378 CLS hosts with modeled ages. $N_\star$ is the number of stars, $N_{\mathrm{eff}}$ the total number of fractional companions, and $Q_{\text{avg}}$ the average completeness over the region of interest: $0.4-50~\Mjup$ (or the equivalent mass ratios) and $0.1-10$ AU. Integrated occurrence rates have units of companions per star. The amplitude parameters $A$, $B$, $C$, and $D$ have units of companions per star per $\Delta\log a$ per $\Delta\log~M_c$ (or per $\Delta\log~q$ for mass ratio fits). We omit the integrated occurrence rate and BIC for the log-linear model because we fit it to only a subset of the catalog. We fit the escarpment model in only two experiments: the full sample in companion mass and the full sample in mass ratio.
}
\end{deluxetable*}
\end{longrotatetable}
\begin{longrotatetable}
\begin{deluxetable*}{ccccccccccccc}
\tablecaption{Histogram Occurrence Rate Densities}
\label{tab:ord_appendix}

\tablehead{
\shortstack{Mass\\Variable} &
\shortstack{Stellar\\Subsample} &
$N_\star$ &
\shortstack{$N_{\mathrm{eff}}$} &
$Q_{\text{avg}}$ &
\shortstack{Integrated\\Rate} &
Bin 1 & Bin 2 & Bin 3 & Bin 4 & Bin 5 & Bin 6 & Bin 7
}
\startdata
 &  &  &  &  &  & 0.4--0.8~\Mjup & 0.8--1.6~\Mjup & 1.6--3.2~\Mjup & 3.2--6.4~\Mjup & 6.4--13~\Mjup & 13--26~\Mjup & 26--50~\Mjup \\
\hline
$M_c$ & All Stars & 719 & 99.1 & 0.77 & \McAllstarsPaperBoundsPiecewiseIntOccBinaZero & $0.05^{+0.02}_{-0.01}$ & $0.08^{+0.01}_{-0.01}$ & $0.08^{+0.01}_{-0.01}$ & $0.057^{+0.010}_{-0.009}$ & $0.027^{+0.007}_{-0.005}$ & $0.010^{+0.003}_{-0.003}$ & $0.003^{+0.003}_{-0.002}$ \\
$M_c$ & High $M_\star$ & 229 & 60.1 & 0.73 & \McHighMstarPaperBoundsPiecewiseIntOccBinaZero & $0.07^{+0.04}_{-0.02}$ & $0.12^{+0.04}_{-0.02}$ & $0.16^{+0.03}_{-0.03}$ & $0.14^{+0.03}_{-0.03}$ & $0.07^{+0.01}_{-0.02}$ & $0.021^{+0.010}_{-0.008}$ & $0.005^{+0.007}_{-0.003}$ \\
$M_c$ & Low $M_\star$ & 490 & 39.0 & 0.79 & \McLowMstarPaperBoundsPiecewiseIntOccBinaZero & $0.05^{+0.02}_{-0.02}$ & $0.047^{+0.018}_{-0.008}$ & $0.038^{+0.009}_{-0.010}$ & $0.021^{+0.006}_{-0.006}$ & $0.012^{+0.003}_{-0.005}$ & $0.005^{+0.003}_{-0.002}$ & $0.003^{+0.003}_{-0.002}$ \\
$M_c$ & High [Fe/H] & 395 & 83.2 & 0.77 & \McHighFeHPaperBoundsPiecewiseIntOccBinaZero & $0.07^{+0.02}_{-0.02}$ & $0.11^{+0.02}_{-0.02}$ & $0.13^{+0.02}_{-0.02}$ & $0.10^{+0.02}_{-0.01}$ & $0.044^{+0.010}_{-0.009}$ & $0.015^{+0.005}_{-0.006}$ & $0.003^{+0.004}_{-0.002}$ \\
$M_c$ & Low [Fe/H] & 324 & 15.7 & 0.78 & \McLowFeHPaperBoundsPiecewiseIntOccBinaZero & $0.022^{+0.017}_{-0.009}$ & $0.023^{+0.011}_{-0.008}$ & $0.018^{+0.006}_{-0.006}$ & $0.012^{+0.005}_{-0.004}$ & $0.009^{+0.005}_{-0.004}$ & $0.005^{+0.005}_{-0.003}$ & $0.003^{+0.006}_{-0.002}$ \\
$M_c$ & Young & 188 & 18.3 & 0.75 & \McHighActPaperBoundsPiecewiseIntOccBinaZero & $0.03^{+0.02}_{-0.01}$ & $0.04^{+0.02}_{-0.01}$ & $0.034^{+0.018}_{-0.009}$ & $0.031^{+0.017}_{-0.009}$ & $0.028^{+0.014}_{-0.009}$ & $0.012^{+0.017}_{-0.005}$ & $0.002^{+0.019}_{-0.002}$ \\
$M_c$ & Old & 190 & 53.0 & 0.81 & \McLowActPaperBoundsPiecewiseIntOccBinaZero & $0.08^{+0.03}_{-0.03}$ & $0.13^{+0.03}_{-0.03}$ & $0.17^{+0.02}_{-0.04}$ & $0.13^{+0.02}_{-0.03}$ & $0.06^{+0.01}_{-0.01}$ & $0.019^{+0.011}_{-0.008}$ & $0.005^{+0.008}_{-0.004}$ \\
\hline
 &  &  &  &  &  & 0.038--0.076\% & 0.076--0.15\% & 0.15--0.31\% & 0.31--0.61\% & 0.61--1.2\% & 1.2--2.5\% & 2.5--4.8\% \\
\hline
$q$ & All Stars & 719 & 101.2 & 0.75 & \QAllstarsPaperBoundsPiecewiseIntOccBinaZero & $0.06^{+0.02}_{-0.01}$ & $0.09^{+0.01}_{-0.01}$ & $0.09^{+0.01}_{-0.01}$ & $0.063^{+0.008}_{-0.013}$ & $0.027^{+0.007}_{-0.005}$ & $0.008^{+0.004}_{-0.003}$ & $0.002^{+0.002}_{-0.001}$ \\
$q$ & High $M_\star$ & 229 & 59.3 & 0.75 & \QHighMstarPaperBoundsPiecewiseIntOccBinaZero & $0.09^{+0.03}_{-0.03}$ & $0.14^{+0.03}_{-0.03}$ & $0.16^{+0.03}_{-0.03}$ & $0.13^{+0.03}_{-0.03}$ & $0.052^{+0.021}_{-0.009}$ & $0.018^{+0.008}_{-0.009}$ & $0.004^{+0.007}_{-0.003}$ \\
$q$ & Low $M_\star$ & 490 & 41.9 & 0.75 & \QLowMstarPaperBoundsPiecewiseIntOccBinaZero & $0.04^{+0.02}_{-0.01}$ & $0.06^{+0.01}_{-0.01}$ & $0.05^{+0.01}_{-0.01}$ & $0.026^{+0.009}_{-0.006}$ & $0.013^{+0.007}_{-0.003}$ & $0.005^{+0.004}_{-0.003}$ & $0.001^{+0.003}_{-0.001}$ \\
$q$ & High [Fe/H] & 395 & 86.1 & 0.75 & \QHighFeHPaperBoundsPiecewiseIntOccBinaZero & $0.09^{+0.02}_{-0.03}$ & $0.12^{+0.03}_{-0.02}$ & $0.13^{+0.02}_{-0.02}$ & $0.10^{+0.02}_{-0.02}$ & $0.043^{+0.012}_{-0.008}$ & $0.013^{+0.007}_{-0.005}$ & $0.003^{+0.004}_{-0.002}$ \\
$q$ & Low [Fe/H] & 324 & 15.1 & 0.75 & \QLowFeHPaperBoundsPiecewiseIntOccBinaZero & $0.020^{+0.020}_{-0.009}$ & $0.03^{+0.01}_{-0.01}$ & $0.021^{+0.010}_{-0.008}$ & $0.012^{+0.008}_{-0.004}$ & $0.008^{+0.004}_{-0.005}$ & $0.002^{+0.004}_{-0.002}$ & $0.0008^{+0.0039}_{-0.0008}$ \\
$q$ & Young & 188 & 18.4 & 0.75 & \QHighActPaperBoundsPiecewiseIntOccBinaZero & $0.03^{+0.02}_{-0.02}$ & $0.04^{+0.02}_{-0.01}$ & $0.034^{+0.023}_{-0.007}$ & $0.033^{+0.017}_{-0.009}$ & $0.03^{+0.01}_{-0.01}$ & $0.02^{+0.01}_{-0.01}$ & $0.003^{+0.020}_{-0.003}$ \\
$q$ & Old & 190 & 53.0 & 0.82 & \QLowActPaperBoundsPiecewiseIntOccBinaZero & $0.08^{+0.03}_{-0.03}$ & $0.14^{+0.03}_{-0.03}$ & $0.15^{+0.04}_{-0.03}$ & $0.11^{+0.03}_{-0.02}$ & $0.05^{+0.02}_{-0.01}$ & $0.017^{+0.010}_{-0.008}$ & $0.004^{+0.008}_{-0.003}$ \\
\hline
\enddata

\tablenotetext{}{%
\textbf{Note.} Each bin lists the posterior mode of the histogram model's occurrence rate density (ORD) and its 68\% highest-density interval, in companions per star per $\Delta\log a$ per $\Delta\log$ of the mass variable. All fits were performed using GP smoothing. Integrated rates have units of companions per star. Mass bin limits are exact; mass ratio bin limits, equal to the corresponding mass limit divided by $1~\Msun$, have been rounded for concision. See Table \ref{tab:model_params_appendix} for the subsample divisions and the definitions of $N_\star$, $N_{\mathrm{eff}}$, and $Q_{\text{avg}}$.
}
\end{deluxetable*}
\end{longrotatetable}
 
\bibliography{main}{}
\bibliographystyle{aasjournalv7}

\end{document}